\documentclass[
a4paper,
reprint,
showpacs,
showkeys,
 amsmath,amssymb,
 aps,
 prc,
floatfix,
]{revtex4-1}

\usepackage[utf8]{inputenc}
\usepackage{graphicx}% Include figure files
\usepackage{color}
\usepackage{qmech-revtex}

\begin{document}

\preprint{}

\title{Photon and dilepton production at the Facility for Antiproton and Ion Research and the beam energy scan program at the Relativistic Heavy-Ion Collider using coarse-grained microscopic transport simulations}

\author{Stephan Endres}
 \email{endres@th.physik.uni-frankfurt.de}
\author{Hendrik van Hees}%
\author{Marcus Bleicher}%
\affiliation{%
Frankfurt Institute for Advanced Studies,
Ruth-Moufang-Straße 1, D-60438 Frankfurt, Germany
}%
\affiliation{
Institut f{\"u}r Theoretische Physik, Universit{\"a}t Frankfurt,
Max-von-Laue-Straße 1, D-60438 Frankfurt, Germany
}

\date{May 2, 2016}

\begin{abstract}
  We present calculations of dilepton and photon spectra for the energy
  range $E_{\text{lab}}=2-35$\,$A$GeV which will be available for the
  Compressed Baryonic Matter (CBM) experiment at the future Facility for
  Anti-Proton and Ion Research (FAIR). The same energy regime
  will also be covered by phase II of the Beam Energy Scan at the 
  Relativistic Heavy-Ion Collider (RHIC-BES). Coarse-grained dynamics from 
  microscopic transport calculations of the Ultra-relativistic Quantum
  Molecular Dynamics (UrQMD) model is used to determine temperature and
  chemical potentials, which allows for the use of dilepton and photon-emission
  rates from equilibrium quantum-field theory calculations. The results
  indicate that non-equilibrium effects, the presence of baryonic matter
  and the creation of a deconfined phase might show up in specific
  manners in the measurable dilepton invariant mass spectra and in the
  photon transverse momentum spectra. However, as the many influences
  are difficult to disentangle, we argue that the challenge for future
  measurements of electromagnetic probes will be to provide a high
  precision with uncertainties much lower than in previous
  experiments. Furthermore, a systematic study of the whole energy range
  covered by CBM at FAIR and RHIC-BES is necessary to discriminate between different 
  effects, which influence the spectra, and to identify possible signatures 
  of a phase transition.
\end{abstract}



\maketitle

%\tableofcontents
%%%%%%%%%%%%%%%%%%%%%%%%%%%%%%%%%%%%%%%%%%%%%%%%%%%%%%%%%%%%%%
%%%%%%%%%%%%%%%%%%%%%%%%%%%%%%%%%%%%%%%%%%%%%%%%%%%%%%%%%%%%%%
\section{\label{sec:Intro} Introduction }

A major goal of the study of heavy-ion collisions at relativistic and
ultra-relativistic collision energies is to explore the properties of
strongly interacting matter at finite temperatures and densities
\cite{Stoecker:1986ci, Danielewicz:2002pu}. When two colliding nuclei
hit each other, the nuclear matter is compressed, and a large amount of
energy is deposited in a small spatial volume. This results in the
creation of a fireball of hot and dense matter \cite{Bjorken:1982qr,
  Stock:2008ru}. The fireball lives for a time span of the order of
several $\fm/c$ until the collective expansion of the matter has driven
the strongly interacting system to a final state of freely streaming
particles.

Today, almost the entire phase diagram governed by Quantum
Chromodynamics (QCD) is accessible for experimental exploration at various
accelerator facilities. The temperature $T$ and
baryochemical potential $\mu_{\text{B}}$ inside the fireball are mainly
determined by the energy which is deposited in the nuclear collision;
more precisely, the collision energy determines the trajectory of the
system in the $T-\mu_{\mathrm{B}}$ plane of the QCD phase diagram. At the highest
currently available energies at RHIC and LHC the reaction is dominated
by high temperatures, significantly above the critical temperature
$T_{c}$, for which the creation of a deconfined state of quarks
and gluons is assumed. At the same time the baryochemical potential is
low or close to zero for the largest part of the fireball
evolution. This situation is similar to the conditions which prevailed
in the universe a short time after the big bang. On the other side, one
finds a complementary situation if considering heavy-ion collisions at
laboratory frame energies of the order of 1 $A\,\GeV$. Here only moderate temperatures are obtained, insufficient to create a
Quark-Gluon Plasma. However, the very high net
baryon densities or baryochemical potentials reached in this case might provide
valuable information about those effects which are not mainly driven by
temperature but by the presence of compressed baryonic matter. This
situation resembles the environments in (super)nova explosions and
neutron stars.

To learn about the different regions of the phase diagram one needs
observables which do not only reflect the diluted final state after the
freeze-out of the system but rather convey information about the entire
fireball evolution. For this purpose electromagnetic probes, i.e.,
photons and dileptons have been for long suggested as ideal probes
\cite{Feinberg:1976ua, Shuryak:1978ij}: Once produced, photons and
dileptons only participate in electromagnetic and weak interactions for
which the mean free paths are much longer than the size and the lifetime
of the fireball. Consequently, they can leave the zone of hot and dense
matter undisturbed. Since electromagnetic probes are emitted in a large
variety of processes over the whole lifetime of the fireball, the
measured spectra reflect the time-integrated evolution of the
thermodynamic properties of the system. While this allows to obtain
convoluted information about the properties of matter it also poses a
serious challenge for the theoretical description. On the one hand, one
needs to identify the relevant microscopic processes that contribute to
dilepton and photon emission and to determine the corresponding
production rates. On the other hand it is important to give a realistic
description of the complete reaction dynamics.

The intense experimental study of photon and dilepton production in the
high-energy regime (at SPS \cite{Adamova:2002kf, Aggarwal:2000th,
  Arnaldi:2006jq}, RHIC \cite{Adamczyk:2013caa, Adamczyk:2013caa,
  Adler:2005ig, Adare:2008ab, Adare:2015ila}, and LHC
\cite{Koehler:2014dba} energies), but also for very low collision
energies as measured at SIS\,18 and BEVALAC \cite{Porter:1997rc,
  Agakichiev:2006tg, Agakishiev:2007ts, Agakishiev:2011vf} in
comparison to theoretical model calculations has significantly
enhanced our knowledge of the reaction dynamics and the properties of
matter in the hot and dense medium created in a heavy-ion reaction. The
importance of partonic emission for the correct theoretical description
of the high-mass region of dilepton invariant mass spectra and the
high-$p_{\mathrm{t}}$ photon spectra has been pointed out
\cite{Bauchle:2010ym, vanHees:2014ida, Linnyk:2015tha} and the various
different hadronic contributions (especially for the photon production
channels) could be identified \cite{Turbide:2003si, Liu:2007zzw,
  Holt:2015cda}. Nevertheless, the most important finding was the large
influence of the baryonic matter on the vector mesons' spectral
shape. Especially in the case of the $\rho$ meson this causes a strong
broadening of the spectral function with small mass shifts
\cite{Rapp:1999ej, Rapp:2013nxa, Endres:2014zua, Endres:2015fna}. This
effect has been observed as an enhancement in the low-mass region of the
dilepton invariant mass spectra and also shows up as a stronger
low-momentum thermal photon yield. Note that the $\rho$ broadening is most
dominant at low collision energies, where one obtains the largest
baryochemical potentials, but even at RHIC energies baryonic
effects are by far not negligible.
 
However, there still remains an up to now unexplored energy window
between the $E_{\mathrm{lab}}=1-2\,A \,\GeV$ dilepton measurements by
the DLS and HADES Collaborations and the CERES results for
$E_{\mathrm{lab}}=40 A\,\GeV$. The future Compressed Baryonic Matter
(CBM) experiment at the Facility for Anti-Proton and Ion Research (FAIR)
with the SIS 100/300 accelerator provides the unique possibility to study
heavy-ion collisions with beam energies from 2 up to $35 A\,\GeV$ and
will therefore enable us to get an insight into exactly that regime of the
phase diagram of highest baryon densities where no dilepton or photon 
measurements have been performed till now \cite{Friman:2011zz,Senger:2015wpa}. 
In addition, also phase II of the Beam Energy 
Scan (BES) program at RHIC will allow to perform measurements in fixed-target 
mode at lab-frame energies of 7.7, 9.1, 11.5, 14.5 and $19.6 A\,\GeV$, 
i.e., in the same collision energy range as FAIR \cite{Odyniec:2015iaa}. 
Further complementary investigations are also planned for the NICA in 
Dubna \cite{Sissakian:2009zza}.

From a theoretical point of view, the handling of this energy range is
quite challenging, as the transition from a purely hadronic fireball at
low collision energies to the creation of a partonic phase is expected
here. Furthermore, at the high baryochemical potentials which still
dominate the fireball at these energies, a first order phase transition
from a hadron gas to the QGP is assumed, in contrast to the situation at
RHIC or the LHC where a cross-over is predicted by Lattice QCD
calculations \cite{Fischer:2011mz}.

Although transport models were applied successfully to describe
electromagnetic observables in heavy-ion collisions
\cite{Cassing:1999es, Schmidt:2008hm, Weil:2012ji}, they generally have
some shortcomings when describing very hot and dense systems. In detail,
problems include the following aspects:
Firstly, while the Boltzmann approach works quite well for
quasi-particles of infinite lifetime, for broad resonances as the $\rho$
meson a correct description is challenging. Furthermore, in dense
matter the intervals between scatterings become extremely short and will
consequently modify the spectral characteristics of the single
particles (collisional broadening). To describe the off-shell dynamics correctly a transport
description with dynamical spectral functions following the description
of Kadanoff and Baym \cite{KadanoffBaym1962} is required. However, a
practical implementation of this is currently not possible. Secondly, in a dense
medium not only binary scatterings will occur but also multi-particle
interactions play a role, which is beyond the capabilities of the common
transport models. And finally, the microscopic models usually
concentrate on either the transport of hadrons or partons. However,
modelling a transition from an initially up-heating hadron gas to a
deconfined phase and the later particlization when the system cools down
is extremely difficult to realize within a transport approach.

There have been several investigations over the last years on these
aspects (see, e.g., Refs.\,\cite{Bratkovskaya:1997mp, Bratkovskaya:2008bf,
  Bratkovskaya:2013vx, Schenke:2005ry, Schenke:2006uh, Weil:2012qh, Berges:2015ixa}),
but a full treatment of all these issues is still beyond
the scope of present investigations.

On the other side, the short mean free paths of particles in a medium
might suggest to treat the reactions from a macroscopic
point-of-view. However, approaches as simple fireball expansion models
\cite{vanHees:2006ng} or hydrodynamics \cite{Vujanovic:2013jpa}, which
have been successfully applied for SPS, RHIC and LHC energies, also have
their shortcomings in the FAIR energy regime for three main reasons:
Firstly, the separation of the fireball expansion from dynamics of the
initial projectile-target dynamics is not applicable; secondly, the often
applied simplification to assume a 2+1-dimensional boost invariant
geometry is not possible; and finally, the time scale necessary for an
approximate thermal equilibration of the fireball will be longer due to
the slower overall evolution of the reaction and the lower temperatures
reached.

To avoid the disadvantages of both pictures the coarse-graining method
has been developed, based on previous studies \cite{Huovinen:2002im}, 
and was successfully applied to describe
dilepton production at SPS and SIS\,18 energies \cite{Endres:2014zua,
  Endres:2015fna, Steinheimer:2016vzu}. The approach represents a combination of the
microscopic picture from the underlying transport simulations with the
resulting description of the dynamics in terms of the macroscopic
quantities temperature and chemical potential. By averaging over many
events one can extract the local energy and baryon densities at each
space-time point from the transport simulations and use an equation of
state to determine the corresponding temperature and baryochemical
potential. With this the calculation method of thermal dilepton and photon
emission by application of full in-medium spectral functions is straightforward,
employing the rates available from equilibrium quantum field theory.

In the present work the coarse-graining approach is used to calculate
photon and dilepton spectra with focus the FAIR energy regime, but naturally the results also serve as a
theoretical prediction for the fixed-target measurements of the RHIC-BES since the prospected collision 
energies of both experimental programs overlap.
Although the details of the future experimental set-ups are not yet determined, the
results shall provide a general baseline calculation for the
interpretation of the measurements to be
conducted. Furthermore, it shall be investigated if and how one can
obtain valuable information on the properties of matter from the
measured spectra and discriminate between several effects that might
influence the dilepton and photon results. In detail, we will concentrate on
the following three aspects: The modification of the thermal emission
pattern by high baryochemical potentials, signals for a phase transition
or the creation of a deconfined phase and possible non-equilibrium
effects on the thermal rates.

This paper is structured as follows. In Section \ref{sec:Model} the
coarse-graining approach will be presented. Thereafter, in Section
\ref{sec:Rates} we will introduce the various microscopic sources for
thermal emission of photons and dileptons and in short discuss the
non-thermal cocktail contributions. In Section \ref{sec:Results} the
results for the fireball evolution and the photon and dilepton spectra
at FAIR energies are shown. The results are used to systematically
analyse in which way it might be possible to discriminate between
different scenarios for the fireball evolution in Section
\ref{sec:Scenario}. We conclude the present work with a summary and an
outlook to subsequent investigations.

%%%%%%%%%%%%%%%%%%%%%%%%%%%%%%%%%%%%%%%%%%%%%%%%%%%%%%%%%%%%%%
\section{\label{sec:Model} The model }

While microscopic transport models describe the reaction dynamics of a
heavy-ion collision in terms of many different degrees of freedom, the
general idea of the coarse-graining approach is that in principle only a
very reduced amount of the provided information is necessary to account
for the thermal production of electromagnetic probes. The microscopic
information about all individual particles and their specific properties
-- such as mass, charge and momentum -- are ignored and the whole
dynamics is reduced to macroscopic quantities which are assumed to fully
determine the local thermodynamic properties: The energy and particle
densities.

The coarse-graining method combines two advantages: On the one hand the
collision dynamics is still based on the microscopic transport evolution
and thereby gives a very nuanced picture of the entire collision
evolution, on the other hand the reduction to macroscopic state
variables enables an easy application of in-medium spectral functions
from equilibrium quantum field theory calculations.

In the following the ingredients of the approach are presented in detail.

\subsection{\label{ssec:UrQMD} Ultrarelativistic Quantum Molecular
  Dynamics }

The underlying microscopic input for the present calculations stems from
the Ultra-relativistic Quantum Molecular Dynamics (UrQMD) approach
\cite{Bass:1998ca, Bleicher:1999xi, Petersen:2008kb}. It is a
non-equilibrium microscopic transport model based on the principles of
molecular dynamics \cite{Bodmer:1977zz,Molitoris:1984xv}. It constitutes
an effective Monte Carlo solution to the relativistic Boltzmann equation
and connects the propagation of hadrons on covariant trajectories with a
probabilistic description of the hadron-hadron scattering processes. To
account for the quantum nature of the particles, each hadron is
represented by a Gaussian density distribution, and quantum statistical
effects such as Pauli blocking are considered \cite{Aichelin:1991xy}.

The model includes all relevant mesonic and baryonic resonances up to a
mass of $2.2 \,\GeV/c^{2}$. Production of particles occurs via resonant
scattering of particles (e.g., $NN \rightarrow N\Delta$ or
$\pi\pi \rightarrow \rho$) or the decay of higher resonances, e.g., the
process $\Delta \rightarrow \pi N$. The individual interaction and decay
processes are described in terms of measured and extrapolated hadronic
cross-sections and branching ratios. For collision energies above
$\sqrt{s_{NN}}=3\,\GeV$ also the excitation of strings is possible.

\subsection{\label{ssec:Coarse} Coarse-graining of microscopic dynamics }

Within the UrQMD model the particle distribution function
$f(\vec{x},\vec{p},t)$ is determined by the space and momentum
coordinates of all the different particles in the system at a
certain time. However, due to the finite number $h$ of hadronic
particles involved and produced in a heavy-ion collision, one needs to
take the average over a large ensemble of events to obtain a smooth
phase-space distribution of the form
\begin{equation}
f(\vec{x},\vec{p},t)=\left\langle \sum_{h} \delta^{(3)}(\vec{x}-\vec{x}_{h}(t)) \delta^{(3)}(\vec{p}-\vec{p}_{h}(t))\right\rangle.
\end{equation}
Note that this distribution is Lorentz invariant if all particles are on
the mass-shell, as provided in our case. Due to the non-equilibrium
nature of the model, one will of course have to extract the
particle distribution function locally. In the present approach, this is done by
the use of a grid of small space-time cells where for each of these
cells we determine the (net-)baryon four-flow and the
energy-momentum tensor according to the relations
\begin{alignat}{2}
  j^{\mu}_{\text{B}} &= \int d^{3}p\frac{p^{\mu}}{p^{0}}f^{\text{B}}(\vec{x},\vec{p},t), \label{eqjB} \\
  T^{\mu\nu}&= \int
  d^{3}p\frac{p^{\mu}p^{\nu}}{p^{0}}f(\vec{x},\vec{p},t). \label{eqTmn}
\end{alignat}
In practice, the integration is done by summing over the $\delta$
functions. As we use cells of finite size, we have
\begin{equation}
\delta^{(3)}(\vec{x}-\vec{x}_{h}(t)) = \begin{cases} \frac{1}{\Delta V}&
  \text{\ if \ } \vec{x}_{h}(t) \in \Delta V, \\ \ \ 0& \text{\ otherwise \ } \end{cases}
\end{equation}
and in the limit for small volumes the density of some observable $\hat{O}$ then becomes
\begin{equation}
%\begin{split}
  \int d^{3}p \ \hat{O}(\vec{x},\vec{p},t) f(\vec{x},\vec{p},t) =
  \frac{1}{\Delta V} \left\langle \sum\limits_{h}^{\vec{x}_{h} \in \Delta 
      V}\hat{O}(\vec{x},\vec{p},t) \right\rangle.
%\end{split}.
\end{equation}
Consequently, Eqns. (\ref{eqjB}) and (\ref{eqTmn}) take the form
\begin{equation}
\begin{split}
 T^{\mu\nu}&=\frac{1}{\Delta V}\left\langle \sum\limits_{i=1}^{N_{h} \in \Delta %
      V} \frac{p^{\mu}_{i}\cdot p^{\nu}_{i}}{p^{0}_{i}}\right\rangle, \\
   j^{\mu}_{\mathrm{B}}&=\frac{1}{\mathrm{\Delta} V}\left\langle
    \sum\limits_{i=1}^{N_{\mathrm{B}/\bar{\mathrm{B}}} \in \Delta
      V}\pm\frac{p^{\mu}_{i}}{p^{0}_{i}}\right\rangle.
\end{split}
\end{equation} 
Having obtained the baryon flow, we can boost each cell into the rest
frame as defined by Eckart \cite{Eckart:1940te}, where
$j^{\text{B}}_{\mu}$ is $(\rho_{\text{B}}, \vec{0})$. The according transformation of the
energy-momentum tensor provides the rest frame energy density.

\subsection{\label{ssec:Nonequilibrium} Non-equilibrium dynamics}

While macroscopic models usually introduce thermal and chemical
equilibrium as an ad-hoc assumption, microscopic simulations -- in the
present case the UrQMD simulations -- are based on the description of
single particle-particle interactions and non-equilibrium will be the
normal case. Consequently, we have to account for these deviations from
equilibrium in such a manner that we can reliably apply equilibrium
spectral functions to calculate the emission of photons and dileptons.

In general it is difficult to really determine to which degree a system
has reached equilibrium. Basically there are two dominant effects, which
may serve as indicators for thermal and chemical equilibration: The
momentum-space anisotropies and the appearance of meson-chemical
potentials.

\subsubsection{\label{sssec{thernoneq}} Thermal non-equilibrium}

Regarding thermal equilibration, it was found in microscopic simulations
that independent of the collision energy the system needs a time of
roughly $10\,\fm/c$ after the beginning of the heavy-ion collision until
the transverse and longitudinal pressures are approximately equal
\cite{Bravina:1999kd}. The pressure anisotropy stems from the initial
strong compression along the beam axis when the two nuclei first hit and
traverse each other. As thermal equilibrium requires isotropy, one will
obtain too high values for the energy density in highly anisotropic
cells. To obtain effective quantities that account for the
thermal properties in the system we apply a description that explicitly
includes the momentum-space anisotropies and in which the energy
momentum-tensor is assumed to take the form \cite{Florkowski:2010cf,
  Florkowski:2012pf}
\begin{equation}
T^{\mu \nu} = \left( \varepsilon  + P_{\perp}\right) u^{\mu}u^{\nu} -
P_{\perp} \, g^{\mu\nu} - (P_{\perp} - P_{\parallel}) v^{\mu}v^{\nu}. 
\label{Tmnaniso}
\end{equation}
where $P_{\perp}$ and $P_{\parallel}$ denote transverse or parallel
pressure components, respectively; $u^{\mu}$ and $v^{\mu}$ are the
cell's four-velocity and the four-vector of the beam direction. The
effective energy density $\varepsilon_{\text{eff}}$ is obtained via the
generalized equation of state for a Boltzmann-like system of the form
\begin{equation}
\varepsilon_{\text{eff}}=\frac{\varepsilon}{r(x)},
\end{equation}
where the relaxation function $r(x)$ and its derivative $r'(x)$ are defined by
\begin{equation} r(x) =\begin{cases} \label{relaxfunc}
\frac{x^{-1/3}}{2}\left(1+\frac{x \artanh \sqrt{1-x}}{\sqrt{1-x}}\right)
\text{for } x \leq 1 \\ \frac{x^{-1/3}}{2}\left(1+\frac{x \arctan
\sqrt{x-1}}{\sqrt{x-1}}\right) \text{for } x \geq 1
    \end{cases},
\end{equation} 
and $x=(P_{\parallel}/P_{\perp})^{3/4}$ denotes the pressure
anisotropy.

As we have shown in our previous investigation at SPS energies
\cite{Endres:2014zua}, with this description the effective energy
density deviates from the nominal one only for the very initial stage of
the reaction, where the pressure components differ by orders of
magnitude. Nevertheless, the effective energy density $\varepsilon_{\text{eff}}$ 
allows to calculate meaningful $T$ and $\mu_{\text{B}}$ values for these cells. After
the very initial collision phase the differences still exist, but have
hardly any influence on the energy density, so that we can assume that
these cells are in approximate local equilibrium.

\subsubsection{\label{sssec{chemnoneq}} Chemical non-equilibrium}

Chemical equilibration is a more difficult problem, but one obvious
deviation in microscopic models is the appearance of meson chemical
potentials, especially for the case of pions as these are the most
abundantly produced particles. As the meson number is not a conserved
quantity in strong interactions (in contrast to, e.g., the baryon
number) meson chemical potentials can only show up if the system is out
of chemical equilibrium. While pion chemical potentials are introduced
in fireball models for the stage after the chemical freeze-out to obtain
the correct final pion yields, in non-equilibrium transport models they
intrinsically appear in the early stages of the reaction when a large
number of pions is produced in many initial scattering processes
\cite{Bandyopadhyay:1993qj}. At higher collision energies this mainly
happens via string excitation. The pion (and kaon) chemical potentials
have a large influence on the photon and dilepton production rates as
an overpopulation of pions increases the reactions in many important channels,
for example $\pi\pi \rightarrow \rho \rightarrow \gamma / \gamma^{*}$
\cite{Koch:1992rs}.

To implement the non-equilibrium effects in the calculations, we extract the pion and kaon
chemical potentials in Boltzmann approximation as \cite{Sollfrank:1991xm}
\begin{equation}
  \mu_{\pi/K}= T \ln \left( \frac{2\pi^{2}n}{gTm^{2}
      \mathrm{K}_{2}\left(\frac{m}{T}\right)} \right),
\end{equation}
where $n$ denotes the cell's pion or kaon density and $\mathrm{K}_{2}$
the Bessel function of the second kind. The degeneracy factor $g$ is 3
in the case of pions and 2 for kaons. Note that the Boltzmann
approximation is in order here, as the mesons in the transport model
also account for Boltzmann statistics and no Bose effects are
implemented. However, while for a Bose gas the chemical potential is
limited to the meson's mass, in principle one can get higher values for
$\mu_{\pi}$ or $\mu_{\mathrm{K}}$ here in rare cases. As such values are
non-physical, we assume that the maximum values to be reached are 140 MeV
for $\mu_{\pi}$ and 450 MeV for $\mu_{\mathrm{K}}$.
%\end{equation}

\subsection{\label{ssec:EoS} Equation of state}

Once the rest-frame properties of each cell are determined, an equation
of state (EoS) is necessary to describe the thermodynamic system of the
hot and dense matter in the cell under the given set of state
variables, i.e., the (effective) local energy density and the local net densities of
conserved charges (for the strong interactions considered here the
baryon number is the relevant quantity). For the present calculations we
apply a hadron gas equation of state (HG-EoS) that includes the same
hadronic degrees of freedom as the underlying transport model
\cite{Zschiesche:2002zr}. The EoS allows us to extract the temperature and
baryochemical potential for an equilibrated hadron gas at
a given energy and baryon density. It is similar to the result
obtained for UrQMD calculations in a box in the infinite time limit,
when the system has settled to an equilibrated state.

However, in the FAIR energy regime a purely hadronic description of the
evolving hot and dense fireball will not be sufficient. As the
temperatures will exceed the critical temperature $T_{\mathrm{c}}$, a
transition from hadronic to partonic matter has to be implemented, and
the dynamic evolution of the created Quark-Gluon Plasma has to be
considered. On the other hand, it is necessary to keep the EoS
consistent with the underlying dynamics which is purely hadronic. In our
previous study at SPS energies \cite{Endres:2014zua}, we supplemented
the HG-EoS with a Lattice equation of state \cite{He:2011zx} for
temperatures above $T_{\mathrm{c}} \approx 170\,\MeV$, in line with the
lattice results. In the range around the critical temperature the
results of the HG-EoS and the Lattice EoS match very well for
$\mu_{\text{B}} \approx 0$, while significantly higher temperatures are
obtained with the latter if one reaches temperatures significantly above
$T_{\mathrm{c}}$.

However, this procedure is problematic for the present study, as the
transition from a hadronic to a partonic phase and back is assumed to
take place at finite values of $\mu_{\text{B}}$ at FAIR energies,
whereas the Lattice EoS is restricted to vanishing chemical
potential. To avoid discontinuities in the evolution, we confine
ourselves to the application of the HG-EoS, but with the assumption that
the thermal emission from cells with a temperature above 170\,MeV stems
from the QGP (i.e., we employ partonic emission rates). This should be in order, as the temperatures will not lie
too much above the critical temperature at the energies considered in
the present work, where the deviations from a full QCD-EoS explicitly
including a phase transition are expected to be rather moderate.

Nevertheless, we once again remind the reader that the underlying
microscopic description is purely hadronic and it remains to be studied
which consequences a phase transition has at the microscopic level of
the reaction dynamics.

%%%%%%%%%%%%%%%%%%%%%%%%%%%%%%%%%%%%%%%%%%%%%%%%%%%%%%%%%%%%%%
\section{\label{sec:Rates} Photon \& dilepton rates}

The mechanisms which contribute to the thermal emission of photons and
dileptons are the same. Any process that can produce a real photon
$\gamma$ can also produce a virtual (massive) photon $\gamma^{*}$,
decaying into a lepton pair. However, due to the different kinematic
regimes probed by photons and dileptons, the importance of the single
processes varies. In the following the various sources of thermal
radiation considered in this work are presented.

Determining quantity for the thermal emission of real photons as 
well as virtual photos (i.e., dileptons) is the imaginary part of 
the retarded electromagnetic current-current correlation function 
$\Pi_{em}^{(ret)}$, to which the rates are directly proportional. 
It represents a coherent summation of the cuts of those Feynman 
diagrams which are describing the different processes contributing 
to thermal $\gamma$ and $\gamma^{*}$ emission, and therefore accounts 
for the photon or dilepton self-energy. In the rest frame, the thermal 
emission can be calculated according to \cite{Rapp:1999ej}

\begin{alignat}{2}
\begin{split}
\label{rate_dil} \frac{\mathrm{d} N_{ll}}{\mathrm{d}^4x\mathrm{d}^4q} =&
-\frac{\alpha_\mathrm{em}^2 L(M)}{\pi^3 M^2} \; f_{\mathrm{B}}(q;T) \\ & \times 
\im \Pi^{(\text{ret})}_\mathrm{em}(M,
\vec{q};\mu_B,T),
\end{split} \\
\begin{split}
\label{rate_ph} q_{0}\frac{\mathrm{d} N_{\gamma}}{\mathrm{d}^4x\mathrm{d}^3q} =&
-\frac{\alpha_\mathrm{em}}{\pi^2} \; f_{\mathrm{B}}(q;T) \\ & \times 
\im \Pi^{T,(\text{ret})}_\mathrm{em}(q_{0}=|\vec{q}|;\mu_B,T).
\end{split}
\end{alignat}
Here $L(M^{2})$ is the lepton phase-space factor (which plays a
significant role only for masses close to the threshold $2 m_{l}$ and
is approximately one otherwise), $f_{\mathrm{B}}$ the Bose distribution
function, and $M$ the invariant mass of a lepton pair. Note that only the
transverse polarization of the current-current correlator enters for the
photon rate, as the longitudinal projection vanishes at the photon
point, i.e., for $M=0$.

\subsection{\label{ssec:ThermalRates} Thermal rates from hadronic matter}
\subsubsection{\label{sssec:vecmesspec} Vector meson spectral functions}

In hadronic matter, all the spectral information of a hadron with
certain quantum numbers is specified in
$\im \Pi^{(\text{ret})}_\mathrm{em}$ \cite{Leupold:2009kz}. Assuming
that Vector Meson Dominance (VMD) \cite{SakuraiBook} is valid, the
correlator can be directly related to the spectral functions of vector
mesons. The important challenge for theoretical models is to consider
the modifications of the particle's self-energy inside a hot and dense
medium. Different calculations of in-medium spectral functions exist
\cite{Klingl:1997kf, Post:2000qi, Cabrera:2000dx, Riek:2004kx,
  Ruppert:2004yg}, however not many of them fully consider the effects
of temperature and finite chemical potential. We here apply a hadronic
many-body calculation from thermal field theory for the spectral
functions of the $\rho$ and $\omega$ mesons \cite{Rapp:1997fs,
  Rapp:1999qu, Rapp:1999us}, which has proven to successfully describe
photon and dilepton spectra from SIS\,18 to LHC energies
\cite{vanHees:2006ng, vanHees:2007th, Turbide:2003si,vanHees:2011vb,
  vanHees:2014ida, Endres:2014zua, Endres:2015fna}. The calculation of
the different contributions to the $\rho$ spectral function takes three
different effects into account: The modification due to the pion cloud,
the direct scattering of the $\rho$ with baryons (nucleons as well as
excited $N^{*}$ and $\Delta^{*}$ resonances) and with mesons
($\pi, K, \rho,\dots$)\cite{Rapp:2009yu}. While the pion cloud effects
also contribute in the vacuum, the scattering processes only show up in
the medium. For the $\omega$, the situation is slightly more complex
since this meson basically constitutes a three-pion state. The vacuum
self-energy is represented by a combination of the decays into
$\rho + \pi$ or three pions, respectively. Further the inelastic
absorption $\omega\pi \rightarrow \pi\pi$ and the scattering processes
with baryons as well as the pion (i.e. $\omega\pi \rightarrow b_{1}$)
are implemented \cite{Rapp:2000pe}. In the same manner as in our
previous dilepton study at SIS\,18 energy \cite{Endres:2015fna}, we
relinquish a treatment of thermal emission from the $\phi$ vector meson
here, for reason of the minor in-medium broadening effects observed for
this hadron and its still low multiplicities at least for lower FAIR
energies. In the case of vanishing invariant mass, i.e., for real
photons, only the $\rho$ vector meson will give a significant
contribution. For the present calculation, we used the parametrization
of the photon rates from the $\rho$ as given in Eqs.\,(2)-(7) in
Ref.\,\cite{Heffernan:2014mla}, while a more advanced parametrization is
necessary for the dilepton rates from the $\rho$ and $\omega$, due to
their dependence on invariant mass and momenta \cite{RappSF}.

Note that presently the photon parametrization of the $\rho$
contribution is limited to baryon chemical potentials lower than
400\,MeV and momenta larger than $0.2\,\GeV/c$. While we can easily
neglect the lowest momentum region in the present study, the restriction
to low $\mu_{\text{B}}$ is a problem for the lowest collision energies
considered here, where one expects values of $\mu_{\mathrm{B}}$ which
significantly exceed this range. The difference will be dominant at
lower momenta, where the influence of baryonic effects is known to be
largest, while the effect of a finite chemical potential is rather small
at higher momenta \cite{Heffernan:2014mla}. However, for the present
work we can assume the photon contribution from the $\rho$ meson as a
lower limit with regard to the baryonic effects.

\subsubsection{\label{sssec:mesongas} Meson gas contributions}

The contribution from vector mesons is not the only hadronic source of
thermal emission. The mass region above the $\phi$ meson, i.e., for
$M > 1\,\GeV/c^{2}$, is no longer dominated by well defined particles,
but here one finds a large number of overlapping broad resonances
constituting multi-meson (mainly four-pion) states, which have a
significant impact on the dilepton yield. We here apply a description
relying on model-independent predictions using a low-temperature
expansion in the chiral-reduction approach \cite{vanHees:2007th}.

While the multi-meson effects only show up for high invariant masses,
i.e., in the time-like kinematic region probed by dileptons, when going
to real photons with $M \rightarrow 0$ several scattering and
bremsstrahlung processes become important, which can be mostly neglected
at finite $M$. While baryonic bremsstrahlung processes such as
$NN \rightarrow NN\gamma$ and $\pi N \rightarrow \pi N \gamma$ are
included in the vector meson spectral functions, meson-meson
bremsstrahlung has to be added in the case of photon emission. The most
dominant part will here come from the meson-meson scatterings
$\pi\pi \rightarrow \pi\pi\gamma$ and $\pi K \rightarrow \pi K\gamma$,
for which we use the rates calculated within an effective
hadronic model \cite{Liu:2007zzw} in form of the parametrization given
by Eqs.\,(8) and (9) in Ref.\,\cite{Heffernan:2014mla}. Note that these
bremsstrahlung processes are mainly contributing at low momenta, whereas
they are rather subleading for photons of higher energy.

Besides the $\pi\pi$ bremsstrahlung, also other mesonic reactions
contribute to the thermal photon production, such as strangeness bearing
reactions and meson-exchange processes. In detail, these are
$\pi\rho \rightarrow \pi\gamma$, $\pi K^{*} \rightarrow K\gamma$,
$\pi K \rightarrow K^{*}\gamma$, $\rho K \rightarrow K\gamma$ and
$K^{*} K \rightarrow \pi\gamma$. The corresponding thermal rates were
calculated for a hot meson gas in Ref.\,\cite{Turbide:2003si}, which are applied
here together with the respective form factors.

Since the $\omega$-$t$-channel exchange was found to give a
significant contribution to thermal photon spectra via the
$\pi\rho \rightarrow \pi\gamma$ process, it has been recently argued
that also other processes including a $\pi\rho\omega$ vertex should be
be considered in the calculations, namely the
$\pi\rho \rightarrow \gamma\omega$, $\pi\omega \rightarrow \gamma\rho$
and $\rho\omega \rightarrow \gamma\pi$ reactions for which the rates
(including the form factors) are parametrized in Appendix B of
Ref.\,\cite{Holt:2015cda}. Consequently, we also add these processes
when calculating thermal photon spectra.

\subsubsection{\label{sssec:fugacity} Influence of meson chemical potentials}

As was already mentioned, we do not restrict ourselves to the
consideration of emission from thermally and chemically equilibrated
matter, but also include non-equilibrium effects in form of finite pion
and kaon chemical potentials $\mu_{\pi}$ and $\mu_{\mathrm{K}}$. It has
been shown that the influence of a non-equilibrium distribution of the
respective mesons can be accounted for by introducing an additional
fugacity factor
\begin{equation}
z^{n}_{M=\pi,K}=\exp\left(\frac{n\mu_{M}}{T}\right)
\end{equation}
in the thermal dilepton and photon rates in Eqs.\,(\ref{rate_dil})
and (\ref{rate_ph}). The exponent $n$ depends on the difference in pion
or kaon number $N_{\pi /K}$ between initial and final state of the
process, i.e. $n=N^{i}_{\pi/K}-N^{f}_{\pi/K}$. Note that while the pion
fugacity enters in most processes, the effects of a finite kaon chemical
potential only play a role for the $\pi K$ bremsstrahlung and
$\pi+K^{*}\rightarrow\pi+\gamma$, $\pi+K\rightarrow K^{*}+\gamma$ and
$K^{*}+K\rightarrow\pi+\gamma$ photon production channels. For the
dilepton channels considered here, $\mu_{\mathrm{K}}$ can be neglected.

While for the single mesonic channels the initial and final state are
always well defined, several different types of processes are included
in the $\rho$ and $\omega$ spectral functions, especially processes with
baryons. For the $\rho$ not only processes with an initial two-pion
state of the type
$\pi\pi \rightarrow \rho \rightarrow \gamma / \gamma^{*}$ are accounted
for, but also reactions including only one or no pion as ingoing
particle (e.g., $\pi N \rightarrow \Delta \rightarrow \gamma N$ or
$N N \rightarrow \gamma NN$). However, as the correct fugacity depends
on the initial pion number, one would obtain different enhancements for
each channel. But as the different processes interfere with each other
it is difficult to determine the exact strength of each channel and
consequently one might hardly be able to account for some average
enhancement factor. Instead we here apply a fugacity factor
$z_{\pi}^{2}$ which would be correct for pure $\pi\pi$ annihilation
processes. This can be interpreted as an upper estimate of the influence
which the meson chemical potential might have on the thermal $\rho$
emission rates. The same procedure is applied for the $\omega$ meson,
where we assume a fugacity of $z_{\pi}^{3}$. Note that while the
multi-pion contribution also accounts for different initial states, they
are each treated separately so that one can apply the correct fugacity
factors here.

A full list of all hadronic contributions considered for the present
calculation of thermal dilepton and photon emission, including the
corresponding fugacity factors which account for the enhancement of the
specific channel due to the meson chemical potentials, is given in Table
\ref{tab:rates}.
%%%%%%%%%%%%%%%%%%%%%%%%%%%%%%%%%%%%%%%%%%%%%%%%%%%%%%%%%%%%%%
%%%%%%%%%%%%%%%%%%%%%%%%%%%%%%%%%%%%%%%%%%%%%%%%%%%%%%%%%%%%%%
\begin{table}
\centering
\begin{tabular}{|c||c|c|c|}
\hline 
{\textbf{Type}} & Rates & Fugacity & Ref.\\ 
\hline
Dilepton   &
\begin{tabular}{c} $\rho$ (incl.~baryon effects)\\ $\omega$ (incl.~baryon effects)\\ Multi-Pion \end{tabular} &
\begin{tabular}{c} $z_{\pi}^{2}$ \\ $z_{\pi}^{3}$  \\ $z_{\pi}^{3}$ / $z_{\pi}^{4}$ / $z_{\pi}^{5}$ \end{tabular} &  
\begin{tabular}{c} \cite{RappSF,Rapp:1999us} \\ \cite{RappSF,Rapp:1999us} \\ \cite{vanHees:2006ng} \end{tabular} \\  
\hline
Photon     &
             \begin{tabular}{c} $\rho$ (incl.~baryon effects) \\ $\pi\pi$ and $\pi K$ Bremsstr. \\ $\pi\rho \rightarrow \gamma\pi$ \\ $\pi K^{*} \rightarrow K\gamma$ \\  $\pi K \rightarrow K^{*}\gamma$ \\ $\rho K \rightarrow K\gamma$ \\  $K^{*} K \rightarrow \pi\gamma$ \\  $\pi\omega \rightarrow \gamma\rho$ \\ $\rho\omega \rightarrow \gamma\pi$ \\ $\pi\rho \rightarrow \gamma\omega$ \end{tabular} &
                                                                                                                                                                                                                                                                                                                                                                                                                \begin{tabular}{c} $z_{\pi}^{2}$ \\ $z_{\pi}^{2}+0.2z_{\pi}z_{K}$ \\ $z_{\pi}^{3}$ \\  $z_{\pi}z_{K}$  \\ $z_{\pi}^{2}z_{K}$  \\ $z_{\pi}$ \\  $z_{K}$\\ $z_{\pi}^{4}$ \\ $z_{\pi}^{5}$ \\ --- \end{tabular}&
\begin{tabular}{c} \cite{Rapp:1999us, Heffernan:2014mla} \\ \cite{Turbide:2003si, Heffernan:2014mla} \\ \cite{Turbide:2003si} \\ \cite{Turbide:2003si} \\ \cite{Turbide:2003si}  \\ \cite{Turbide:2003si} \\  \cite{Turbide:2003si} \\  \cite{Holt:2015cda} \\ \cite{Holt:2015cda} \\ \cite{Holt:2015cda} \end{tabular} \\ \hline
\end{tabular} \\ 
\caption{Summary of the different dilepton contributions considered in the present calculations.}
\label{tab:rates}
\end{table}

\subsection{\label{ssec:QGP} Quark-Gluon Plasma}

For the thermal emission of electromagnetic probes from the Quark-Gluon
Plasma one is again confronted with the problem that different processes
govern the dilepton production on the one side and photon emission on
the other. Consequently, one has to apply two different descriptions for
thermal rates, which are presented in the following.

In case of photon emission from a partonic phase of quarks and gluons
the two main contributions stem from quark-antiquark annihilation
($q\bar{q} \rightarrow g \gamma$) and Compton scattering processes
($qg \rightarrow q \gamma$ or $\bar{q}g \rightarrow \bar{q}\gamma$)
\cite{Kapusta:1991qp}. However, it was shown that these processes are
not sufficient to describe the production mechanism correctly and that
it is necessary to (a) include Feynman diagrams accounting for
bremsstrahlung and inelastic annihilation processes which are enhanced due to
near-collinear singularities and (b) to implement the
Landau-Pomeranchuk-Migdal effect
\cite{Aurenche:1998nw,Aurenche:1999tq}. The results of a full
calculation of the photon emission, to leading order in
$\alpha_{\mathrm{em}}$ and the QCD coupling $g\left(T\right)$, was
evaluated by Arnold, Moore and Yaffe and takes the following form
\cite{Arnold:2001ms}:

\begin{equation}
\begin{split}
\label{rate_qgp} q_{0}\frac{\mathrm{d} R_{\gamma}}{\mathrm{d}^3q} = &
-\frac{\alpha_\mathrm{em}\alpha_{\mathrm{s}}}{\pi^2} \, T^{2}\left(\frac{5}{9}\right) f_{\text{B}}\left(q;T\right)   
\\  &\times
\left[\ln\left(\sqrt{\frac{3}{4\pi\alpha_{\mathrm{s}}}}\right) +\frac{1}{2}\ln\left(\frac{2q}{T}\right) + C_{\text{tot}} \right]
\end{split}
\end{equation}
with
\begin{eqnarray}
C_{\text{tot}}&=& C_{2\leftrightarrow 2}\left(\frac{q}{T}\right)+C_{\text{annih}}\left(\frac{q}{T}\right)+C_{\text{brems}}\left(\frac{q}{T}\right), \\
\alpha_{\text{s}}&\approx&\frac{6\pi}{27\ln(T/0.022)}. \label{alphas}
\end{eqnarray}
The functions $C_{2\leftrightarrow 2}$, $C_{\text{annih}}$ and
$C_{\text{brems}}$ are approximated by the phenomenological fits given
in Eqs.\,(1.9) and (1.10) in Ref.\,\cite{Arnold:2001ms}. Note that this
calculation assumes the chemical potential to be vanishing. However, the
overall effect of finite values of a quark chemical potential (i.e.,
non-equal numbers of quarks and anti-quarks in the QGP phase) is known
to be rather small.

In case of the thermal dilepton emission from the QGP, the leading order
contribution is the electromagnetic annihilation of a quark and an
anti-quark into a virtual photon, $q\bar{q} \rightarrow \gamma^{*}$.
This process is irrelevant in the light-cone limit for
$M \rightarrow 0$, as the annihilation of two massive quarks into a
massless photon is kinematically forbidden. The pure perturbative
quark-gluon plasma rate was calculated for the mentioned leading order
process \cite{Cleymans:1986na} as
\begin{equation}
\begin{split} 
  \frac{\dd R_{ll}}{\dd^4p} = & \frac{\alpha_{em}^2}{4\pi^4}
  \frac{T}{p} f^{\text{B}}(p_0;T) \sum\limits_q e_q^2 \\
  & \times \ln \frac{\left(x_-+y\right) \left(
      x_++\exp[-\mu_q/T]\right)} {\left(x_++y\right) \left(
      x_-+\exp[-\mu_q/T]\right)}
\label{qqrate} 
\end{split}
\end{equation}
with $x_\pm=\exp[-(p_0\pm p)/2T]$ and $y=\exp[-(p_0+\mu_q)/T]$. The
quark chemical potential $\mu_{q}$ which shows up here is equal to
$\mu_{\text{B}}/3$. This calculation approximates the full QCD results quite well
at high energies, but for soft processes of the order $g_{\text{s}}(T)$,
i.e., for dileptons with low masses and momenta, the one-loop calculation
is not sufficient and hard-thermal-loop (HTL) corrections to the result
as given in Eq.(\ref{qqrate}) have to be considered
\cite{Braaten:1990wp}. It was found that the rate for soft dileptons is then
by orders of magnitude larger than the simple leading order calculation
\cite{Rapp:2013nxa}.

Recent calculations from thermal lattice-QCD suggest an even stronger
enhancement of the rates for low-mass dileptons
\cite{Ding:2010ga}. These results, which are applied in the present
work, have been extrapolated to finite three-momenta by a fit to the
leading-order pQCD rates such that the correlation function takes the
form \cite{Rapp:2013nxa}
\begin{equation}
\label{latrate}
-\im \Pi_{\text{EM}} =\frac{C_{\text{EM}}}{4\pi}M^{2}
\left(\hat{f}_{2}\left(q0,q;T\right)+Q_{\text{LAT}}^{tot}\left(M,q\right)\right),
\end{equation}
where
\begin{equation}
\label{qtotlat}
\begin{split}
  Q_{\text{LAT}}^{\text{tot}}\left(M,q\right) = &
  \frac{2\pi\alpha_{\text{s}}}{3}\frac{T^{2}}{M^{2}}\left(2+\frac{M^{2}}{q_{0}^{2}}\right)
  \\ & \times KF\left(M^{2}\right)
  \ln\left[1+\frac{2.912}{4\pi\alpha_{\text{s}}}\frac{q_{0}}{T}\right]
\end{split}
\end{equation}
with a form factor $F\left(M^{2}\right)=4T^{2}/\left(4T^{2}+M^{2}\right)$
and a factor $K=2$ to better fit the full lQCD rates. Note that in
contrast to the pQCD result the rate in Eqs.\,(\ref{latrate}) and
(\ref{qtotlat}) is calculated for $\mu_{q}=0$ only, as a calculation for
finite chemical potential is still beyond the current lattice
calculations.

%%%%%%%%%%%%%%%%%%%%%%%%%%%%%%%%%%%%%%%%%%%%%%%%%%%%%%%%%%%%%%
\begin{figure*}
\includegraphics[width=1.03\columnwidth]{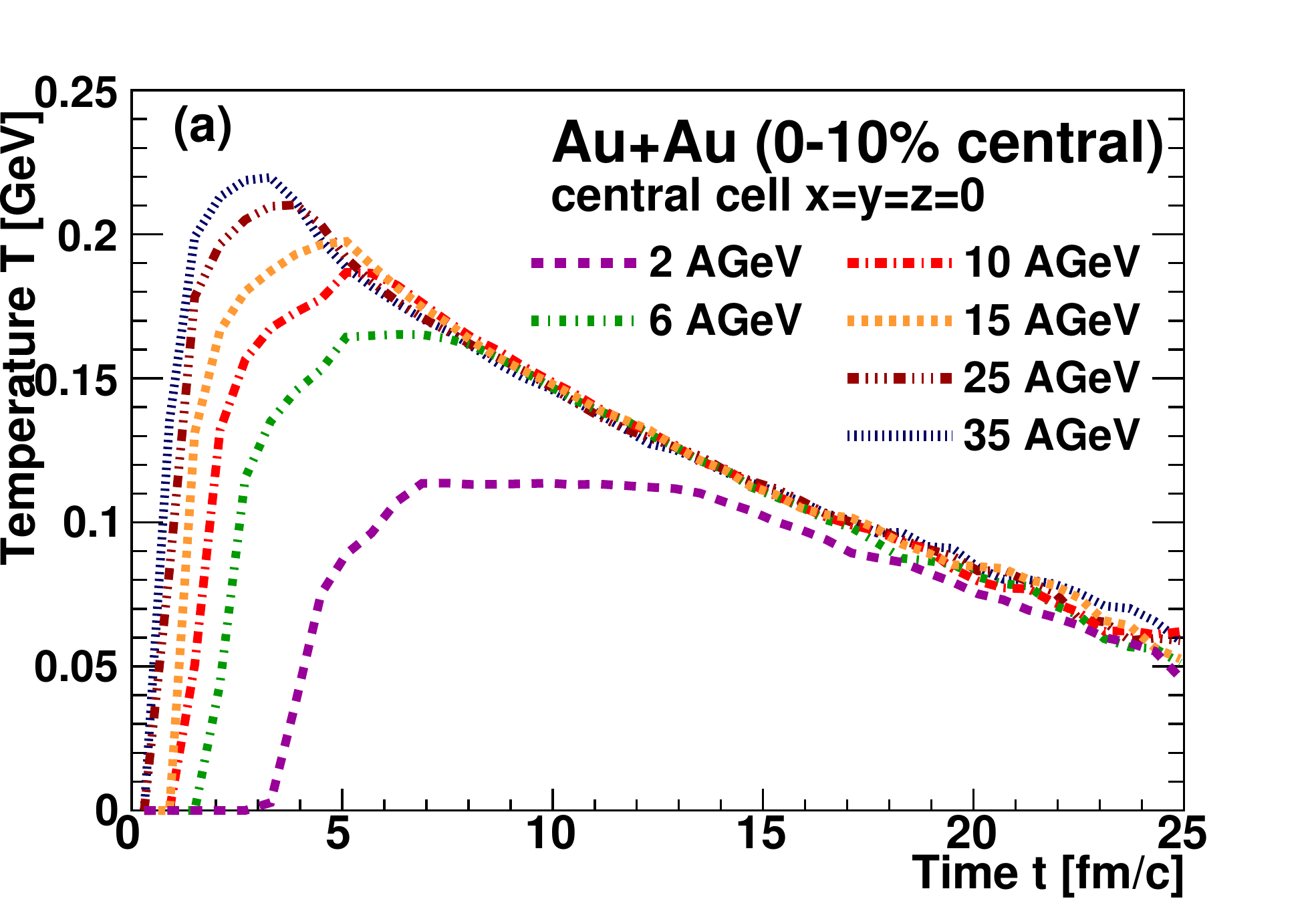}
\includegraphics[width=1.03\columnwidth]{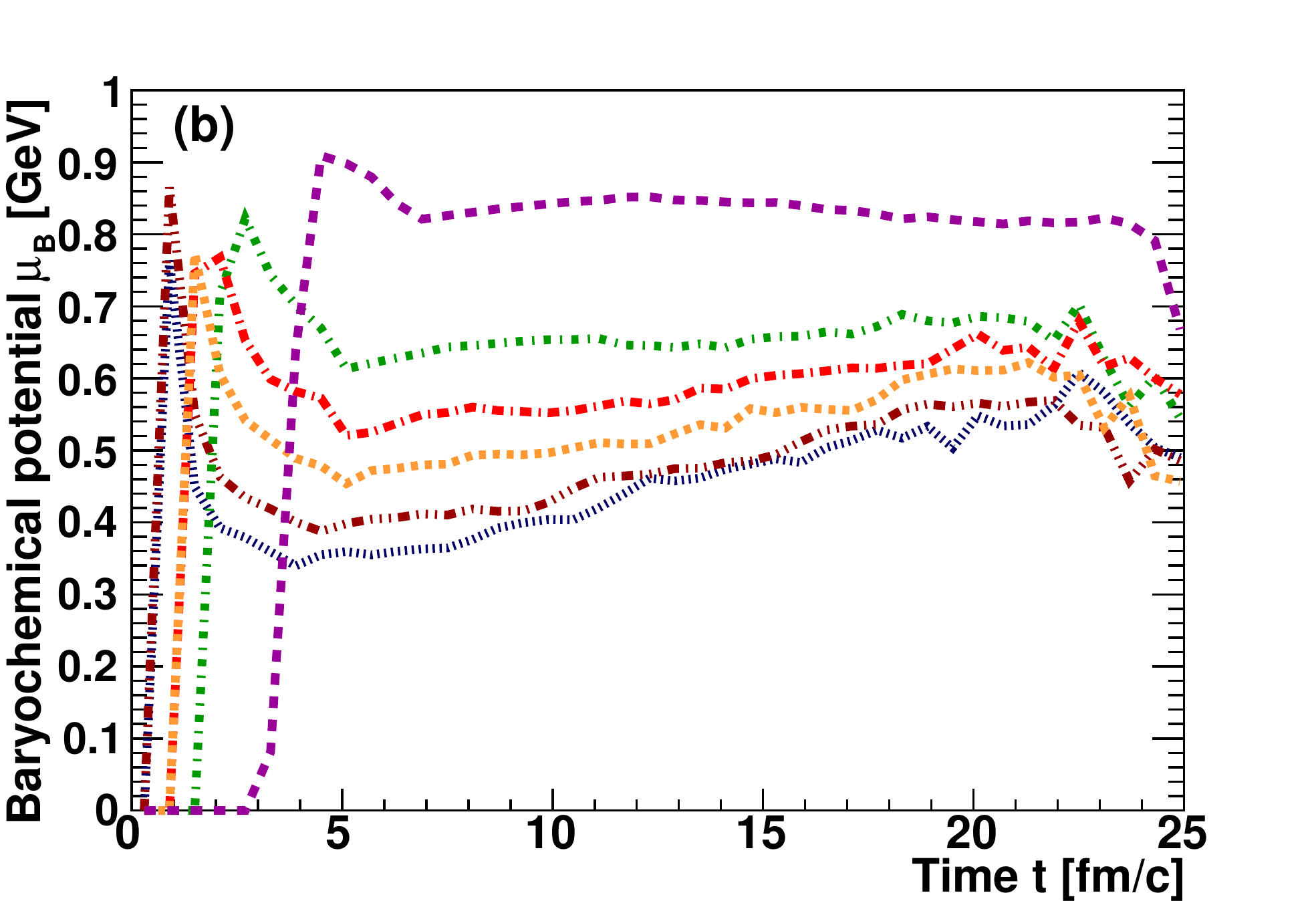}
\caption{(Color online) Time evolution of (a) temperature $T$ and (b)
  baryochemical potential $\mu_{\text{B}}$ for the central cell of the
  coarse-graining grid for different beam energies
  $E_{\text{lab}}=2-35 \,A\GeV$.  The results are obtained for the
  0-10\% most central collisions in Au+Au reactions.}
\label{tmubev}
\end{figure*}
%%%%%%%%%%%%%%%%%%%%%%%%%%%%%%%%%%%%%%%%%%%%%%%%%%%%%%%%%%%%%%
\subsection{\label{ssec:Hadronic} Hadronic decay contributons}

While we restrict the calculation of photon yields to the thermal
contribution, since all decay photons from long-lived hadronic
resonances are usually subtracted from the experimental results, a full
description of the dilepton spectra requires to take also the
non-thermal contributions from the decay of pseudo-scalar and vector
mesons into account. We here follow the same procedures as in our
previous work for SIS\,18 energies. In detail, we determine the following non-thermal
dilepton contributions:
\begin{enumerate}
\item The Dalitz decays of the pseudo-scalar $\pi^{0}$ and $\eta$
  mesons. To determine their contribution, we assume that each final
  state particle contributes with a weight of
  $\Gamma_{M\rightarrow \mathrm{e}^+ \mathrm{e}^-}/\Gamma_{\text{tot}}$.
\item The direct decay of the $\phi$ meson into a lepton pair. As the
  lifetime of the $\phi$ is relatively short, we apply a shining
  procedure which takes absorption and re-scattering processes inside
  the medium into account.
\item Finally, we restricted the calculation of thermal dileptons to
  those cells where the temperature is larger than 50\,MeV, as otherwise
  a thermal description becomes questionable. However, in principle one
  will of course also find $\rho$ and $\omega$ mesons at lower
  temperatures. To account for this, in the mentioned cases we calculate
  a ``freeze-out'' contribution from the $\rho$ and $\omega$ decays
  using the UrQMD results for these mesons.
\end{enumerate}
For a more detailed description of the non-thermal hadronic
contributions the reader is referred to Ref.\,\cite{Endres:2015fna}. We
refrain from an extensive reproduction of the procedure here, as the
cocktail contributions will not play a significant role in the present
investigations.

%%%%%%%%%%%%%%%%%%%%%%%%%%%%%%%%%%%%%%%%%%%%%%%%%%%%%%%%%%%%%%
\section{\label{sec:Results} Results}
In the following we present the results of calculations with the
coarse-graining approach for Au+Au collisions in the energy range of
$E_{\text{lab}}=2-35\,A$GeV. We restrict the analysis to the 10\% most
central reactions, as the medium effects will be largest here. In terms
of the microscopic UrQMD results, this roughly corresponds to an impact
parameter range of $b=0-4.5\,\fm$. For the coarse-graining we use
ensembles of 1000 microscopic events each. The length of the time-steps
is chosen as $\Delta t=0.6\,\fm/c$, and the size of the spatial grid
is $\Delta x= \Delta y= \Delta z= 0.8\,\fm$. These grid parameters are
similar to the ones used for the previous studies at SIS\,18 and SPS
energies and constitute a good compromise between resolution and a
sufficiently large hadron number per cell. To obtain enough statistics,
especially for the non-thermal contributions, several runs with
different ensembles are necessary.
%%%%%%%%%%%%%%%%%%%%%%%%%%%%%%%%%%%%%%%%%%%%%%%%%%%%%%%%%%%%%%
\subsection{\label{ssec:ReacDyn} Reaction dynamics}
As the dilepton and photon production is directly related to the
space-time evolution of the thermodynamic properties of the system, it
seems natural to start with a study of the reaction dynamics obtained
with the coarse-graining of UrQMD input.

In Fig.\,\ref{tmubev} the time evolution of temperature and
baryochemical potential in the central cell of the grid is depicted for
different beam energies. The evolution shows a significant increase of
the temperature maxima from slightly above 100\,MeV for $2\,A\GeV$ up to
roughly 225\,MeV for top SIS\,300 energy. While the temperature is
clearly below the critical temperature of 170\,MeV for the lower
energies, the highest energies covered by FAIR can probe also this
deconfinement region of the phase diagram. The thermal lifetime of the
central cell, i.e., the time for which it rests at temperatures above 50
MeV, increases slightly with increasing collision energy. This is mainly
due to an earlier onset of thermalization after
the first hadron-hadron collisions, which define the origin of the time
axis. However, it is interesting that in the later phase of the
collision the temperature curves for all energies show the same
monotonous decrease and even lie on top of each other. A somewhat different
behavior is observed for the evolution of the baryochemical potential
$\mu_{\text{B}}$. For all energies it shows a clear peak with values
between 700 and 900 MeV at the beginning of the collision, which is due
to the high baryon densities reached in the central cell when the two
nuclei first come into contact. Afterwards $\mu_{\text{B}}$ decreases
and then remains on a plateau level for a significant fraction of the
reaction time for the lower energies, while one observes a slight increase
for the higher beam energies towards later times. If we neglect the peak in the early
reaction stage, the chemical potential shows a clear decrease with
increasing collision energy. While for $2\,A\GeV$ the baryochemical
potential remains around 900\,MeV for the whole thermal lifetime,
$\mu_{\mathrm{B}}$ is only 350\,MeV at $t=5\,\fm$ for $35\,A\GeV$ and slowly
increases up to 600\,MeV after $t=20\,\fm$. 
%%%%%%%%%%%%%%%%%%%%%%%%%%%%%%%%%%%%%%%%%%%%%%%%%%%%%%%%%%%%%%
\begin{figure*}
\includegraphics[width=1.0\columnwidth]{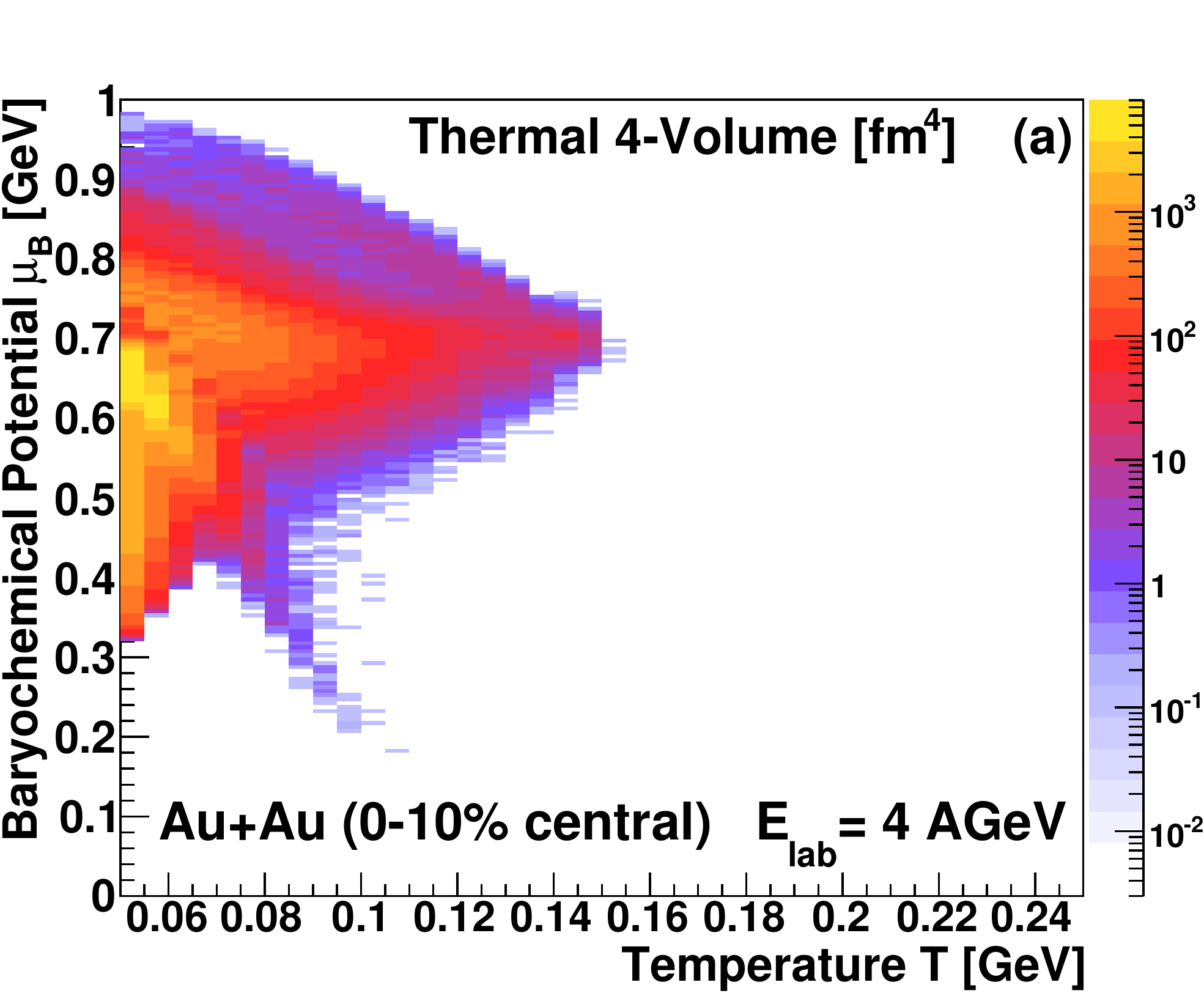}
\includegraphics[width=1.0\columnwidth]{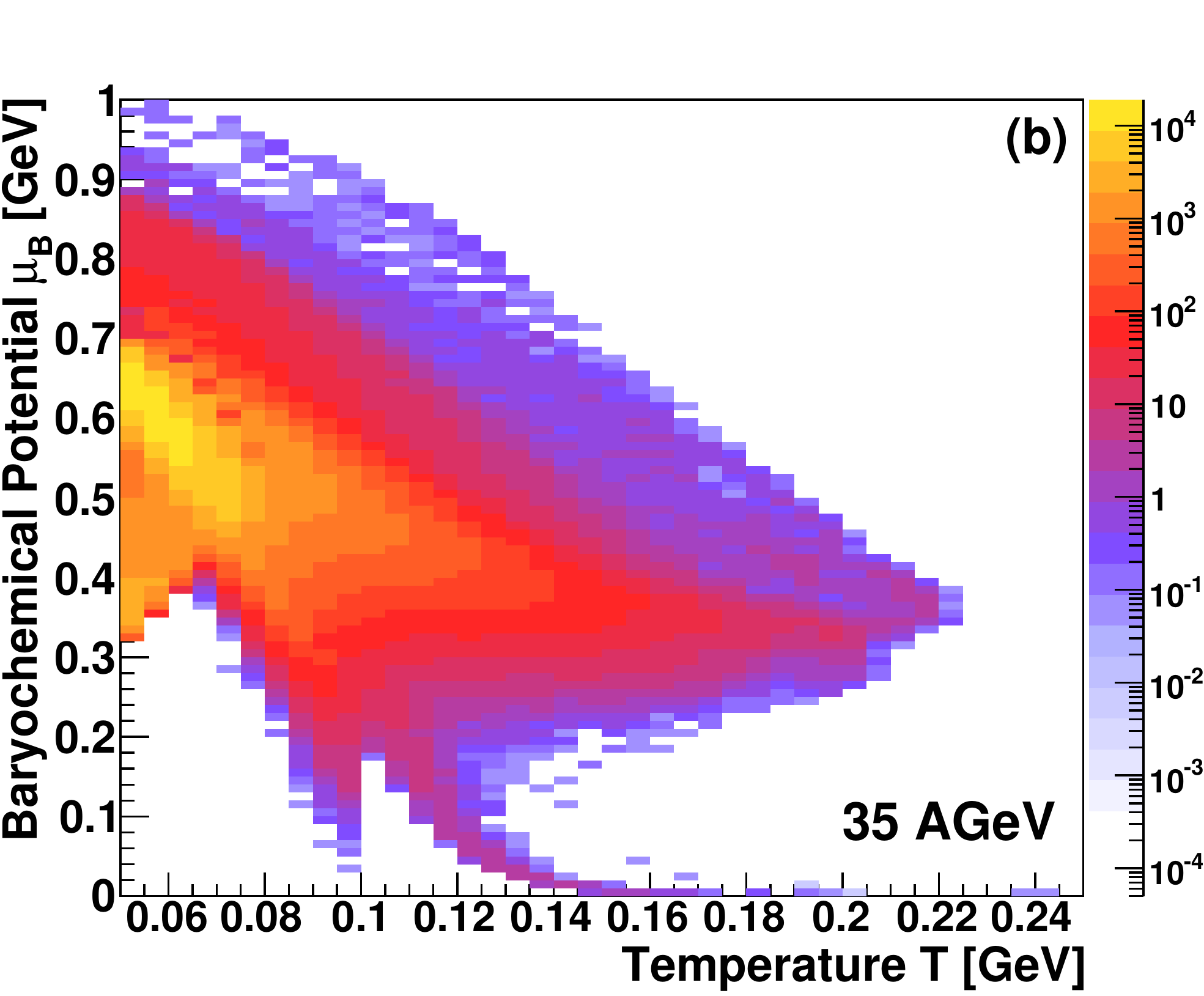}
\caption{(Color online) Thermal four-volume in units of $\fm^{4}$ from
  the coarse-grained transport calculations in dependence of temperature
  $T$ and baryochemical potential $\mu_{\mathrm{B}}$. Results are shown
  for $E_{\text{lab}}= 4 \,A\GeV$ (a) and $35\,A\GeV$ (b) in central Au+Au
  collisions.}
\label{fourvolev}
\end{figure*}
%%%%%%%%%%%%%%%%%%%%%%%%%%%%%%%%%%%%%%%%%%%%%%%%%%%%%%%%%%%%%%
%%%%%%%%%%%%%%%%%%%%%%%%%%%%%%%%%%%%%%%%%%%%%%%%%%%%%%%%%%%%%%
\begin{figure}[b]
\includegraphics[width=1.0\columnwidth]{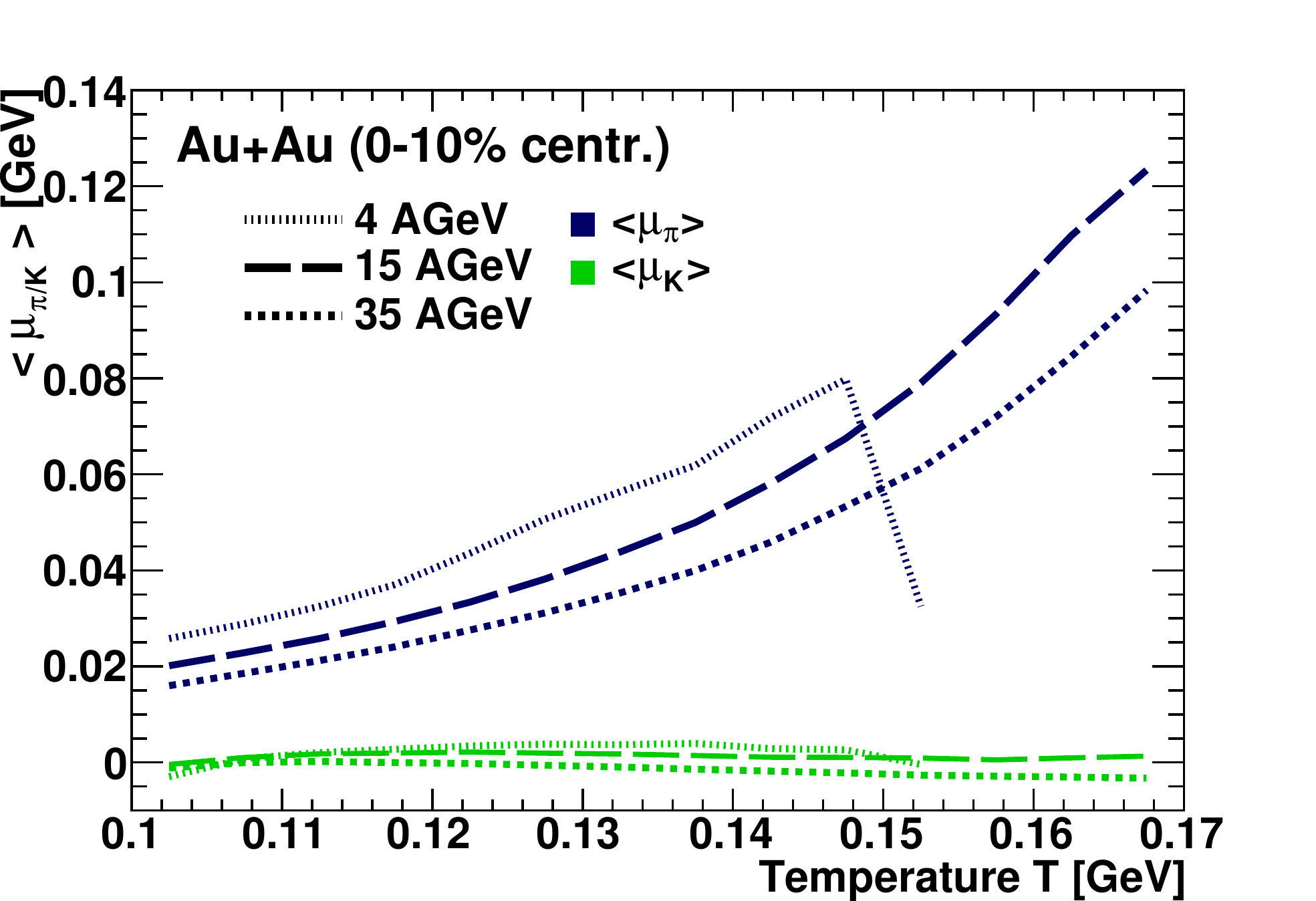}
\caption{(Color online) Average values of the pion chemical potential
  $\mu_{\pi}$ (blue lines) and the kaon chemical potential $\mu_{\mathrm{K}}$ (green lines) in
  dependence on the cell temperature. Results are shown for central
  Au+Au collisions at three different collision energies,
  $E_{\text{lab}}=4,\ 15 \text{ and } 35\,A\GeV$.}
\label{mupikev}
\end{figure}
%%%%%%%%%%%%%%%%%%%%%%%%%%%%%%%%%%%%%%%%%%%%%%%%%%%%%%%%%%%%%%
%%%%%%%%%%%%%%%%%%%%%%%%%%%%%%%%%%%%%%%%%%%%%%%%%%%%%%%%%%%%%%
\begin{figure*}
\includegraphics[width=1.0\columnwidth]{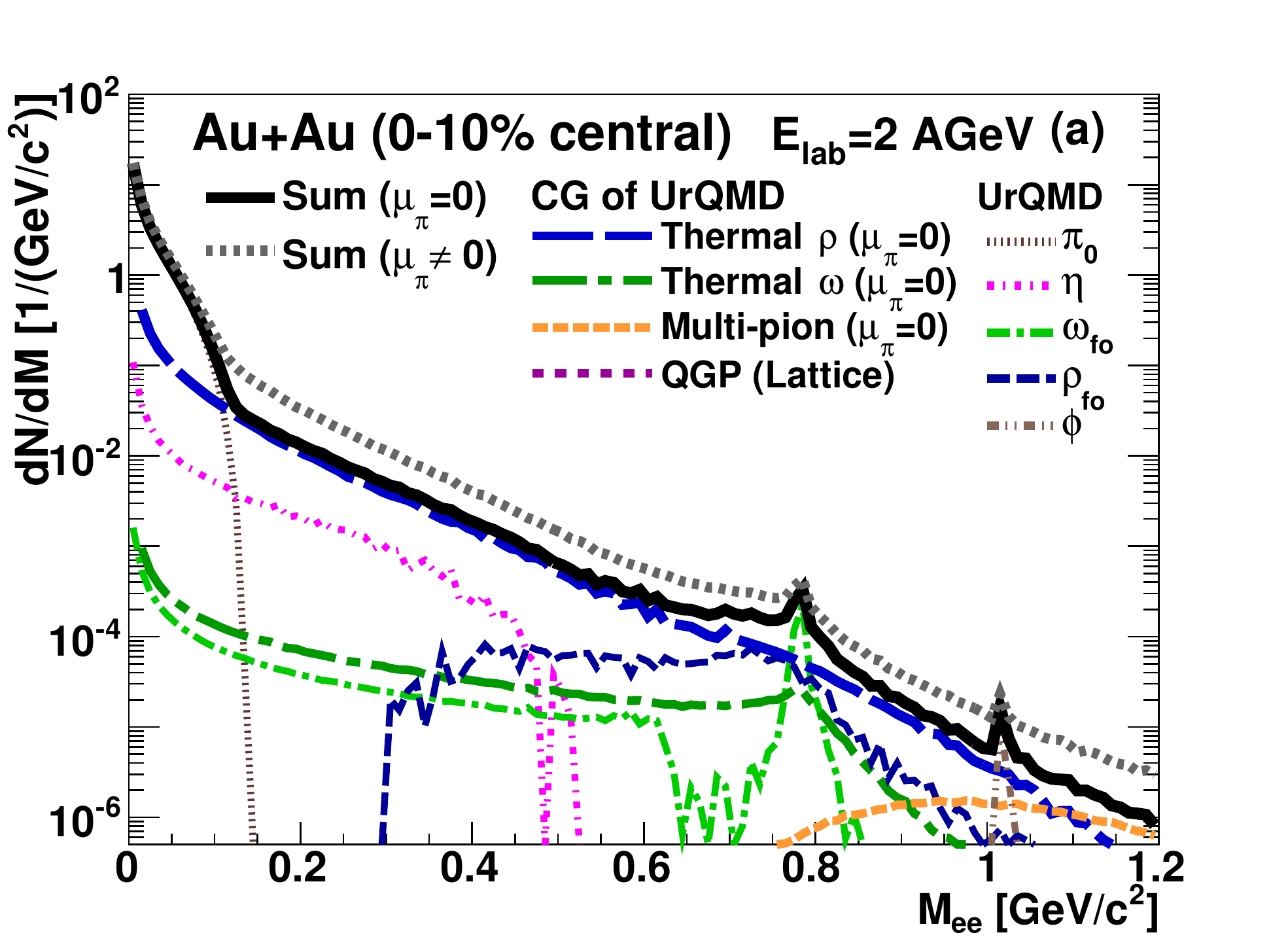}
\includegraphics[width=1.0\columnwidth]{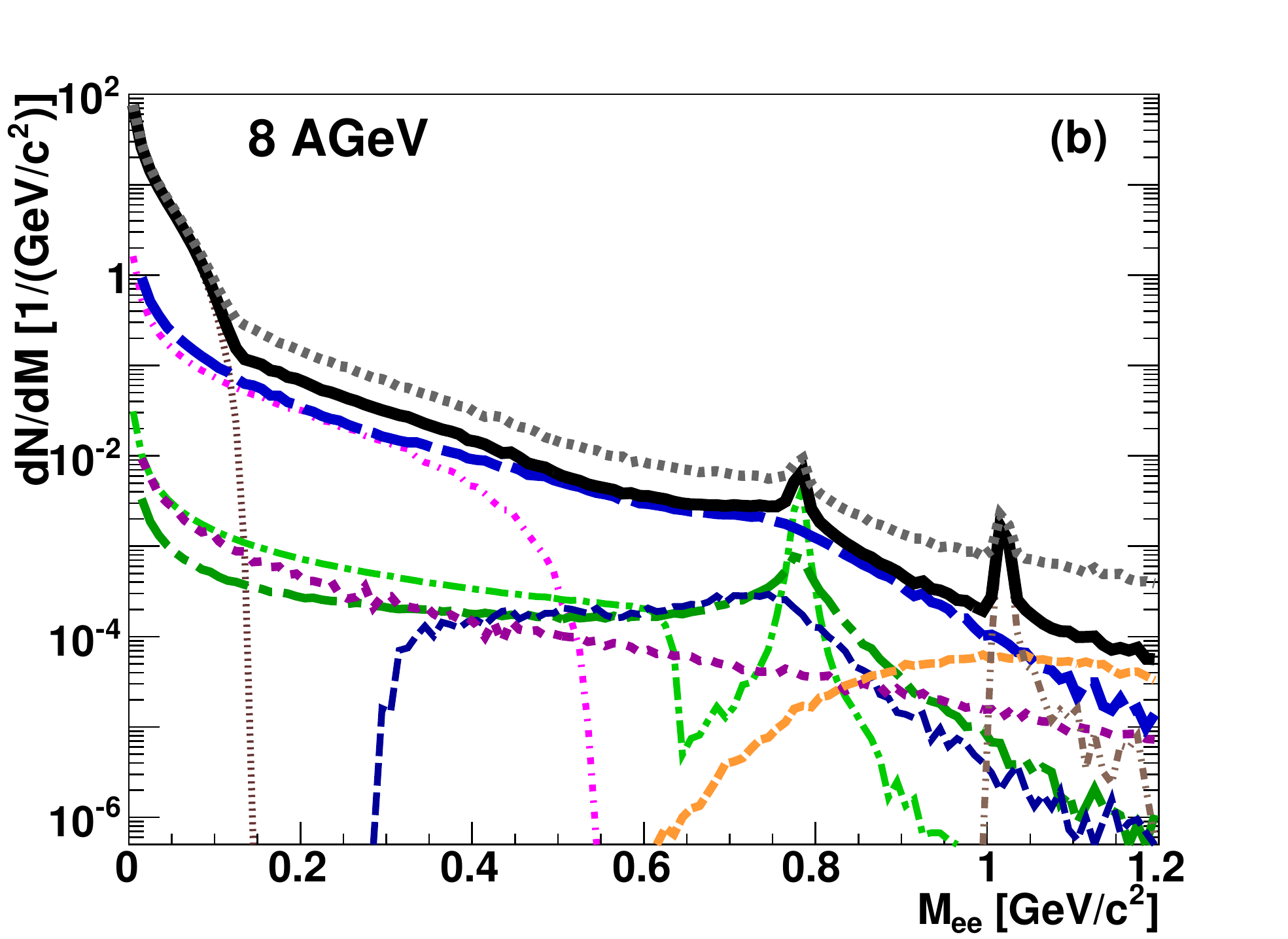}
\\
\includegraphics[width=1.0\columnwidth]{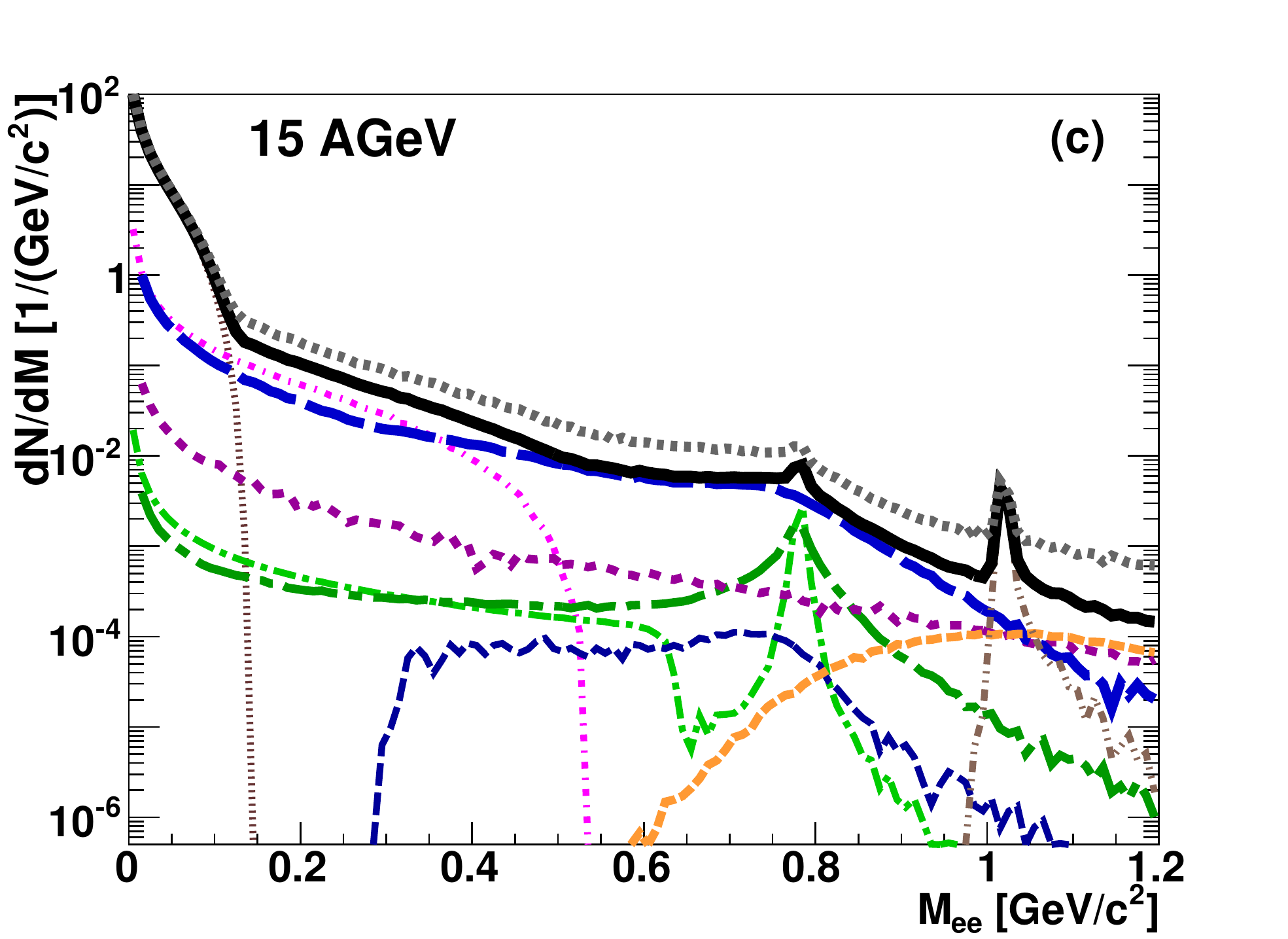}
\includegraphics[width=1.0\columnwidth]{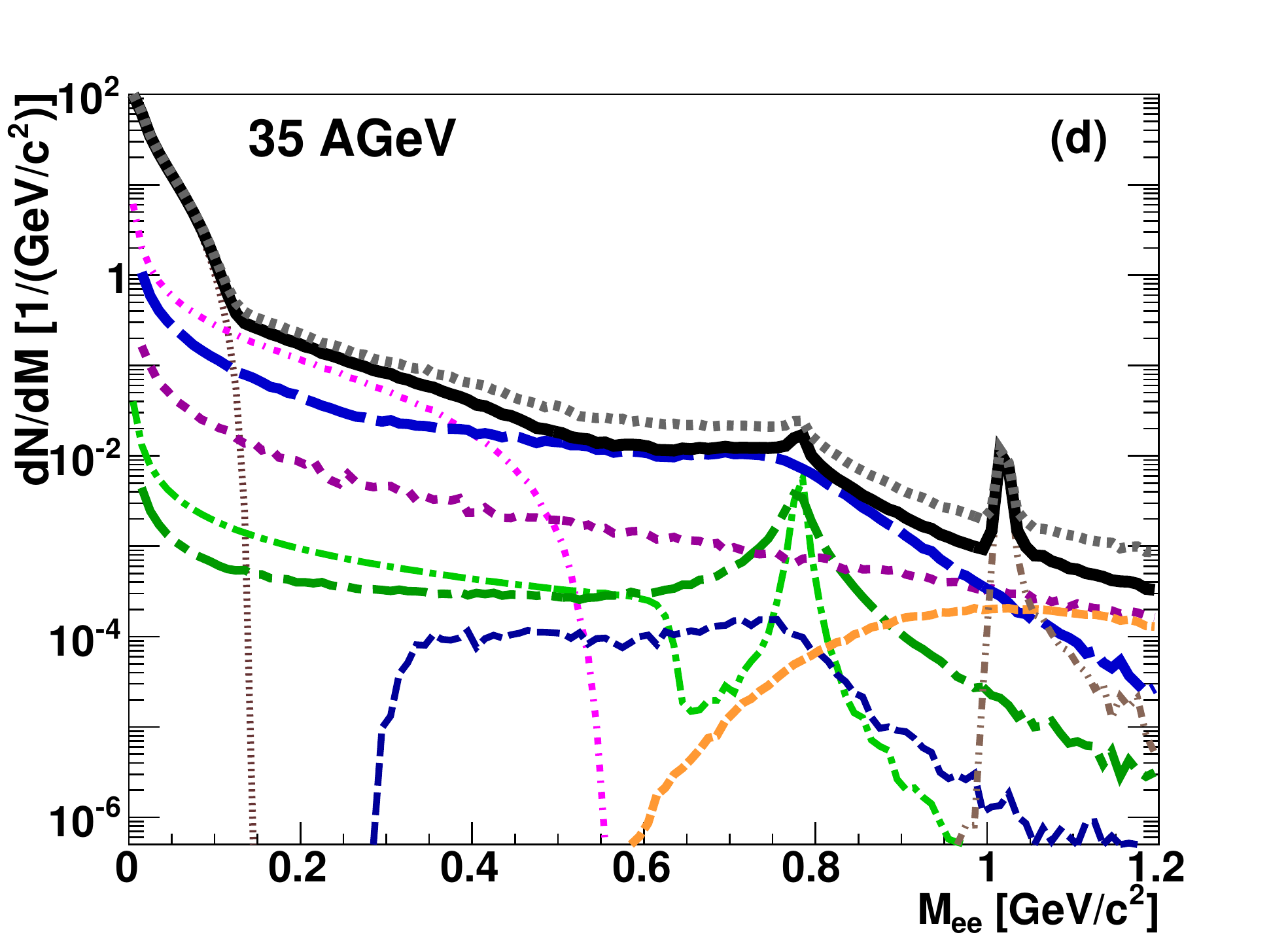}
\caption{(Color online) Dilepton invariant mass spectra for Au+Au
  reactions at different energies $E_{\text{lab}}=2 - 35\,A\GeV$ within
  the centrality class of 0-10\% most central collisions. The resulting spectra include thermal contributions from the coarse-graining of the microscopic simulations (CG of UrQMD) and the non-thermal contributions directly extracted from the transport calculations (UrQMD).  The
  hadronic thermal contributions are only shown for vanishing pion
  chemical potential, while the total yield is plotted for both cases,
  $\mu_{\pi}=0$ and $\mu_{\pi}\neq 0$.}
\label{dilinvmass}
\end{figure*}
%%%%%%%%%%%%%%%%%%%%%%%%%%%%%%%%%%%%%%%%%%%%%%%%%%%%%%%%%%%%%%
%%%%%%%%%%%%%%%%%%%%%%%%%%%%%%%%%%%%%%%%%%%%%%%%%%%%%%%%%%%%%%
\begin{figure*}
\includegraphics[width=1.03\columnwidth]{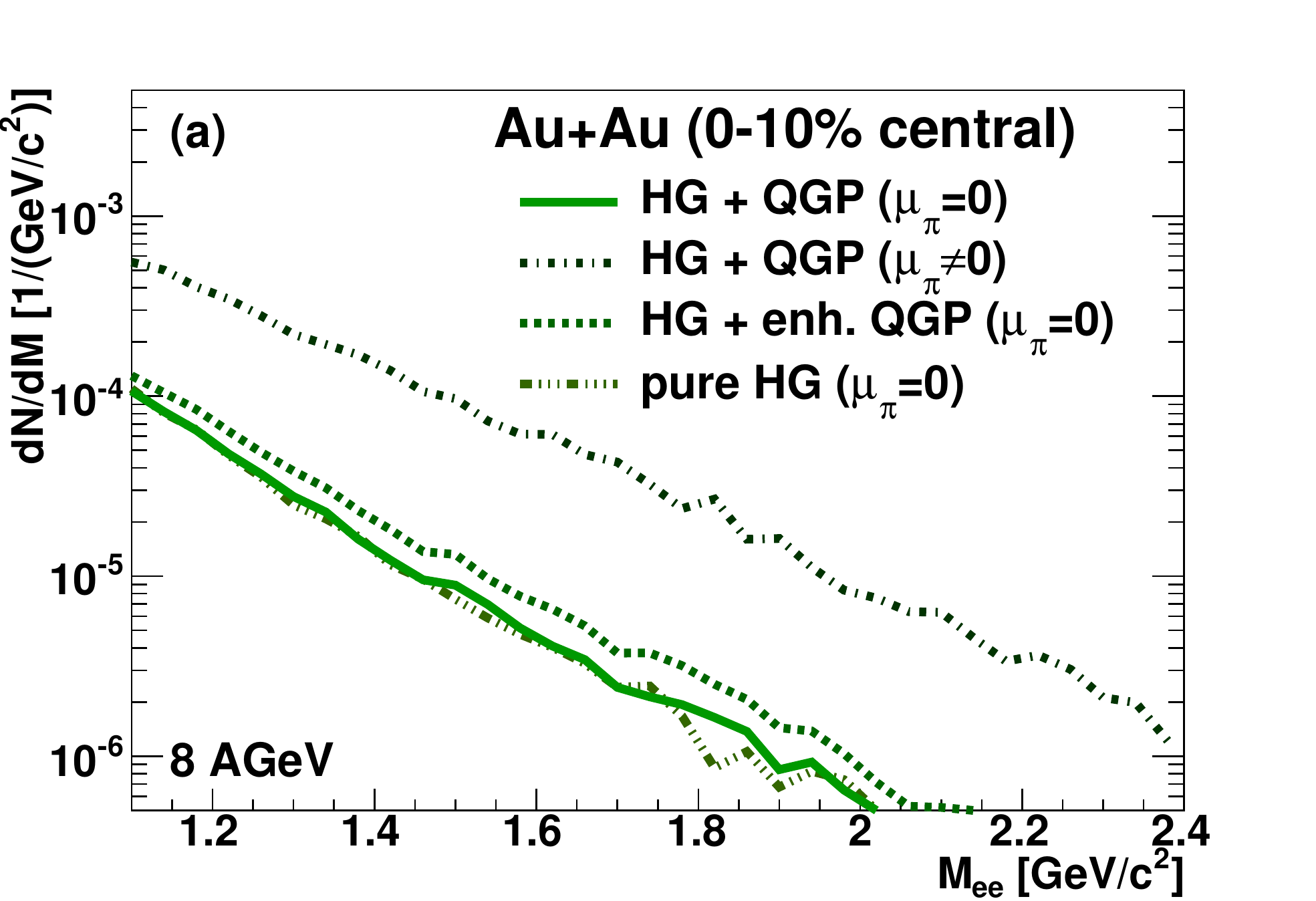}
\includegraphics[width=1.03\columnwidth]{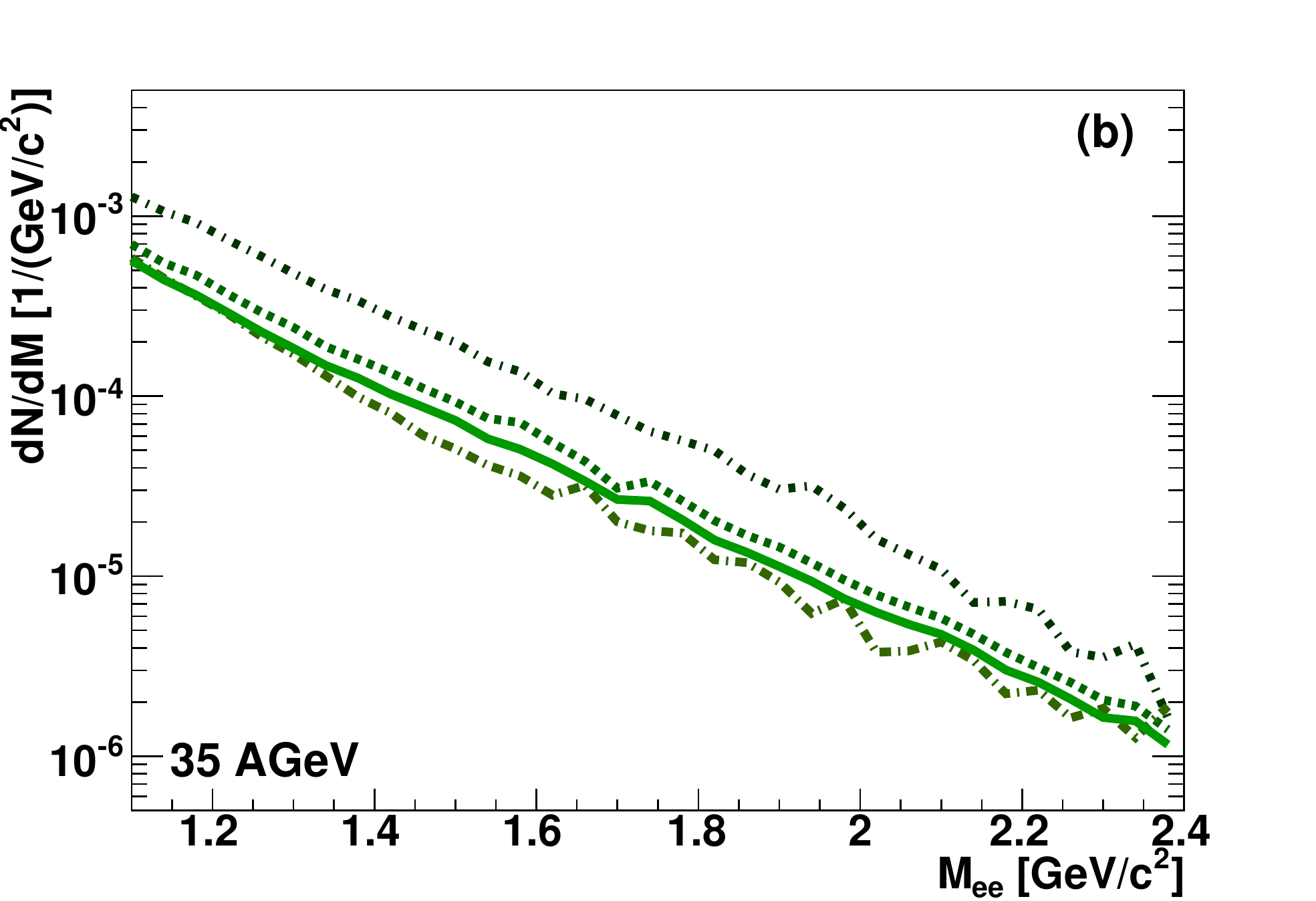}
\caption{(Color online) High-mass region of the dilepton
  $M_{\mathrm{e}^+ \mathrm{e}^-}$ spectrum for central Au+Au collisions
  at $E_{\text{lab}}=8\,A\GeV$ (a) and $35\,A\GeV$ (b), assuming
  different emission scenarios. The the baseline calculation assumes
  hadronic emission up to 170\,MeV with vanishing $\mu_{\pi}$ and
  partonic emission for higher temperatures (full line). Besides the
  results with a five times enhanced emission around the critical
  temperature (dashed line) and for a pure hadron gas scenario with
  emission at all temperatures from hadronic sources
  (dashed-triple-dotted line) are shown. Again, we also show the
  baseline result including finite pion chemical potential
  (dashed-dotted line).}
\label{dilhighmass}
\end{figure*}
%%%%%%%%%%%%%%%%%%%%%%%%%%%%%%%%%%%%%%%%%%%%%%%%%%%%%%%%%%%%%%

Note that the results shown in Fig.\,\ref{tmubev} are \textit{only} for one single cell at the center of the collision. The evolution in other cells of the grid may differ largely in dependence on their location (e.g., one finds in general lower temperature and chemical potential in more dilute peripheral cells). But yet it clearly depicts the influence of collision energy on $T$ and $\mu_{\mathrm{B}}$. One finds two effects when going from the lowest to the highest FAIR energies: An
\emph{increasing} temperature combined with a \emph{decreasing}
baryochemical potential. This behavior is not specific for the
central cell but reflected by the whole space-time evolution, as can be
seen from Fig.\,\ref{fourvolev}. The two plots show the
thermal four-volume in dependence of temperature $T$ (X-axis) and
baryochemical potential $\mu_{\mathrm{B}}$ (Y-axis). Results are shown for
$E_{\text{lab}}= 4\,A\GeV$ (a) and $35\,A\GeV$ (b). We see that for the
lower energy the largest part of the four volume is concentrated at
values of the baryochemical potential between 500 and 800\,MeV, while
the temperature remains below 160\,MeV for all cells. In contrast, for
$E_{\text{lab}} = 35\,A\GeV$ the four-volume distribution extends to
higher temperatures up to $T=240\,\MeV$ while at the same time the
distribution is shifted to lower baryochemical potentials especially for
higher temperatures while the lower-temperature cells are mainly
dominated by high values of $\mu_{\text{B}}$. Interestingly, especially for the higher collision energy of 35\,AGeV one finds some cells in a separate region with moderate to high temperature and very low baryochemical potential $\mu_{\mathrm{B}} \approx 0$. These cells are mainly found in the more peripheral regions of the collision, where the baryon density (and particle density in general) is rather low and where nevertheless in some cases hadrons with large momenta are found, resulting in high energy density for these cells. However, compared to the large overall total thermal volume the relevance of these low-$\mu_{\mathrm{B}}$ cells is negligible.

It is important to bear in
mind that the dilepton and photon spectra will directly reflect the
four-volume evolution in the $T-\mu_{\mathrm{B}}$ plane, as presented here.
The results show that at FAIR energies the region of the QCD phase
diagram with temperatures above the critical temperature \emph{and}
large $\mu_{\mathrm{B}}$ can be probed, in contrast to the situation at LHC or
RHIC, where the transition from hadronic matter to a deconfined phase is
assumed to happen at $\mu_{\text{B}} \approx 0$. However, note that the
results presented here are obtained with a purely hadronic equation of
state which does not include any effects of the phase transition
itself. For an improvement the description one might need to implement
the transition properly, to account, e.g., for the latent heat which would
cause the cells to remain for a longer time at temperatures around
$T_{\mathrm{c}}$. Nevertheless the present results can serve as a lower limit
baseline calculation, assuming that we have a smooth crossing from
hadronic to QGP emission. Significant deviations from this assumption
might then show up in the photon and dilepton spectra. We will discuss
this later.

As was pointed out before, the effects of chemical non-equilibrium show
up in the form of finite meson chemical potentials for the $\pi$ and K;
and $\mu_{\pi}$ and $\mu_{\text{K}}$ can have a significant effect on the population of
several photon and dilepton production channels. The mean values of
the pion chemical potential $\mu_{\pi}$ and the kaon chemical potential
$\mu_{\mathrm{K}}$ in dependence on the cell's temperature for different
collision energies are shown in Figure\,\ref{mupikev}. Note that the results for the chemical potentials here are obtained by averaging the values of $\mu_{\pi}$ and $\mu_{\mathrm{K}}$ over all space-time cells with a specific temperature. The study indeed indicates that
regarding the pion density the system will be clearly out of equilibrium
during the collision evolution. The value of $\mu_{\pi}$
increases with temperature, which is not surprising since a
large part of the pion production in the microscopic simulation takes
place in initial scatterings and via string formation at the
beginning of the reaction, when the system still heats up. At all temperatures one finds that the $\mu_{\pi}$ decreases with increasing collision energy, which may indicate a faster and stronger equilibration of the system if more energy is
deposited in the system. In addition, for top SIS\,300 energies the initial emission is dominated by QGP radiation at temperatures above $T_{c}$ and consequently a larger fraction of cells with $T<170$\,MeV is found later in the course of the fireball evolution, when the system is in a more equilibrated condition compared to the very beginning of the collision. For the higher collision energies we get average values up to $\mu_{\pi}=100-120$\,MeV around the
critical temperature of 170\,MeV. Note that for $E_{\mathrm{lab}}=4\,A$GeV the maximum temperature found in the evolution is around 155\,MeV, which explains the drop in the corresponding curve around this temperature. 

In contrast to the large pion chemical potential, no such dominant off-equilibrium effect is observed for the kaons, where $\mu_{\mathrm{K}} \approx 0$ at all energies and temperatures. This is not surprising, as in the underlying microscopic simulations any inelastic reaction results in the creation of a $\pi$ whereas the cross-section for kaon production is rather low in the cases considered here (and especially for the lower FAIR energies). Consequently, the kaon production is a slow process which seems to happen synchronously with the equilibration of the system while a large amount of pions is produced in the initial hard nucleon-nucleon scatterings before any equilibration could take place. 

The present results for the pion chemical potential are quite different
from other model descriptions. For example, in fireball parametrizations
the particle numbers are fixed at the chemical freeze-out of the
system and consequently meson chemical potentials develop when the
system cools down. However, in the fireball model this is just an ad-hoc
assumption, as such macroscopic models are based on a presumed equilibrium within the system. In contrast,
the overpopulation of pions is an intrinsic result stemming from the
microscopic simulation in the case of the coarse-graining approach.
Nevertheless, the very high
pion chemical potentials in the temperature region close to the phase
transition might be questionable, as one would assume that the
transition from the Quark-Gluon Plasma to a hadronic phase should produce
a system where the mesons are in an equilibrium state.
A fully satisfying description of the chemical off-equilibrium
evolution is not feasible within the present approach and would require
a microscopic and dynamical description of the phase transition and its underlying
dynamics.

%%%%%%%%%%%%%%%%%%%%%%%%%%%%%%%%%%%%%%%%%%%%%%%%%%%%%%%%%%%%%%
\subsection{\label{ssec:DilSpec} Dilepton spectra}
The dilepton invariant mass spectra in the low-mass range up to
$M_{\mathrm{e}^+ \mathrm{e}^-}=1.2 \,\GeV/c^{2}$ for four different beam
energies ($E_{\text{lab}}=2,8,15\text{ and }35\,A\GeV$) are presented in
Fig.\,\ref{dilinvmass}. The comparison shows some interesting
similarities and differences: While the very low masses up
$0.15\,\GeV/c^{2}$ are generally dominated by the Dalitz decays of
neutral pions, the region beyond the Dalitz peak up to
the pole masses of the $\rho$ and $\omega$ mesons (i.e.,
$\approx 770 \,\MeV/c^{2}$) is dominated by a strong thermal $\rho$
contribution. The thermal yield shows an absolute
increase with $E_{\text{lab}}$, but its importance decreases relative to
the non-thermal $\eta$ yield. This means that the thermal low-mass
enhancement of the dilepton yield above a hadronic vacuum cocktail
decreases with increasing collision energy. This observation is explained by
the decrease of the baryon chemical potential at higher collision
energies, as has been mentioned in the previous section. In contrast,
the increasing temperature leads to a significantly flatter shape of
especially the $\rho$ distribution in the invariant mass spectrum. While for low energies the thermal yield
decreases strongly when going to higher invariant masses, at the top
FAIR energies this effect is less prominent and the population of
high masses is enhanced. This is can be clearly seen by the fact
that the multi-pion yield shows a strong rise.

While at the lowest of the four energies the whole system is well below the critical
temperature $T_{\mathrm{c}}$, we know from the temperature evolution in Fig.\,\ref{tmubev}\,(a)
that the region around $T \approx 170$\,MeV from is reached
$E_{\text{lab}} = 6-8\,A\GeV$ on. Consequently, in the dilepton invariant mass 
spectra of Fig.\,\ref{dilinvmass} the resulting QGP contribution is very
small at $8\,A\GeV$, but even at $35\,A\GeV$ the partonic yield is suppressed by
roughly an order of magnitude compared to the leading contributions in
the mass range up to $1\,\GeV/c^{2}$.

The hadronic thermal yields in Fig.\,\ref{dilinvmass} are shown for the case
of vanishing pion chemical potential. However, we also compare the result for $\mu_{\pi}=0$ with the total
yield assuming finite values of $\mu_{\pi}$. One can see that chemical
non-equilibrium can increase the overall dilepton yield in the low-mass
range up to a factor of two. For the region above $1\,\GeV/c^{2}$ the effect
can be even larger as the fugacity factor enters the thermal rate with a
power of four for the multi-pion contribution. It is important to bear
in mind that this result should rather be seen as
an upper estimate, as the approximation $\mu_{\rho}=2\mu_{\pi}$ is
only correct for the rate $\pi\pi \rightarrow \rho$, which
represents only one of the many processes included in the $\rho$ spectral
function. Furthermore, as UrQMD has no intrinsic description of the
phase transition, the pion chemical potential might be overestimated in
vicinity of the critical temperature. Nevertheless, the results show
that a deviation from pion equilibrium has a huge impact on the thermal
dilepton rates.
%%%%%%%%%%%%%%%%%%%%%%%%%%%%%%%%%%%%%%%%%%%%%%%%%%%%%%%%%%%%%%
\begin{figure*}
\includegraphics[width=1.0\columnwidth]{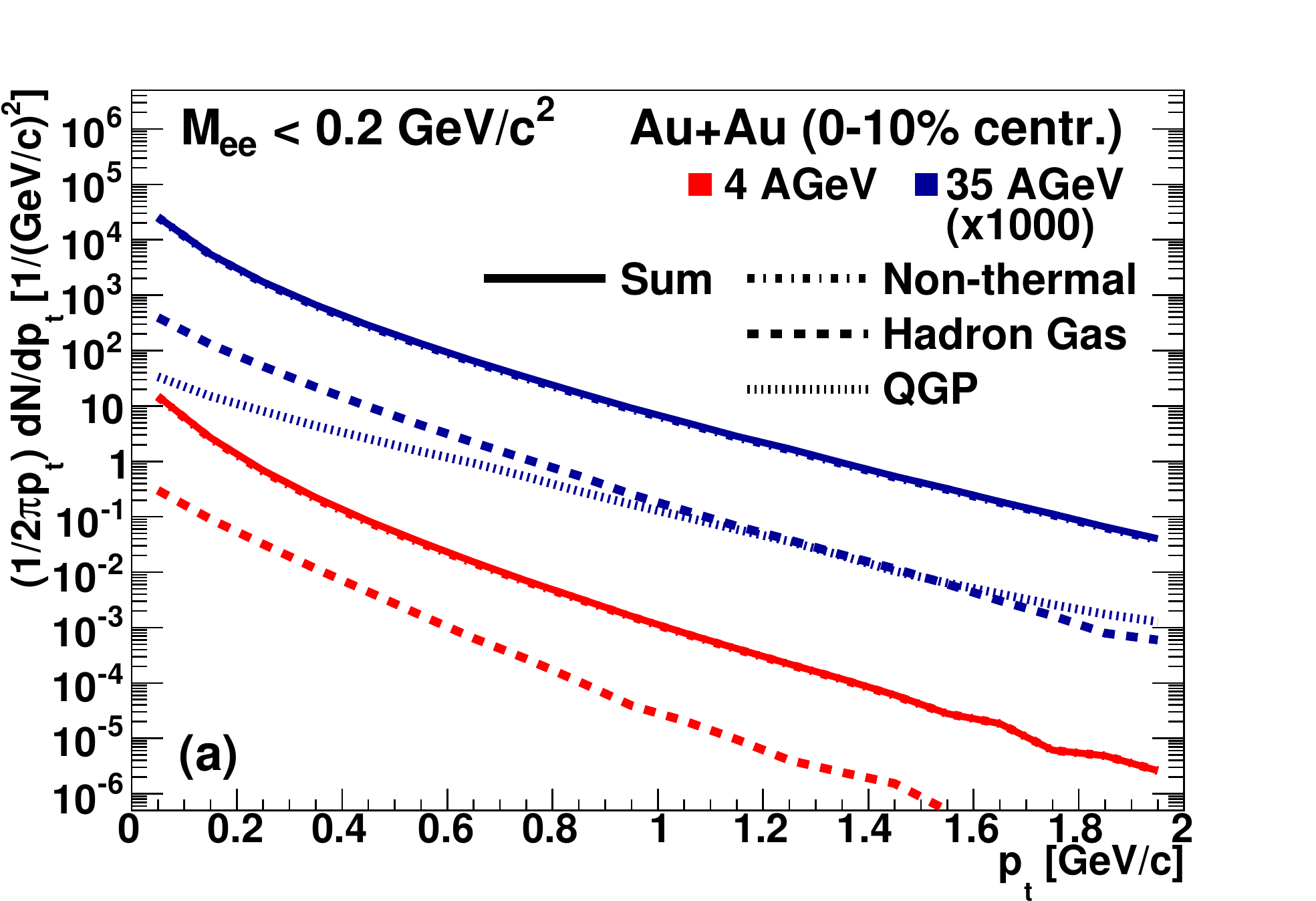}
\includegraphics[width=1.0\columnwidth]{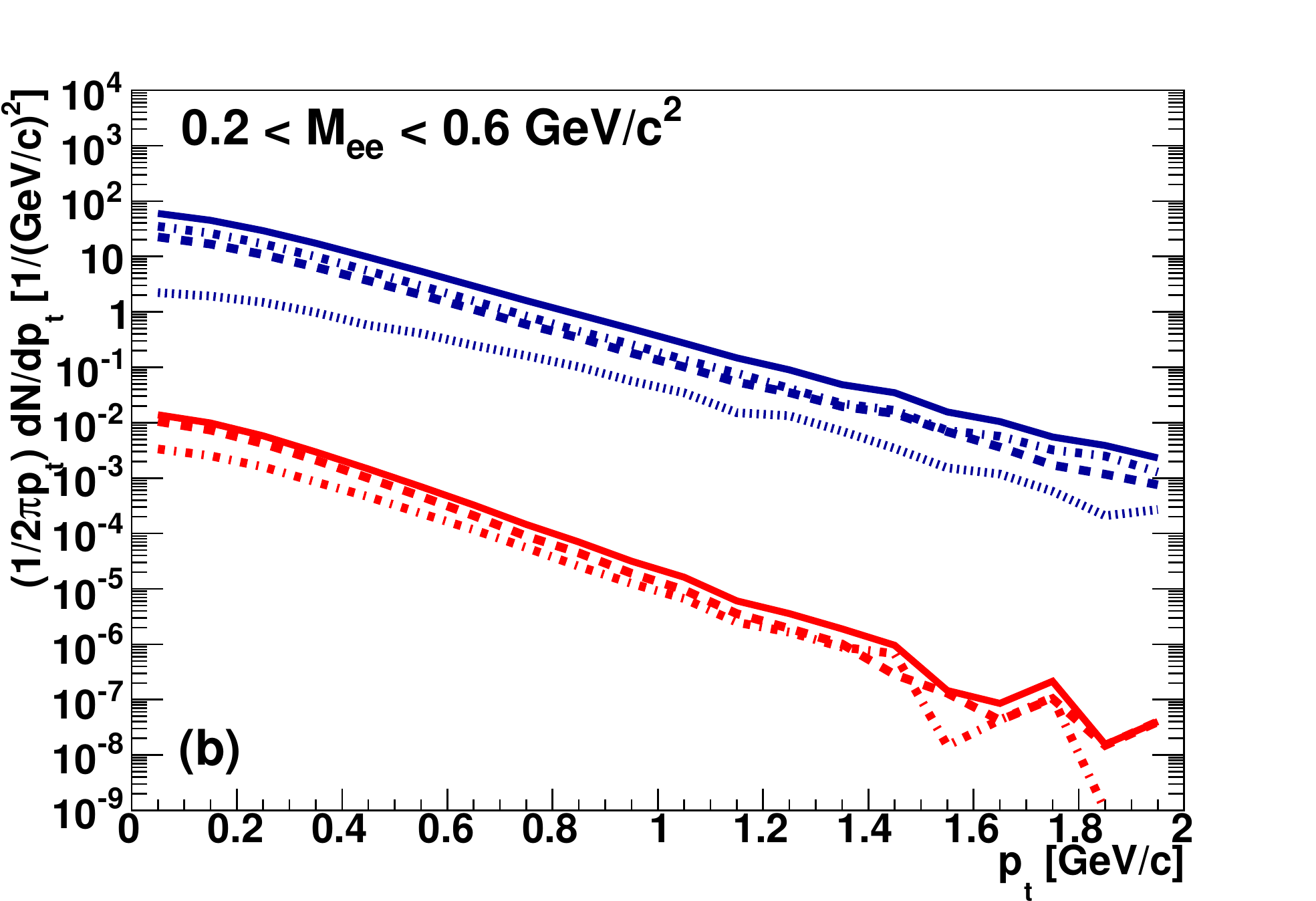}
\\
\includegraphics[width=1.0\columnwidth]{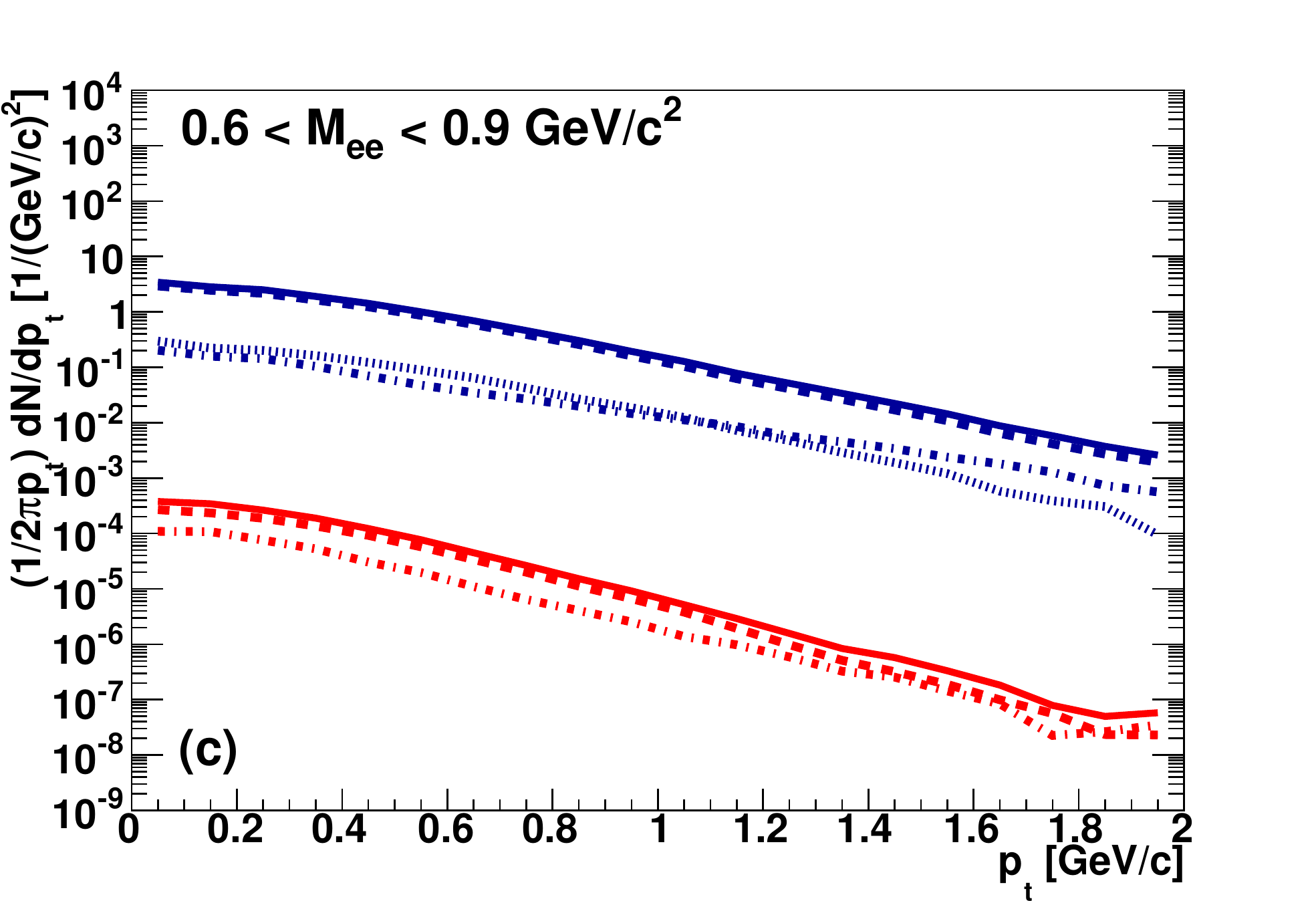}
\includegraphics[width=1.0\columnwidth]{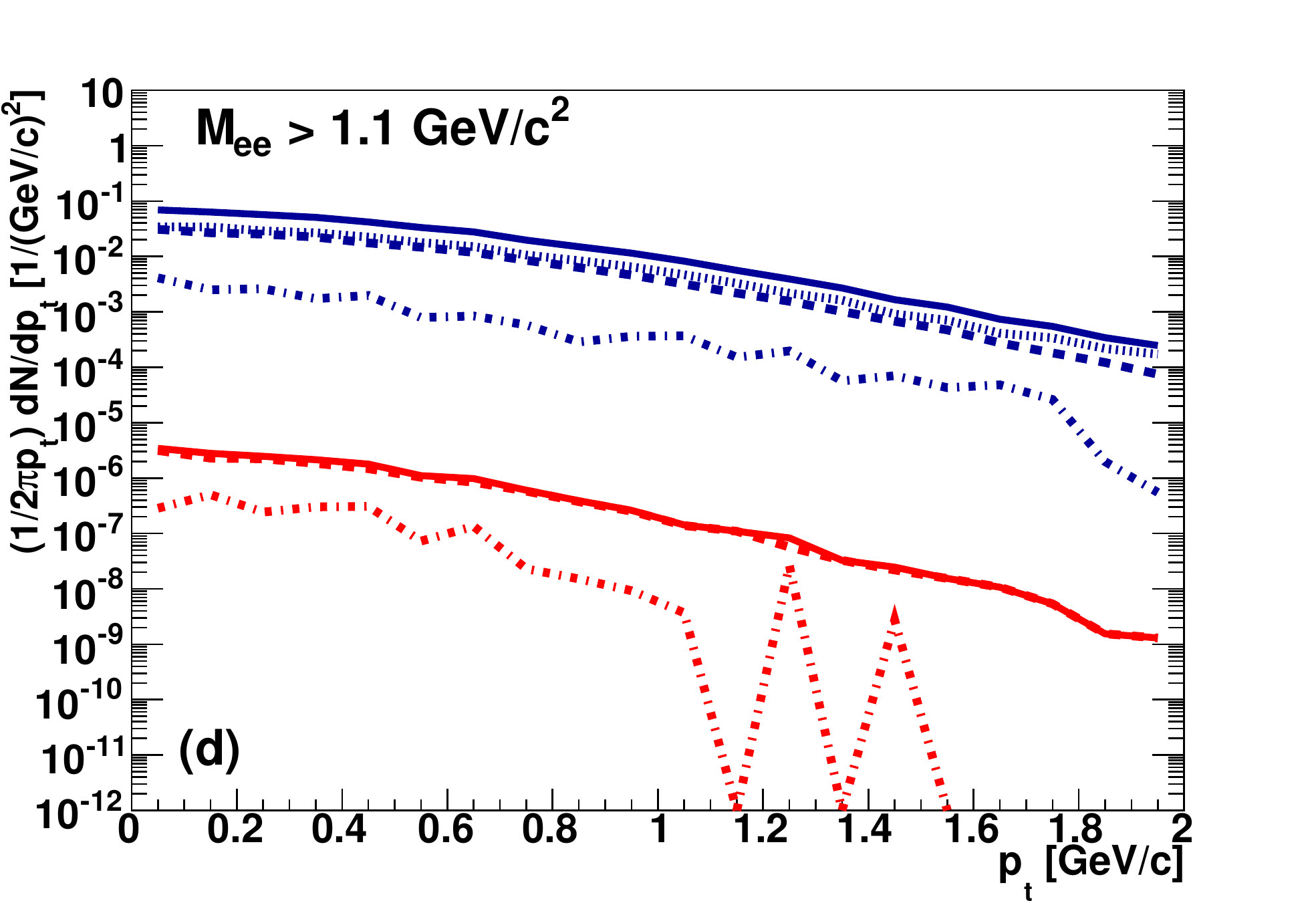}
\caption{(Color online) Transverse-momentum spectra of the
  dilepton yield in central Au+Au reactions $E_{\text{lab}}=4\,A\GeV$
  (red) and $35\,A\GeV$ (blue). The results are shown for four different
  invariant mass bins:
  $M_{\mathrm{e}^+ \mathrm{e}^-} < 0.2 \,GeV/c^{2}$ (a),
  $0.2 < M_{\mathrm{e}^+ \mathrm{e}^-} < 0.6\,\GeV/c^{2}$ (b),
  $0.6 < M_{\mathrm{e}^+ \mathrm{e}^-} < 0.9 \,\GeV/c^{2}$ (c) and
  $M_{\mathrm{e}^+ \mathrm{e}^-} > 1.1\,\GeV/c^{2}$ (d).}
\label{dilpt}
\end{figure*}
%%%%%%%%%%%%%%%%%%%%%%%%%%%%%%%%%%%%%%%%%%%%%%%%%%%%%%%%%%%%%%

Considering possible signatures for a phase transition and the creation
of a deconfined phase, the low-mass region is rather unsuited due to the
dominance of the hadronic cocktail contributions and hadronic thermal
emission from the vector mesons. Consequently, it might be more
instructive to explore the mass range above the pole mass of the $\phi$,
where one has a continuum dominated by thermal radiation. In this region the hadronic cocktail contributions can be
neglected and thermal sources will dominate the spectrum. In
previous works \cite{vanHees:2006ng, Endres:2014zua} it has been shown
for SPS energies that the dilepton invariant mass spectrum at very high
masses $M_{l^+l^-} > 1.5\,\GeV/c^{2}$ could only be explained by including thermal
radiation from the Quark-Gluon Plasma. In Fig.\,\ref{dilhighmass} the
higher invariant mass region for
$M_{\mathrm{e}^+ \mathrm{e}^-} > 1.1\,\GeV/ c^{2}$ is shown for the two
collision energies $E_{\text{lab}}=8\text{ and }35\,A\GeV$. Here we
compare four different scenarios to study whether the high-mass
invariant mass spectrum might help to identify the creation of a
Quark-Gluon Plasma. Besides the two standard scenarios (hadron gas +
partonic emission above $T_{\mathrm{c}}=170\,\MeV$) for (i) finite and
(ii) vanishing $\mu_{\pi}$, we include a scenario with (iii) a 5-times
enhanced emission from the partonic phase around the transition
temperature to simulate the effect of a critical slowdown of the system
due to a first order phase transition and, finally, (iv) a pure hadron
gas scenario, where we assume all thermal radiation (also for
$T > 170\,\MeV$) to stem from hadronic sources. For (iii) and (iv)
$\mu_{\pi}=0$ is assumed, too. The comparison shows that the spectral
shape of the total yield is very similar for all scenarios, at both
energies considered here. While the results for a purely hadronic
scenario and including QGP emission from temperatures above $T_{\mathrm{c}}$ give quite
the same results within 10\% deviation, also the artificially enhanced QGP
emission does not significantly increase the overall yield. In contrast, one observes
a very strong enhancement due to a finite pion chemical potential, which
shows up in our calculation by an overall increase by a factor of 5 at
$8\,A\GeV$ and still a factor of 2 at $35\,A\GeV$. The results indicate
that it will be difficult to draw unambiguous conclusions from single
measurements of the higher mass region at a specific energy, as
according to our calculations a stronger QGP yield and less hadronic
contribution can finally result in the same overall dilepton
spectrum. Furthermore, the non-equilibrium effects may lead to much
larger modifications of the spectrum than caused by the dynamics of the
phase transition.
%%%%%%%%%%%%%%%%%%%%%%%%%%%%%%%%%%%%%%%%%%%%%%%%%%%%%%%%%%%%%%
\begin{figure*}
\includegraphics[width=1.0\columnwidth]{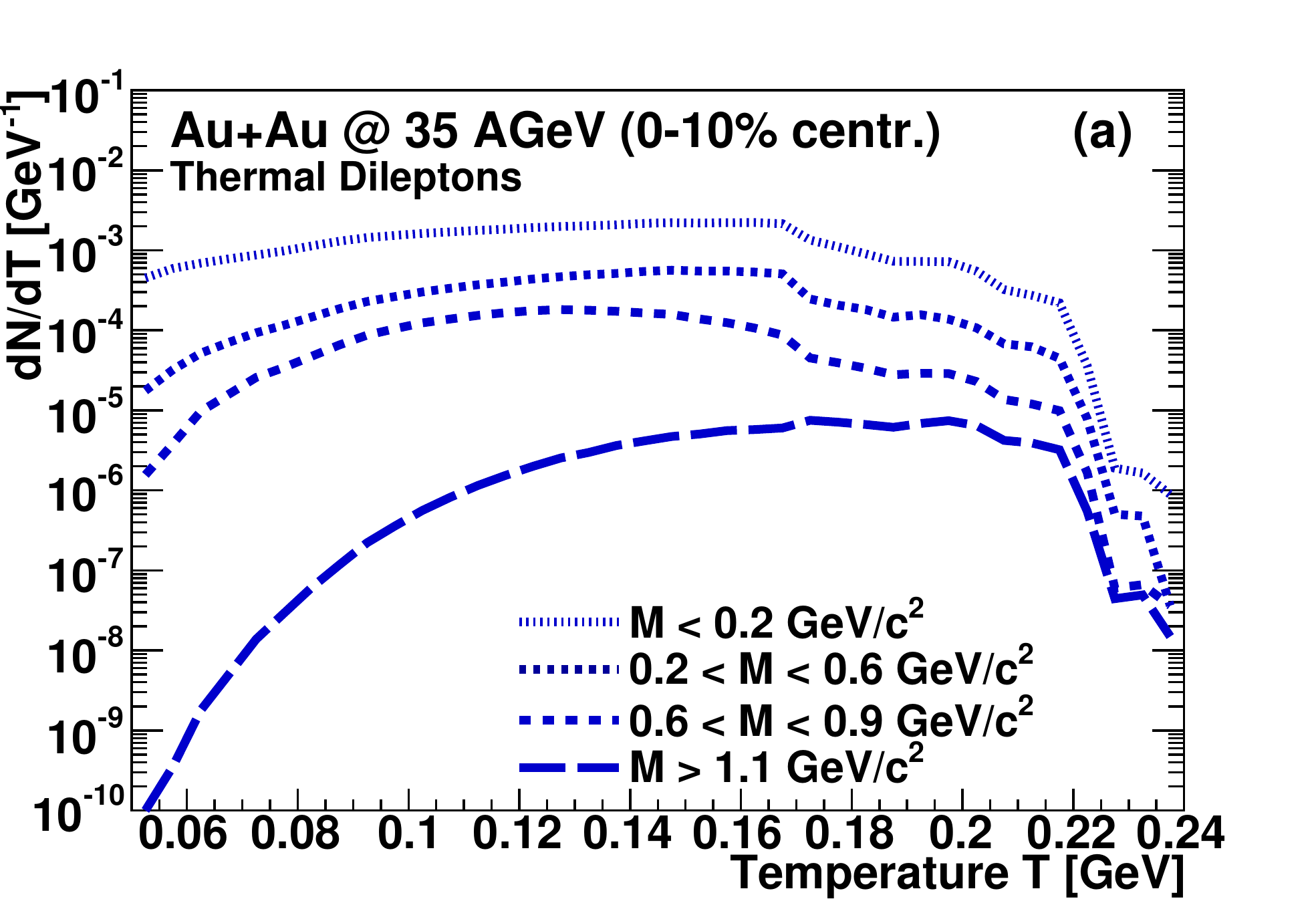}
\includegraphics[width=1.0\columnwidth]{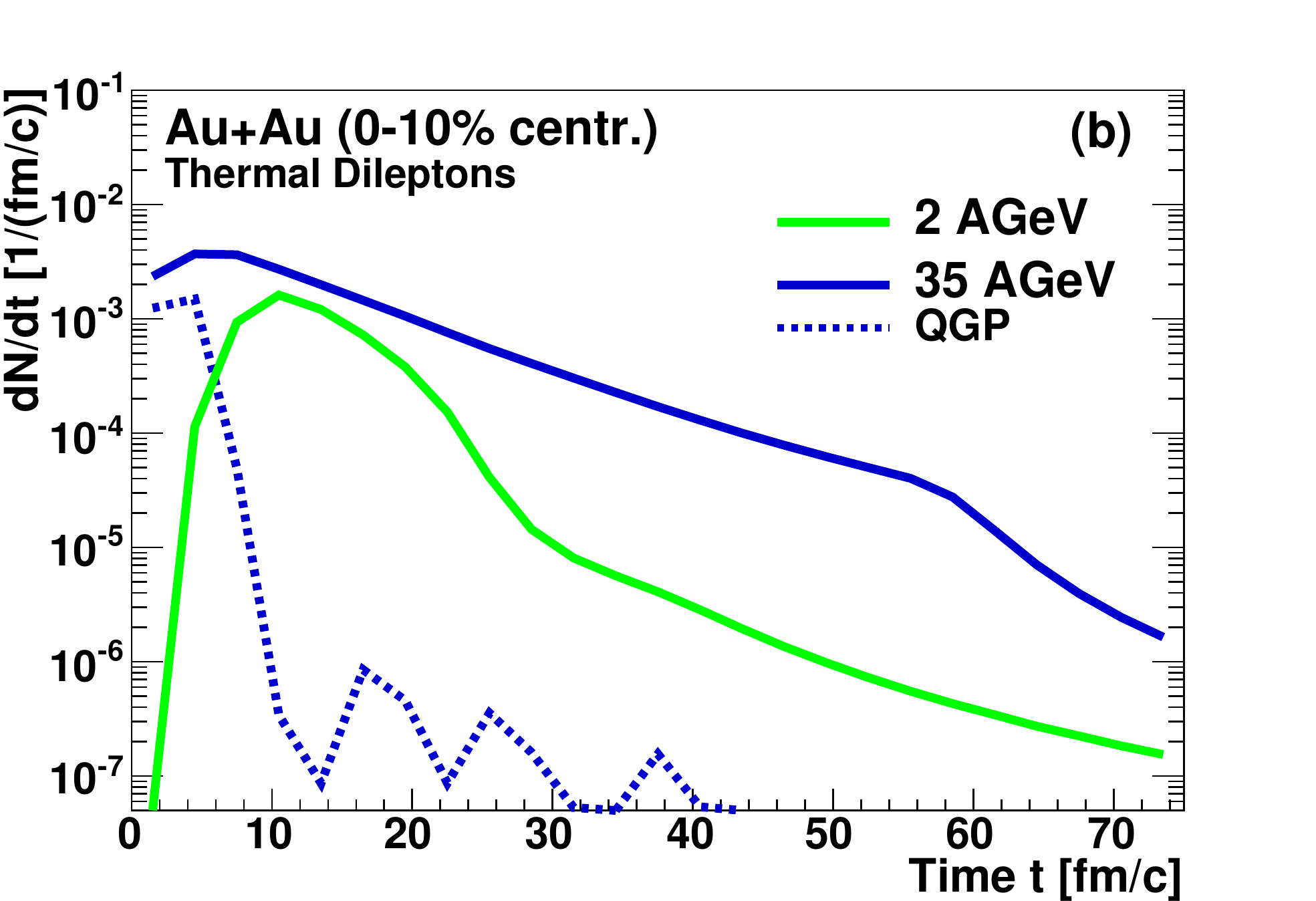}
\caption{(Color online) (a) Temperature dependence of dilepton emission $\mathrm{d}N/\mathrm{d}T$ from thermal sources for central Au+Au at $E_{\mathrm{lab}}=35$\,AGeV,
 i.e., the lepton pairs directly extracted from the
  hadronic cocktail as calculated with UrQMD are not included. The
  results are shown for four different invariant mass bins:
  $M_{\mathrm{e}^+ \mathrm{e}^-} < 0.2 \,\GeV/c^{2}$,
  $0.2 < M_{\mathrm{e}^+ \mathrm{e}^-} < 0.6 \,\GeV/c^{2}$,
  $0.6 < M_{\mathrm{e}^+ \mathrm{e}^-} < 0.9\,\GeV/c^{2}$ and
  $M_{\mathrm{e}^+ \mathrm{e}^-} > 1.1\,\GeV/c^{2}$; (b) Time evolution   $\mathrm{d}N/\mathrm{d}t$ of
  thermal dilepton emission for 2 (green) and 35\,AGeV (blue). The dashed line
  shows the emission from the QGP for the top FAIR energy.}
\label{diltempbin}
\end{figure*}
%%%%%%%%%%%%%%%%%%%%%%%%%%%%%%%%%%%%%%%%%%%%%%%%%%%%%%%%%%%%%%

The transverse-momentum spectra, plotted for two different
energies in different invariant mass bins in Fig.\,\ref{dilpt} underline
the previous finding. Again the slopes of the curves for hadronic and
partonic emission are very similar for high masses, especially for
$M_{\mathrm{e}^+ \mathrm{e}^-}>1.1\,\GeV$ they are virtually
identical. On the contrary we find the partonic contribution to be
harder (i.e., having a stronger relative yield at high transverse
momenta) than the hadronic contribution at lower masses. Together with
the very similar invariant-mass spectra at high masses obtained with and
without a partonic phase in the reaction evolution, the result confirms
previous studies which showed a duality of emission rates in the
transition-temperature region \cite{Rapp:2013nxa}. However, note that
this is no longer true if one goes to a temperature regime significantly
above $T_{\mathrm{c}}$. For this case clear differences between the hadronic
and the partonic emission are observed. Unfortunately, even at
the top SIS\,300 energy only few cells reach temperature maxima above 200\,MeV and one
will not see clear differences between the partonic and hadronic
emission patterns even at high $p_{\mathrm{t}}$ and high
$M_{\mathrm{e}^+ \mathrm{e}^-}$.

The reason for the duality showing up only at high invariant masses (and
momenta, respectively) is twofold: On the one hand, the low mass region
is governed by the vector mesons with their specific spectral shapes and
the baryonic effects on them. This effect has been called the ``duality
mismatch'' \cite{Rapp:1999ej} since the hadronic rates show an increase for finite
baryochemical potentials, while the partonic emission rates are quite
insensitive with regard to $\mu_{q}$. On the other hand, while the spectra at low
masses and momenta are populated by thermal emission at all
temperatures, the production of dileptons for masses above
$1\,\GeV/c^{2}$ and for higher values of $p_{\mathrm{t}}$ is strongly
suppressed at low temperatures. This is visible from
Fig.\,\ref{diltempbin}\,(a), where the temperature dependent dilepton yield
from thermal sources for Au+Au at $E_{\text{lab}}=35\,A\GeV$ is shown
for different invariant-mass bins. (Note that the non-thermal lepton
pairs directly extracted from the hadronic cocktail as calculated with
UrQMD are not included here). The yields shown in this plot represent the sum of the contributions from all cells at a certain temperature. While for the lowest mass bin $M_{\mathrm{e}^+ \mathrm{e}^-} < 0.2\,\GeV/c^{2}$ the total thermal
dilepton yield is built up by roughly equal fractions stemming from the
whole temperature range, with slight suppression of emission from
temperatures above $T_{\mathrm{c}}$, one can see that the mass region
above $M_{\mathrm{e}^+ \mathrm{e}^-}=1.1\,\GeV/c^{2}$ is dominated by
emission from temperatures between 140 to 220 MeV, which is exactly the
assumed transition region between hadronic and partonic
emission. Dilepton emission at lower temperatures is strongly suppressed
in this mass range. Furthermore, one finds a smooth behavior of the
thermal emission in the highest mass bin, but at lower masses one
observes a slight kink in the rates at $T_{\mathrm{c}}=170\,\MeV$. The
finding indicates that for lower masses the partonic and hadronic rates
do not perfectly match, as was discussed above. Another
observation is the dominance of emission from the temperature range
$T=100-140\,\MeV$ for the mass region from 0.6 to
$0.9\,\GeV/c^{2}$. This result is in contrast to the general trend of a
shift of the emission to higher temperatures when going to higher
masses. However, the very mass region covers the pole masses of the
$\rho$ and $\omega$ meson. As the peak structures show a melting
especially for finite baryon densities, one will get the largest yields
in this mass range from cells for which $\mu_{\text{B}}\approx 0$. A
comparison with the $T$-$\mu_{\text{B}}$ distribution of the cells at this
energy in Fig.\,\ref{fourvolev}\,(b) shows that the largest fraction of
cells for which the baryochemical potential is below 200\,MeV lies
exactly in the temperature range from 100 to 140\,MeV.
%%%%%%%%%%%%%%%%%%%%%%%%%%%%%%%%%%%%%%%%%%%%%%%%%%%%%%%%%%%%%%
\begin{figure*}
\includegraphics[width=1.0\columnwidth]{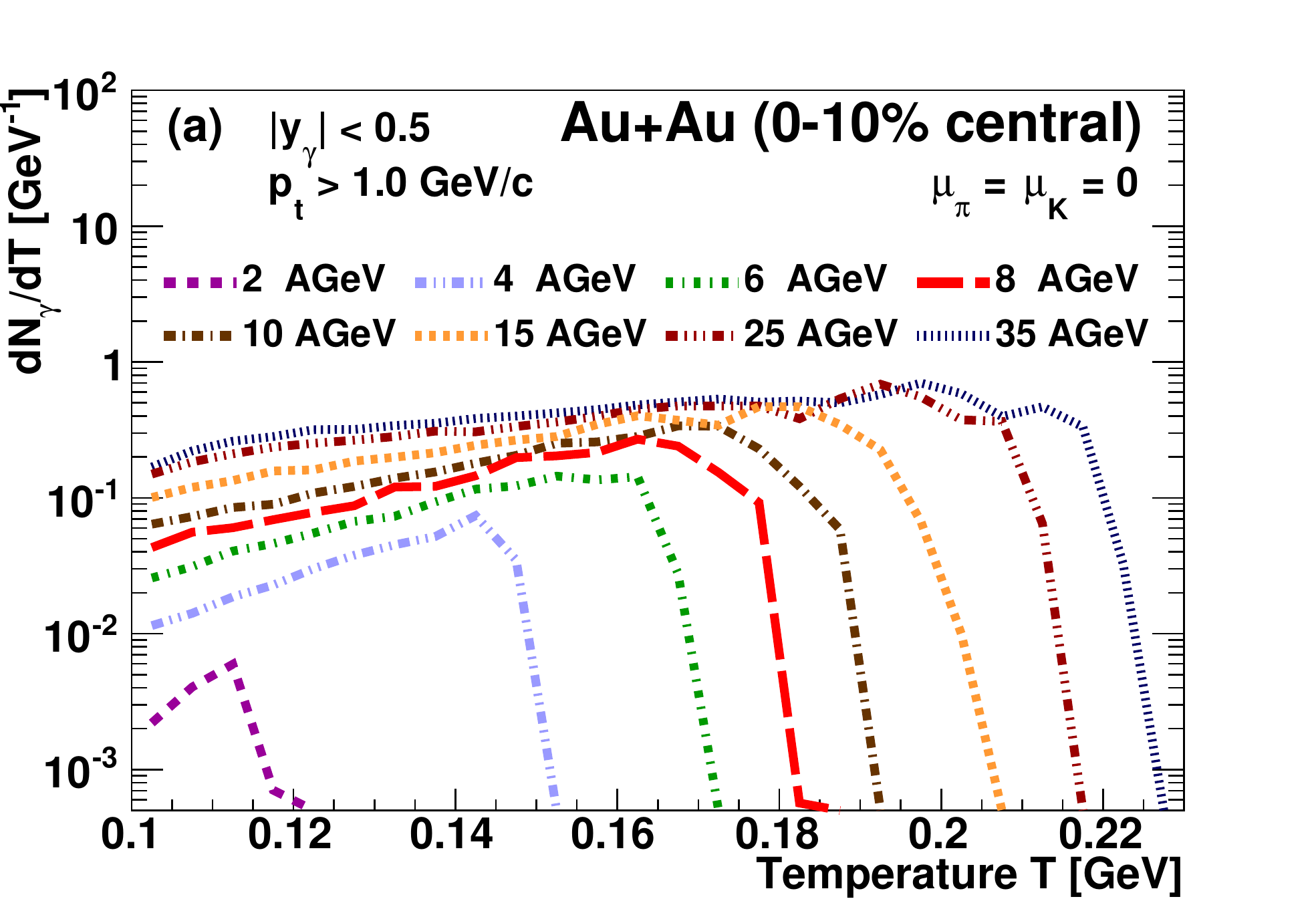}
\includegraphics[width=1.0\columnwidth]{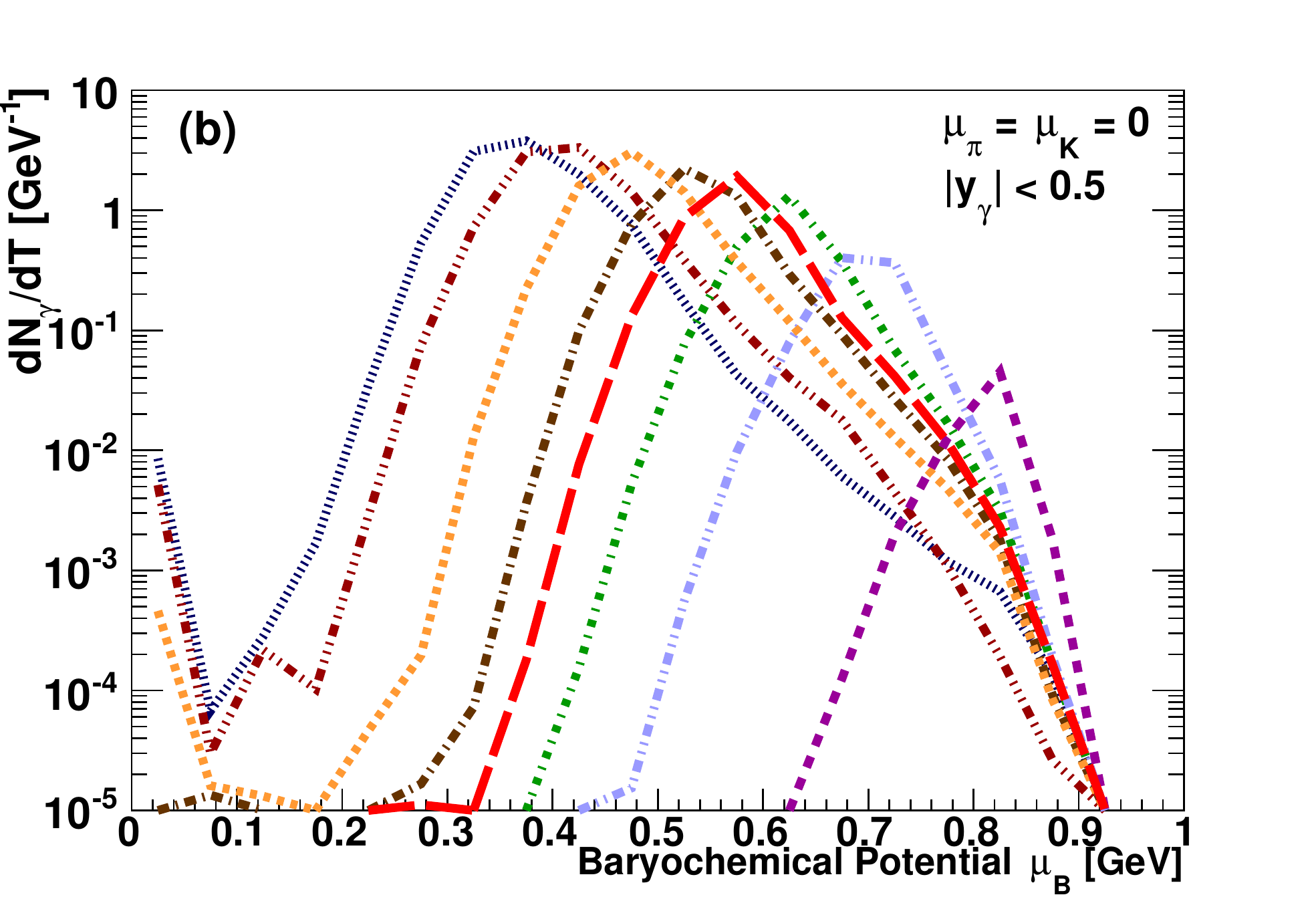}
\caption{(Color online) Thermal photon yield at mid-rapidity in 
  dependence on (a) temperature, $\mathrm{d}N_{\gamma}/\mathrm{d}T$, and (b) baryochemical 
  potential, $\mathrm{d}N_{\gamma}/\mathrm{d}\mu_{\text{B}}$, for central Au+Au collisions at different beam energies
  $E_{\text{lab}}=2-35\,A\GeV$. The results are shown for the case of 
  vanishing $\mu_{\pi}$ and $\mu_{\text{K}}$.}
\label{phtempbar}
\end{figure*}
%%%%%%%%%%%%%%%%%%%%%%%%%%%%%%%%%%%%%%%%%%%%%%%%%%%%%%%%%%%%%%

Finally, it is instructing not only to look at the temperature but also at the time dependence of thermal dilepton emission as presented in Figure\,\ref{diltempbin}\,(b). The total dilepton emission per timestep $\mathrm{d}N/\mathrm{d}t$ as sum from all cells is shown for $E_{\mathrm{lab}}=2$ and 35\,AGeV. In principle, the results reflect the findings from Fig.\,\ref{tmubev} and are similar to the temperature evolution depicted there. The system shows a faster heating for the top SIS\,300 energy with higher temperatures, resulting in a larger number of emitted dileptons; partonic emission from cells with $T > T_{c}$ is found for the first $5-10$\,fm/$c$ and only very sporadically thereafter. At lower energies the evolution is retarded and $T$ remains below the critical temperature. However, in  contrast to the slow heating of the system (which is simply due to the fact that the nuclei are moving slower) the thermal emission drops much earlier for 2\,AGeV (compared to 35\,AGeV) and only few cells with thermal emission are found for $t>30$\,fm/$c$. The higher energy deposited in the system with increasing $E_{\mathrm{lab}}$ obviously results not only in higher initial temperatures, but also in an enhanced emission at later stages, as it takes the system longer to cool down. Interestingly, for both energies one finds some sparse cells with thermal emission even after 60-70\,fm/c; however, their contribution to the overall result is suppressed by 3-4 orders of magnitude compared to the early reaction stages. 
%%%%%%%%%%%%%%%%%%%%%%%%%%%%%%%%%%%%%%%%%%%%%%%%%%%%%%%%%%%%%%
\subsection{\label{ssec:PhSpec} Photon spectra}

While we have two kinematic variables (momentum and invariant mass)
which can be probed for virtual photons, real (i.e., massless) photons
only carry a specific energy. In this sense, dileptons are the more
versatile probes of the hot and dense medium and carry additional
information, especially regarding the spectral modifications of the vector
mesons. Nevertheless, the correct description of the experimental photon
spectra has been a major challenge for theory at SPS and RHIC
energies. In the kinematic limit
$M \rightarrow 0$ several processes become
dominant which are negligible in the time-like region probed by
dileptons, as was discussed in Sec.\,\ref{sec:Rates}. Consequently, the
study of photon production can provide complementary information for the
theoretical description of thermal emission rates and the reaction
dynamics.

As the parametrized photon emission rates have some restrictions with
regard to the $\mu_{\text{B}}$ range of their applicability, it will be
instructive to find out at the beginning under which thermodynamic
conditions the photons are emitted at FAIR. Fig.\,\ref{phtempbar} shows the
dependence of thermal photon emission (at mid-rapidity) on temperature in the
left plot (a) and on baryochemical potential in the right plot (b). As in Fig.\,\ref{diltempbin}, the yields are the sum of the thermal contributions from all cells with a certain temperature or baryochemical potential, respectively. For
both results we consider the case of vanishing meson chemical
potentials. Note that for the temperature dependence we consider only
the thermal emission at higher transverse momentum values
$p_{t} > 1$\,GeV/$c$, as here the duality between hadronic and partonic
rates should be approximately fulfilled, which is indeed visible from
the continuous trend of the thermal photon emission around
$T_{\mathrm{c}}$. One can see that especially for lower collision
energies the thermal emission is dominated by the cells which reach the
maximum temperature, whereas the curves become flatter at higher
energies. Even at the top energy of $35\,A\GeV$ with maximum
temperatures above 220\,MeV still a significant amount of emission also
stems from the cells with temperatures around 100 MeV.
%%%%%%%%%%%%%%%%%%%%%%%%%%%%%%%%%%%%%%%%%%%%%%%%%%%%%%%%%%%%%%
\begin{figure*}
\includegraphics[width=1.0\columnwidth]{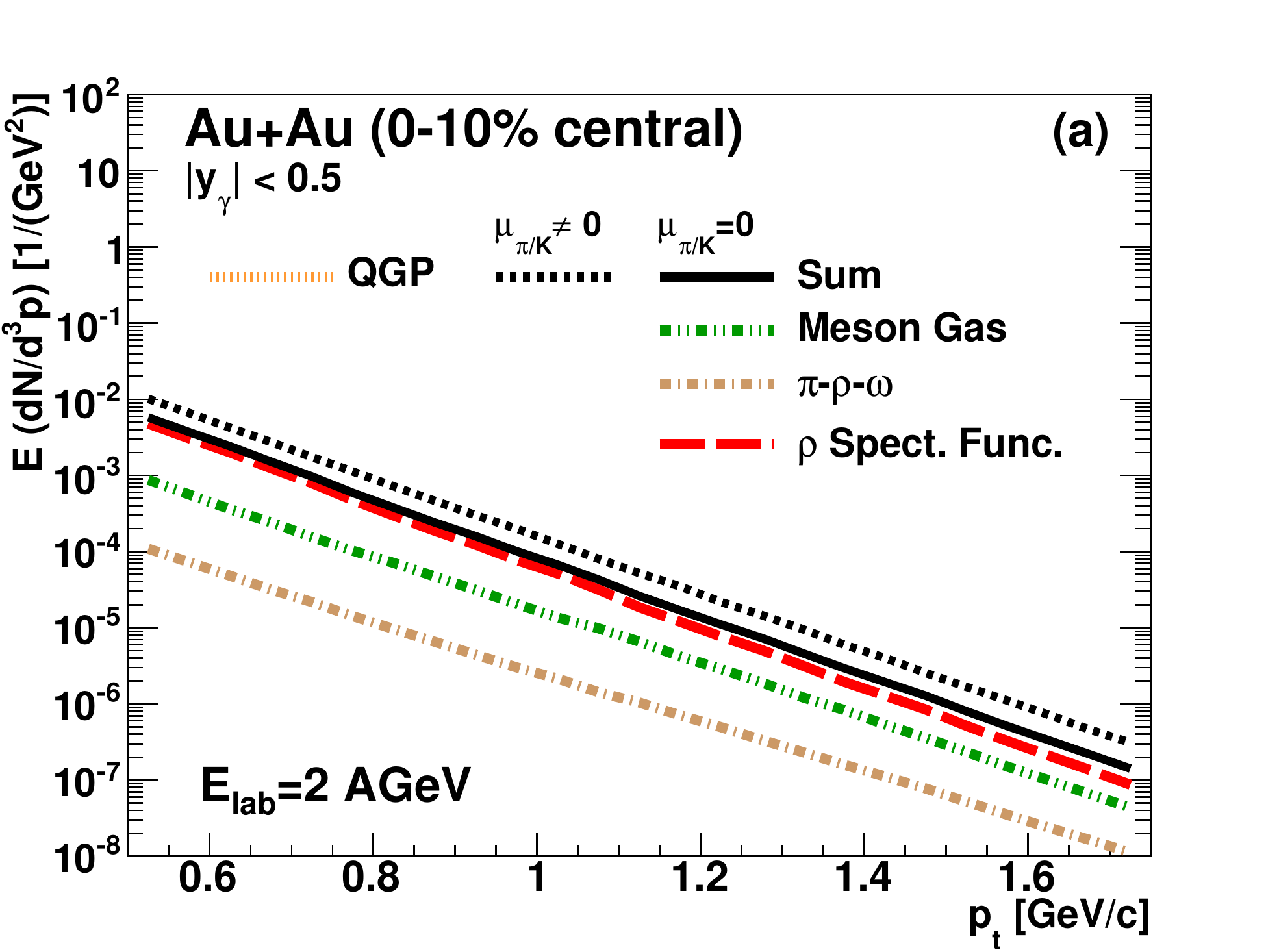}
\includegraphics[width=1.0\columnwidth]{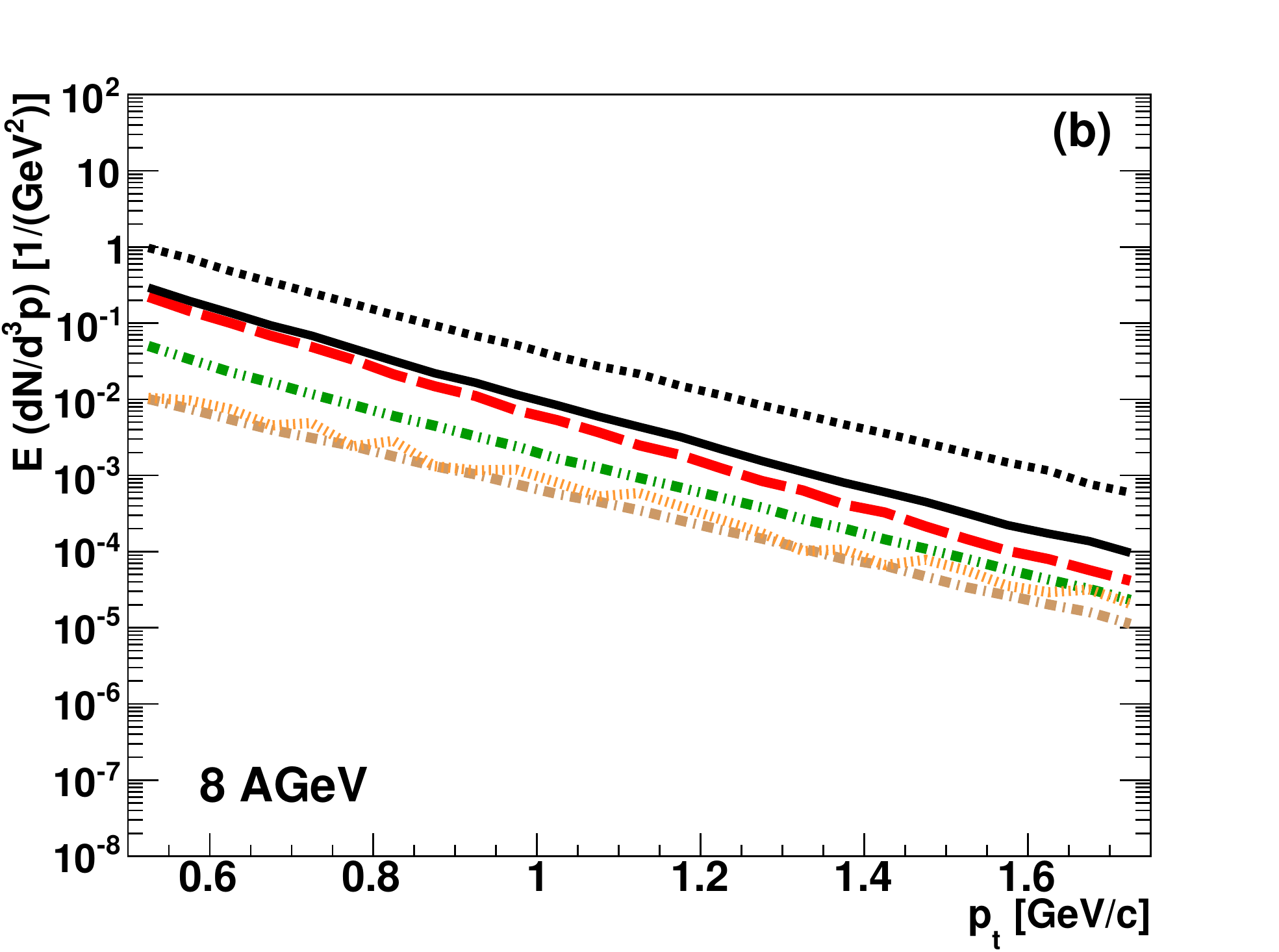}
\\
\includegraphics[width=1.0\columnwidth]{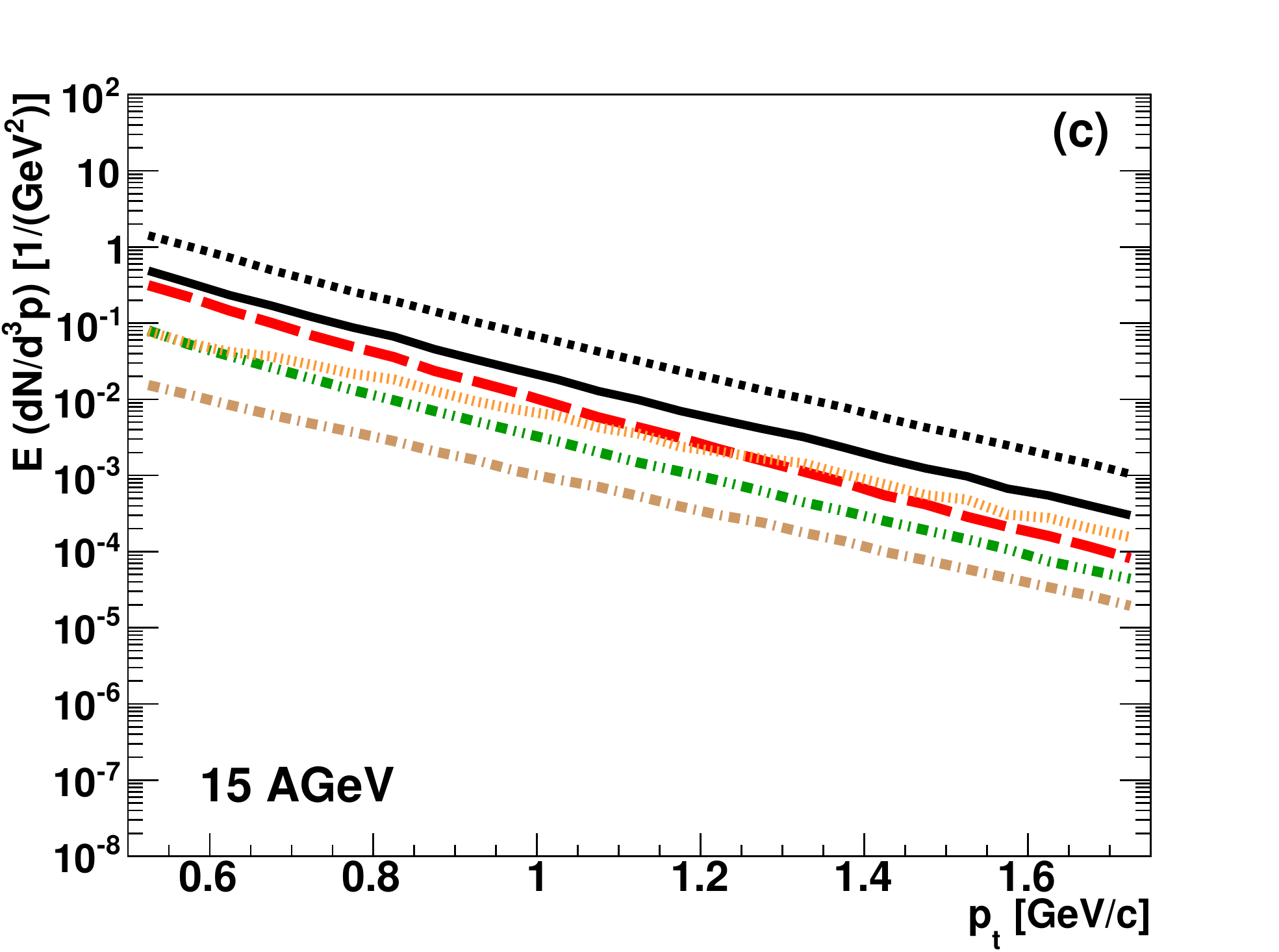}
\includegraphics[width=1.0\columnwidth]{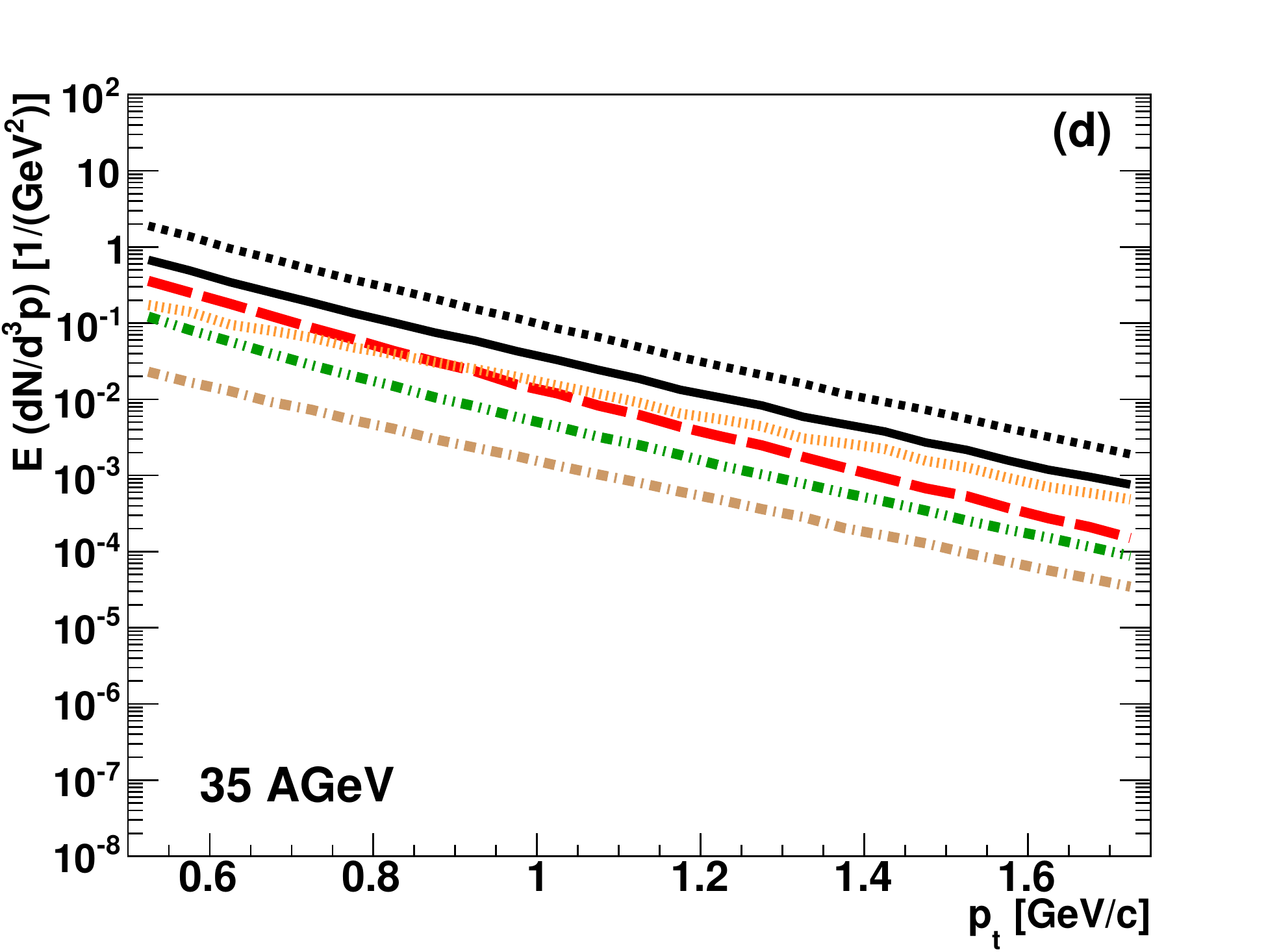}
\caption{(Color online) Transverse-momentum spectra at mid-rapidity
  ($|y_{\gamma}|<0.5$) of the thermal photon yield for central Au+Au
  reactions at $E_{\text{lab}}=2\,A\GeV$ (a), $8\,A\GeV$ (b),
  $15\,A\GeV$ (c) and $35\,A\GeV$ (d). The total yields are plotted for
  both cases $\mu_{\pi}=0$ (full black line) and $\mu_{\pi}\neq 0$
  (dashed line). The single hadronic contributions from the $\rho$
  spectral function (red, long dashed), the meson gas (green,
  dashed-double-dotted) and the $\pi-\rho-\omega$ complex (beige,
  dashed-dotted) are only shown for vanishing pion chemical
  potential. The partonic contribution from the QGP is plotted as orange
  short-dashed line.}
\label{photonpt}
\end{figure*}
%%%%%%%%%%%%%%%%%%%%%%%%%%%%%%%%%%%%%%%%%%%%%%%%%%%%%%%%%%%%
\begin{figure*}
\includegraphics[width=1.0\columnwidth]{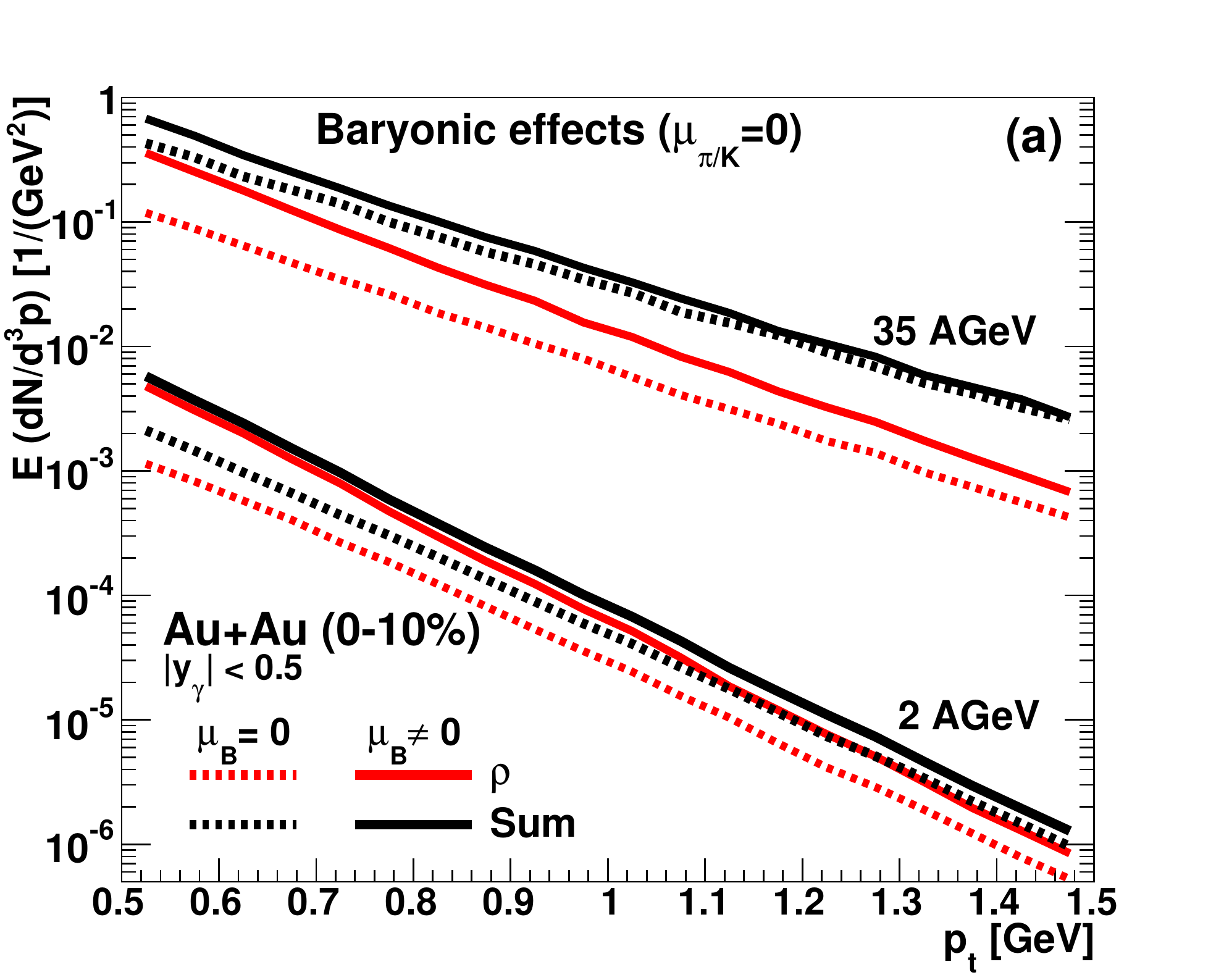}
\includegraphics[width=1.0\columnwidth]{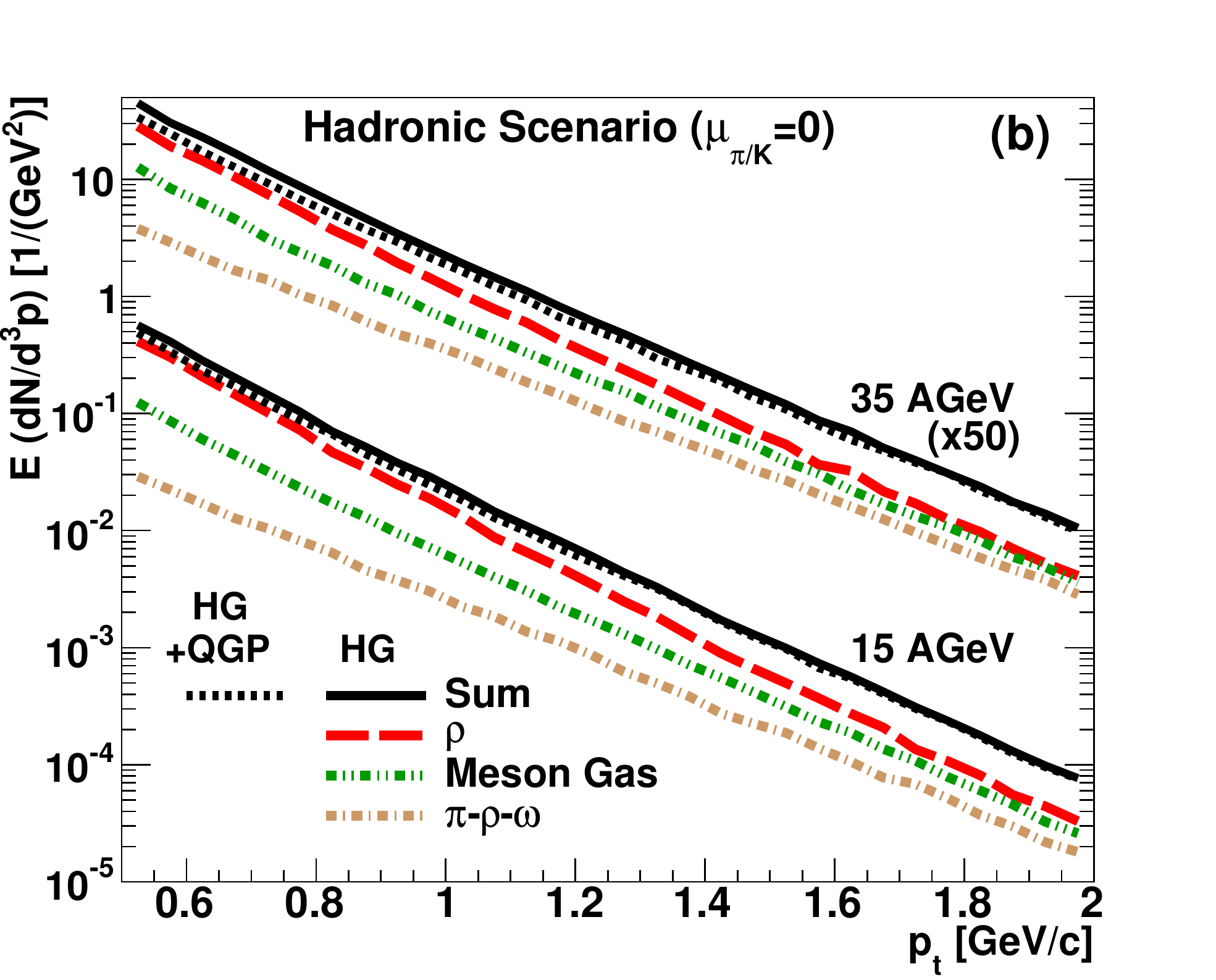}
\caption{(Color online) Comparison of the transverse-momentum spectra at mid-rapidity
  ($|y_{\gamma}|<0.5$) of the thermal photon yield for central Au+Au
  reactions resulting from different emission scenarios. In (a) the effect of a finite baryon chemical
  potential $\mu_{\mathrm{B}}$ on the transverse momentum spectrum for the $\rho$
  contribution and the total yield is shown for $2\,A\GeV$ (lower
  results) and $35\,A\GeV$ (upper results), comparing the standard
  scenario with $\mu_{\mathrm{B}}\neq 0$ (full lines) with the results for
  $\mu_{\mathrm{B}}=0$ (dashed lines). Plot (b) shows the results for a purely
  hadronic scenario, i.e., for emission from the hadron gas also for
  $T > 170\,\MeV$ and no partonic contribution. The single contributions
  are plotted as in Fig.\,\ref{photonpt}; for comparison the total yield for the
  standard result including hadronic + partonic emission is shown (black
  dashed).}
\label{photonptcomp}
\end{figure*}

%%%%%%%%%%%%%%%%%%%%%%%%%%%%%%%%%%%%%%%%%%%%%%%%%%%%%%%%%%%%%%

Regarding the photon emission related to baryochemical potential as
presented in Fig.\,\ref{phtempbar} (b), a clear energy dependent trend
is visible: While at low energies the largest fraction of emission stems
from cells with very high values of $\mu_{\text{B}}$ around 900\,MeV,
the emission weighted average chemical potential drops continuously to
300--400\,MeV at $35\,A\GeV$. However, the emission from cells with
higher values of the baryochemical potential is by far not
negligible. In consequence, the findings once again underline that the
strongest baryonic modifications of the spectral functions will be
present at low energies. The results also show that to fully account for
the baryonic effects on the photon emission the spectral functions
should be able to reliably cover the whole $\mu_{\text{B}}$ region from
0 to the nucleon mass (i.e. $\approx 900 \,\MeV$). The presently used
parametrization will provide only a lower limit for the photon yield,
especially for the lower collision energies.

In Fig.\,\ref{photonpt}(a)-(d) the transverse momentum spectra of
thermal photons for four different collision energies are presented. As
for the dilepton invariant mass spectra, we show the results with (full
black line) and without meson chemical potentials (dashed black); once
again the two calculations provide a lower and upper boundary for the
off-equilibrium influence on the thermal yields, respectively. Two
observations can be made when comparing the results for the different
energies: An overall increase of the photon $p_{\mathrm{t}}$-yield with
increasing energy and, secondly, a simultaneous hardening of the spectra,
i.e., one gets a stronger relative contribution for higher momenta. This
is similar to the dilepton invariant-mass spectra, where the yield in
the higher mass region is suppressed for lower collision energies due to
the lower overall temperatures in the fireball. (A more explicit
comparison of the energy dependence of the results will be undertaken in
Sec.\,\ref{ssec:CompPhoDil}). Furthermore, one can see that at all
energies the contribution from the $\rho$ meson dominates above the
other hadronic contributions especially for low $p_{\mathrm{t}}$, while
the relative dominance of the $\rho$ decreases for higher momenta. The
contribution from the Quark-Gluon Plasma is visible for
$E_{\text{lab}}=8\,A\GeV$ and higher energies, giving an
increasing fraction of the overall yield. Note the similarity between
the low-mass dilepton and photon $p_{t}$ spectra for $35\,A\GeV$: In
both cases the slope of the (virtual or real) photons emitted from 
the QGP stage is significantly harder than the contribution
from hadronic sources. Furthermore, looking only at the $\rho$ and the
partonic contribution, one finds that the first is stronger for
$p_{t}<1\,\GeV/c$, and the latter dominates for higher momenta - for
both dileptons and photons. This behavior is expected as
the real photon represents just the
$M_{\mathrm{e}^+ \mathrm{e}^-} \rightarrow 0$ limit of virtual photon
production. This is another requirement of consistency for the thermal
rates.

Although finite values of $\mu_{\pi}$ and $\mu_{\mathrm{K}}$ have an even more
pronounced effect on the photon rates than on the dilepton
rates, as several processes to be considered are very sensitive to an
overpopulation of pions, it is remarkable that the overall effect leaves
the shape of the photon $p_{\mathrm{t}}$ spectra mostly unchanged: The
yields are enhanced by the same factor at all
transverse momenta. This is interesting as the effect of
$\mu_{\pi/K} \neq 0$ on the different contributions is varying in
strength. For example, the yield from the $\pi-\rho-\omega$ system shows
a much stronger enhancement than the $\rho$ contribution (compare
Table\,\ref{tab:rates}).

But not only meson chemical potentials influence the photon spectra,
similar to the case of dileptons one expects also an enhancement of the
$\rho$ contribution in the presence of baryonic matter. In
Fig.\,\ref{photonptcomp}\,(a) the effect of a finite baryon chemical
potential $\mu_{\mathrm{B}}$ on the transverse momentum spectrum for the $\rho$
contribution and the total yield is shown for $2\,A\GeV$ and
$35\,A\GeV$, comparing the standard scenario with $\mu_{\mathrm{B}}\neq 0$ (full
lines) with the results for $\mu_{\mathrm{B}}=0$ (dashed lines). The comparison
shows that especially at lower momenta the $\rho$ contribution is
significantly increased for finite baryochemical potential, while this
effect is less dominant at larger $p_{\mathrm{t}}$. Furthermore the
effect is stronger for lower collision energies, where one obtains
larger average values of $\mu_{\text{B}}$. However, one should bear in
mind that the parametrization for the photon emission rates is limited
to chemical potentials below 400\,MeV, so that one can not fully account
for the very large chemical potentials in this case. In
consequence, one can expect an even larger enhancement in the
experimental measurements than in the present calculation.

To conclude the study of the different influences on the photon spectra,
we also consider whether the possible creation of a deconfined phase
has any effect on the thermal emission pattern. Plot (b) of
Fig.\,\ref{photonptcomp} shows the results for a purely hadronic scenario,
i.e., for emission from the hadron gas also for $T > 170\,\MeV$ and no
partonic contribution. For comparison also the total yield for the
standard scenario including hadronic + partonic emission is shown. (For
both cases the meson chemical potentials are assumed to be zero.) Again,
as for the high-mass dileptons (compare Fig.\,\ref{dilhighmass}), the
differences between the two scenarios are negligible, especially compared
to the effect of the meson and baryochemical potentials. Only a very
slight enhancement of the yield at low momenta is obtained for the pure
hadron gas scenario, reflecting the different sensitivity of partonic
and hadronic rates to finite $\mu_{\text{B}}$.
%%%%%%%%%%%%%%%%%%%%%%%%%%%%%%%%%%%%%%%%%%%%%%%%%%%%%%%%%%%%%%
\begin{figure*}
\includegraphics[width=1.0\columnwidth]{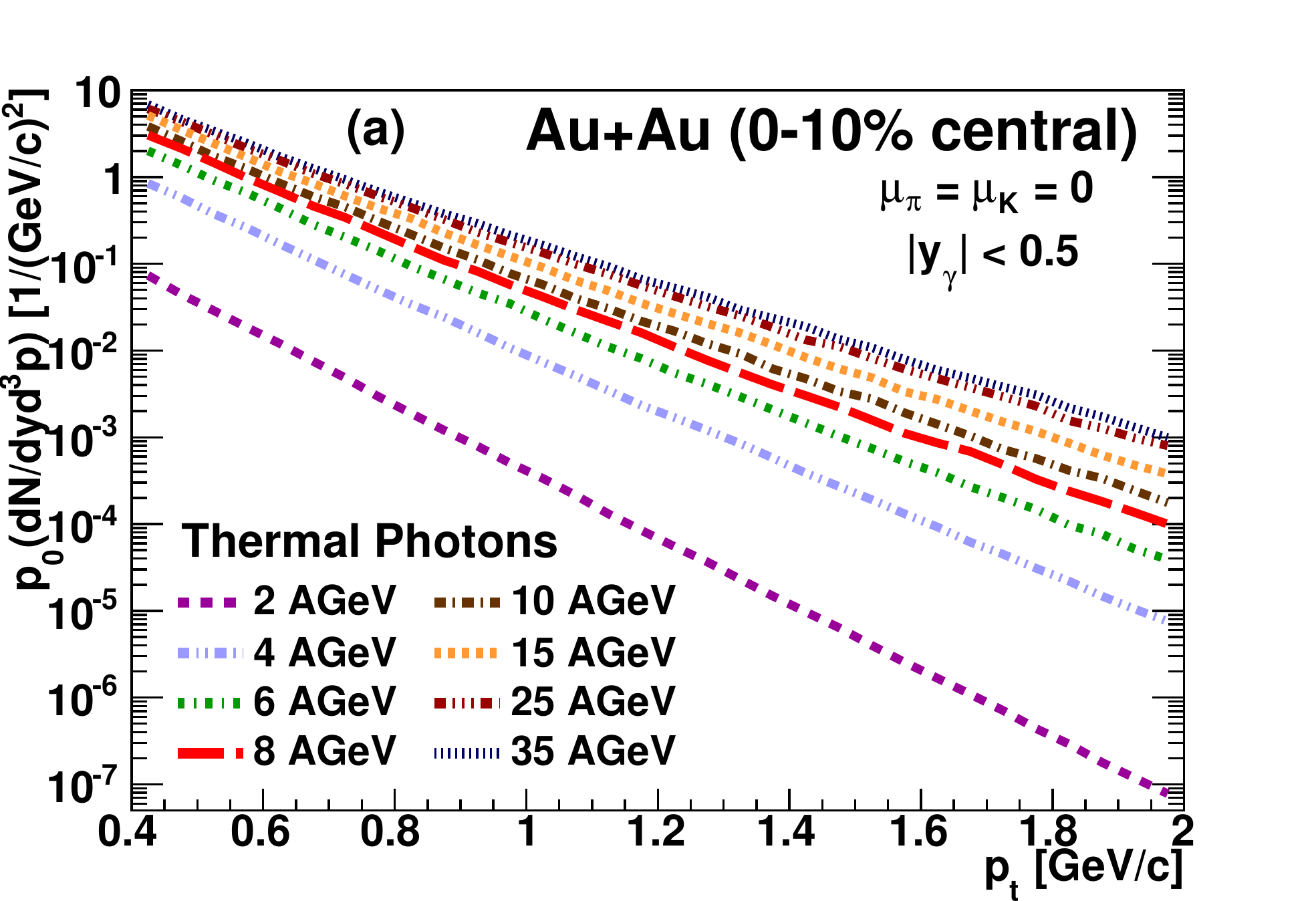}
\includegraphics[width=1.0\columnwidth]{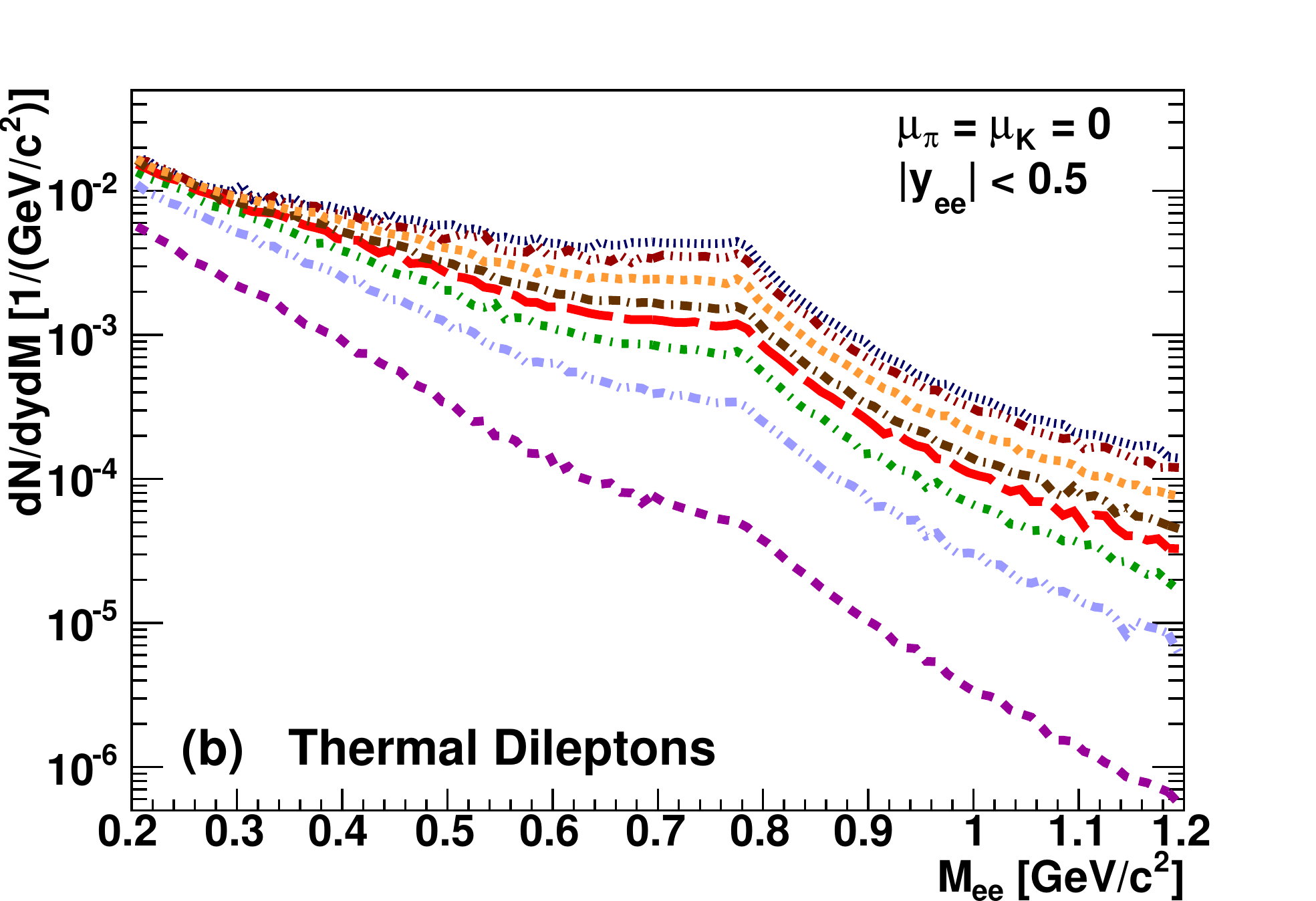}
\caption{(Color online) Comparison of the overall photon transverse
  momentum spectra (a) and the
  thermal dilepton invariant mass spectra (b) at mid-rapidity $|y|<0.5$ for different energies in
  the range $E_{\text{lab}}=2-35\,A\GeV$. The results are shown for
  vanishing meson chemical potentials $\mu_{\pi}=\mu_{\mathrm{K}}=0$.}
\label{theryield}
\end{figure*}
%%%%%%%%%%%%%%%%%%%%%%%%%%%%%%%%%%%%%%%%%%%%%%%%%%%%%%%%%%%%%%

\subsection{\label{ssec:CompPhoDil} Excitation function of photon and dilepton yields}
In the previous sections several differences and similarities between
dilepton and photon spectra have already been discussed. However, it is
instructing to do this in more detail and to compare the energy
dependence of the emission patterns for photons and dileptons. Considering
experimental measurements, an advantage of studying the excitation
function of thermal yields might be that the trends and results
obtained hereby at different energies are more robust and less sensitive
to errors of measurement. It reflects the results of several different
measurements in contrast to single spectra at a specific energy. For reason of comparison and because the baryonic effects are strongest in this case, all results in the following will be considered for mid-rapidity $|y|<0.5$.

\begin{figure*}
\includegraphics[width=1.0\columnwidth]{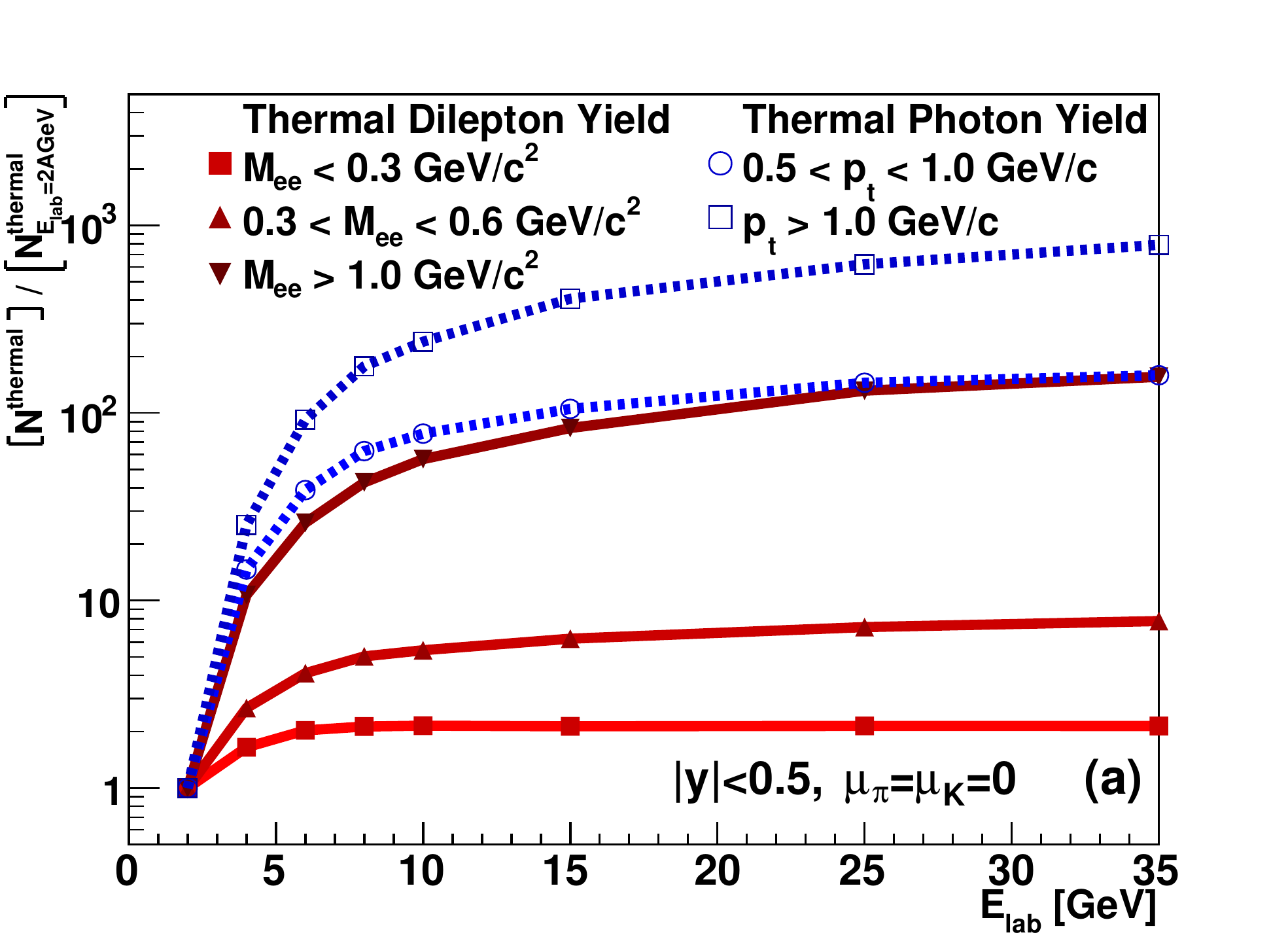}
\includegraphics[width=1.0\columnwidth]{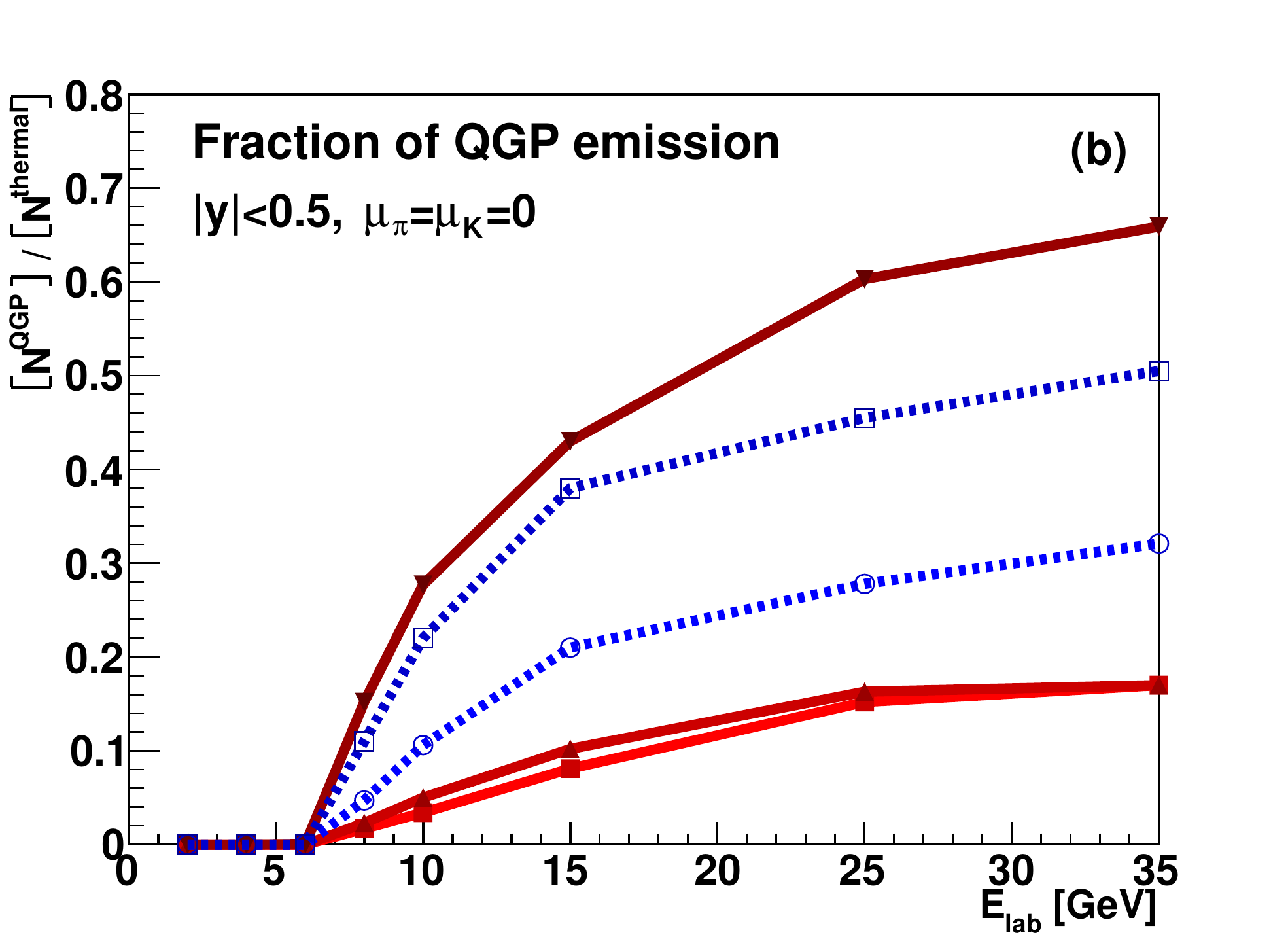}
\caption{(Color online) (a) Energy dependence of the thermal photon and
  dilepton yield in different invariant mass or transverse momentum bins,
  respectively. The yields are normalized to the result at
  $E_{\text{lab}}=2\,A\GeV$. (b) Fraction of thermal QGP emission in relation to the
  total thermal yield of photons of dileptons in different
  $M_{\mathrm{e}^+ \mathrm{e}^-}$ and photon-$p_{\mathrm{t}}$ bins. 
  All results in this figure are shown for mid-rapidity and vanishing meson chemical potentials.}
\label{yieldratios}
\end{figure*}

In Fig.\,\ref{theryield} the total thermal photon $p_{\mathrm{t}}$ and
dilepton $M_{\mathrm{e}^+ \mathrm{e}^-}$ spectra for eight different
collision energies in the range $E_{\text{lab}}=2-35\,A$GeV are shown. It was already mentioned that---besides the hadronic
structures due to the direct connection of the dilepton spectrum with
the spectral function of the light vector mesons---the two spectra are
strikingly similar. And also the change of the spectra with increasing
collision energy is alike. At high masses/momenta the yield
shows a stronger increase with $E_{\text{lab}}$ than in the
low-mass/ -momentum region. More quantitatively, this can be seen in
Fig.\,\ref{yieldratios}\,(a), where the relative increase of the thermal
photon and dilepton yield for different transverse momentum or invariant
mass regions, respectively, is shown. The results are normalized to 1
for $E_{\text{lab}}=2\,A\GeV$. One observes that the relative increase
is stronger for high momenta and masses than for the lower
$p_{\mathrm{t}}$ or $M_{\mathrm{e}^+ \mathrm{e}^-}$ bins. For example,
the total dilepton yield for masses below $300\,\MeV/c^{2}$ increases
only by a factor of 2 from $E_{\text{lab}}=2 \text{ to } 8\,A$GeV and remains nearly constant thereafter,
whereas in the high mass region above $1\,\GeV/c^{2}$ the yield increases
by a factor 300 when going to the top SIS\,300 energy of 35\,$A$GeV. A similar behavior is found for the photons,
where the yield shows a more pronounced rise at high momenta. As was
pointed out already before, one reason for this is the fact that much
energy is needed to produce a dilepton at high masses or a photon with
high momentum. Their production is strongly suppressed at the rather
moderate temperatures obtained at lower collision energies. Note that in
general the overall increase in the photon spectra is slightly stronger
than for the dilepton production. This might be due to the limitation of
the photon parametrization to temperatures above 100\,MeV and
baryochemical potentials below 400 MeV, which might somewhat
underestimate the photon yield at the lowest collision energies. Besides, one should keep in mind that in detail the processes contributing to the thermal emission rates for dileptons and photons differ, which may also explain some differences between the results.
%%%%%%%%%%%%%%%%%%%%%%%%%%%%%%%%%%%%%%%%%%%%%%%%%%%%%%%%%%%%%%
%%%%%%%%%%%%%%%%%%%%%%%%%%%%%%%%%%%%%%%%%%%%%%%%%%%%%%%%%%%%%%
\begin{figure}[b]
\includegraphics[width=1.0\columnwidth]{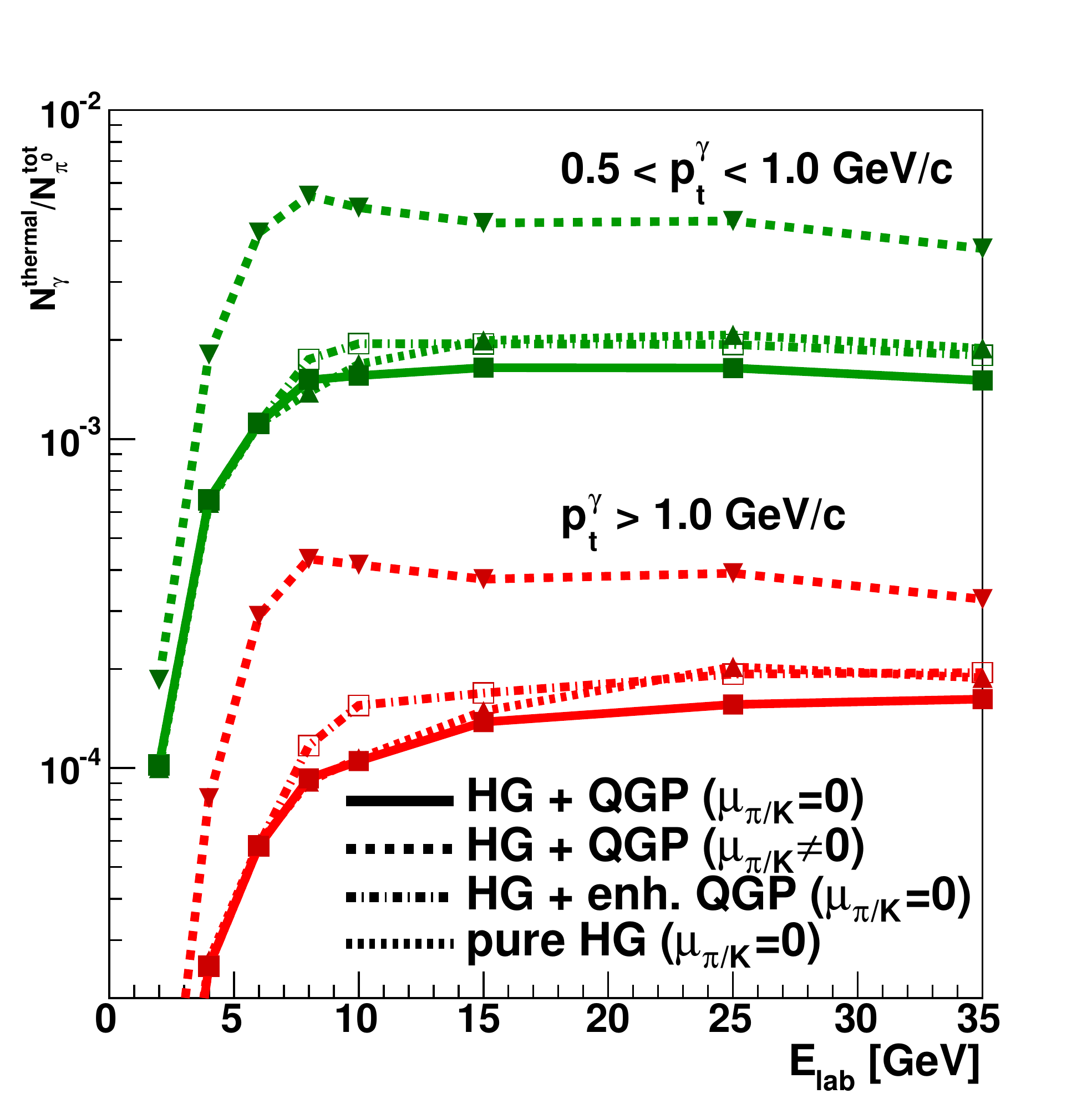}
\caption{(Color online) Energy dependence for the ratio of thermal
  photon yield $N^{\text{thermal}}_{\gamma}$ at mid-rapidity ($|y|<0.5$) to the overall number of
  neutral pions $N_{\pi^{0}}$. The results are shown for two different
  regions of the photon transverse momenta,
  $0.5 < p_{\mathrm{t}} < 1.0\,\GeV/c$ (green) and
  $p_{\mathrm{t}} > 1.0$\,GeV/$c$ (red). The baseline calculation
  assumes hadronic emission up to 170\,MeV with vanishing $\mu_{\pi/K}$
  and partonic emission for higher temperatures (full line). In addition, the results with a five times enhanced emission around the critical
  temperature (dashed-dotted line), for a pure hadron gas scenario with
  emission at all temperatures from hadronic sources (short dashed line)
  and including meson chemical potentials (dashed line) are shown.}
\label{ratiophpi}
\end{figure}
%%%%%%%%%%%%%%%%%%%%%%%%%%%%%%%%%%%%%%%%%%%%%%%%%%%%%%%%%%%%%%

In contrast, we find that the fraction of the QGP yield compared to the
total thermal emission is larger for the high-$M_{\mathrm{e}^+ \mathrm{e}^-}$ dileptons compared to the high-$p_{\mathrm{t}}$
photons. The first significant QGP contribution is found for $E_{\text{lab}}=8\,A\GeV$, and
the fraction continuously increases up to roughly 70\% at $35\,A\GeV$
for dilepton masses above $1\,\GeV/c^{2}$ and 50\% for photon momenta over
$1\,\GeV/c$. At lower masses or momenta, respectively, the hadronic
contribution becomes more dominant. This is not surprising, as one could
already conclude that we probe lower temperatures at lower masses and
momenta (compare Fig.\,\ref{diltempbin}). Furthermore the baryonic
influence increases here. In the dilepton spectra the direct connection
to the vector meson spectral functions makes a comparison between the
photon and dilepton results for low momenta and masses difficult.

Finally, we consider also the thermal photon yield in relation to the
number of (neutral) pions which are produced in the heavy-ion collision,
as presented in Fig.\,\ref{ratiophpi}. This ratio is of theoretical and 
experimental interest: For the experimental study of photons, decays 
of neutral pions are the major background in the analysis. On the 
theoretical side, the number of pions gives an estimate  
of the freeze-out volume and is
not sensitive to the details of the reaction evolution. While the
electromagnetic emission takes place over the whole lifetime of the
fireball and therefore reflects the evolution of the system in the phase
diagram, the pion yield allows to scale cut trivial dependences.
Previously, the thermal dilepton yield was found to scale with
$N_{\pi}^{4/3}$ if one compares different system sizes at SIS\,18
energies \cite{Endres:2015fna}. However, the situation is more complex
here as we consider a large range of energies, which will be covered by
FAIR. Several effects play a role, e.g., the lifetime of the
fireball, the temperatures and baryochemical potentials which are
reached and the different processes which contribute at different
temperatures. While in our study at SIS\,18 energies only the system size
was modified (and all the other parameters could be assumed to remain
quite constant, as one single energy was considered), in the
present study \emph{only} the size of the colliding nuclei is constant.

The investigation of the $N_{\gamma}/N_{\pi^{0}}$ ratio is here combined with a comparison of the different scenarios
for the conditions of thermal emission, which were already studied in
case of the photon and dilepton spectra (see Figs.\,\ref{dilhighmass}
and \ref{photonpt}). Varying
$E_{\text{lab}}$ (and, in consequence, $T$ and $\mu_{\mathrm{B}}$) might
result in distinct excitation functions of the ratio
$N_{\gamma}/N_{\pi^{0}}$ for the various scenarios, in contrast to the spectra for one specific energy where no unambiguous distinction was possible. In
Fig.\,\ref{ratiophpi}, one can see a strong increase of the photon to
pion ratio for the lowest energies for both the lower $p_{\mathrm{t}}$
range from 0.5 to 1\,GeV/$c$ and the high transverse momentum region
above 1\,GeV/$c$ in case of the baseline scenario with QGP emission for
$T>170$\,MeV and $\mu_{\pi}=0$. However, for higher collision energies
above 10\,$A$GeV we still observe a further increasing ratio for the
higher momentum range, whereas at low $p_{\mathrm{t}}$ the ratio remains
relatively constant and even decreases for the highest collision
energy. One can understand the decreasing ratio for lower
$p_{\mathrm{t}}$ reviewing again the energy dependence of $T$ and
$\mu_{\text{B}}$ as shown in Fig.\,\ref{tmubev}. The rise of temperature
becomes less intensive for higher collision energies, while the
baryochemical potential decreases for higher collision energies, causing
a less pronounced increase of the thermal yield (compare also
Fig.\,\ref{yieldratios}). Besides, the effects due to finite
$\mu_{\text{B}}$ are more pronounced at low momenta, explaining the
different trends for the two $p_{\mathrm{t}}$ regions. Including the
finite meson chemical potentials, we observe a strong increase of the
ratio by factors 2-5 at all collision energies. The strongest effect in
the present calculation is seen around $8\,A\GeV$, so that a slight peak
structure builds up. However, as mentioned already several times, this
scenario can only be seen as an upper limit for the non-equilibrium
effects, most probably the increase will be smaller.

When comparing the two scenarios including enhanced QGP emission
around $T_{\mathrm{c}}$ on the one side and a pure hadronic scenario on
the other side it is interesting that both cases lead to a similar
result, namely an increase of the $N_{\gamma}/N_{\pi^{0}}$ ratio at
higher collision energies. The effects show up more dominantly at high
momenta, as this region is more sensitive with regard to emission from
high temperatures. Note, however, that there are also significant
differences between the two cases. The scenario with enhanced QGP
emission around $T_{\mathrm{c}}$ shows the most prominent increase at
$E_{\text{lab}}=8-10\,A\GeV$ whereas this enhancement becomes smaller
again for higher energies. This can be explained by the fact that the
relative fraction of emission from temperatures around 170-175\,MeV is
largest at those collision energies where the transition to a partonic
phase is just reached. At higher $E_{\text{lab}}$ the corresponding
higher temperatures may outshine any effects from the transition
region. On the contrary, the enhancement over the baseline scenario
increases with energy for the case of a pure hadron gas. However, for
highest collision energies the difference seems to remain stable or even
to drop again.

We remind again that for the experimental measurement the ratio of thermal photons
from Fig.\,\ref{ratiophpi} is of importance, as almost all of the
$\pi^{0}$ mesons decay into a photon. Therefore the vast majority will
be decay photons, not stemming from direct (thermal or prompt) emission
processes. Their spectra have to be subtracted in experiment to draw
conclusions about the direct photons from thermal sources. This might be
relatively difficult for the lowest energies available at FAIR,
as here the ratio is suppressed by up to an order of magnitude compared
to the higher collision energies.

\section{\label{sec:Scenario} Discriminating different scenarios}

It has so far become clear that one can extract only limited information
regarding the properties of the hot and dense fireball from single
photon and dilepton spectra, as there are usually several different
effects that might interfere and finally lead to the same invariant-mass
or transverse-momentum yields. However, the picture might be quite
different if---in addition---the results in distinct mass or momentum
regions, respectively, are systematically compared for several collision
energies. In this case one might be able to discriminate the hadronic and
partonic effects from each other. The FAIR energy regime will be
ideally suited for such a study, as the transition from pure hadronic
fireballs to the creation of a deconfined phase will take place
somewhere around $E_{\text{lab}}=6-8\,A\GeV$, as our results
suggest. Nevertheless, there is no single observable that seems to allow for
unambiguous conclusions on the details of the reaction evolution. On the
contrary, it will still be necessary to carefully compare theoretical
calculations and experimental results.

Based on the findings of the present work, one may consider the
following scheme which might help to determine the strength of the
different effects on the thermal rates and discriminate between the
contributions:
\begin{enumerate}
\item The influence of baryonic matter leads to an enhancement which is
  most dominant for low transverse momenta and low masses. In general,
  it steepens the $p_{\mathrm{t}}$ slope of the overall yield. A large
  advantage is that today the spectral function of the $\rho$ meson is
  quite well known from previous experimental and theoretical
  studies. Detailed and precise photon (dilepton) measurements for low
  momenta (low masses) in the FAIR and RHIC-BES energy regime might give further
  constraints for the spectral function in the region of extremely high
  baryon densities and can, vice versa, help to see whether the models
  correctly describe the fireball evolution in terms of
  $\mu_{\mathrm{B}}$.
\item In contrast to the baryonic effects on the emission rates,
  non-equilibrium effects caused by finite pion (and kaon) chemical
  potentials will show up as enhancement in the dilepton and photon
  spectra at all masses and momenta, and will be visible at all
  collision energies. The effect should be slightly more dominant in the
  high invariant mass or high $p_{\mathrm{t}}$ region, as here the
  multi-meson contributions become more pronounced. Ideally, one can
  discriminate between the $\mu_{\text{B}}$- and $\mu_{\pi}$-driven
  effects by comparing the modification of the slope and the overall
  enhancement in relation to baseline calculations.
\item If the baryon and non-equilibrium effects are under
  control, one might be able to find signals from the partonic phase in
  the dilepton and photon spectra for high $p_{\mathrm{t}}$ and
  $M_{\mathrm{e}^+ \mathrm{e}^-}$. In general, the dilepton and photon
  rates do not differ much from each other around $T_{\mathrm{c}}$, but
  effects such as a critical slowdown of the evolution might lead to an
  increased yield from the Quark-Gluon Plasma. On the other hand the
  ``duality mismatch'' might lead to a relative decrease of the yield,
  as hadronic rates are sensitive to finite baryon and meson chemical
  potentials while the QGP rates show hardly any modification. However,
  any effects connected to a phase transition can only show up if the
  obtained temperatures are large enough. Consequently, we would observe
  subsequent modifications of the spectra only for energies larger than
  $E_{\text{lab}}=6-8\,A\GeV$, in contrast to non-equilibrium and
  baryonic effects which also appear at lower temperatures. Significant
  differences from calculations which \emph{only} show up for the higher
  energies might then indicate the creation of a deconfined phase.
\item Furthermore, these effects should be dominant in the regions which
  are most sensitive to QGP formation: The region for
  $M_{\mathrm{e}^+ \mathrm{e}^-} > 1 \,\GeV$ in the dilepton spectra,
  and for high-$p_{\mathrm{t}}$ in the photon respectively low-mass
  dilepton spectra (provided, it is possible to get control over the
  hadronic decay background). Another advantage in these regions is that
  they are relatively insensitive to the finite baryon chemical
  potential.
\end{enumerate}
The different issues are not easy to disentangle and
several interdependencies exist. Another aspect, which is not explicitly
considered in our work but might further complicate the situation, is
the influence of different EoS on the thermal yields. In the present
work we use a Hadron-Gas EoS to provide consistency with the
underlying microscopic model. However, previous investigations in a
transport + hydrodynamics hybrid model showed that an MIT Bag Model EoS
or a chiral EoS lead to different evolutions in the hydrodynamic phase
compared to the HG-EoS, resulting in higher temperatures and,
consequently, an increase of the emission rates.

\section{\label{sec:Conclusion} Conclusions \& Outlook}

We have presented photon and dilepton spectra for the collision energy
range $E_{\text{lab}}=2-35\,A\GeV$, which will be covered by the future
FAIR facility (and, in parts, by phase II of the BES at RHIC). The
calculations were performed using a coarse-graining approach with
transport simulations from the UrQMD model as input. In this approach
local particle and energy densities are extracted from an ensemble
average of the microscopic transport calculations, and an equation of
state is used to calculate the corresponding values for temperature and
chemical potential. The local thermodynamic properties are then used to
determine the thermal emission rates.

The resulting spectra show a strong influence of finite baryochemical
potentials and an enhancement due to non-equilibrium effects caused by
finite meson chemical potentials. Regarding the search for signals of
the deconfinement phase transition, there is no clear signal from which
the creation of a partonic phase can be unambiguously inferred.
Similarly, the results suggest that it is also hard to identify the type
of the transition, whether it is a cross-over or a first-order phase
transition, as effects due to a critical slowing down might be small
compared to other influences on the spectra. The main difficulty is the
dual connection between hadronic and partonic emission rates in the
transition region around the critical temperature $T_{\mathrm{c}}$,
resulting in very similar slopes in the invariant-mass and
transverse-momentum spectra.

For a clarification of the open issues, experimental input is
needed. Our results suggest that one needs very precise and detailed
measurements, as different evolution scenarios for the nuclear
collisions are modifying the dilepton and photon spectra in a quite
subtle manner. Systematic studies of several collision energies in the
future FAIR energy range from $E_{\text{lab}}=2\,A\GeV$ to $35\,A\GeV$
are required to get more insights into the structure of the phase
diagram of QCD matter and especially to find clues for the creation of a
deconfined phase. Besides the experimental efforts, it will be similarly
important to intensify the theoretical studies. \\

%%%%%%%%%%%%%%%%%%%%%%%%%%%%%%%%%%%%%%%%%%%%%%%%%%%%%%%%%%%%%%
\begin{acknowledgments} 
  The authors especially thank Ralf Rapp for providing the
  parametrizations of the spectral functions and many fruitful
  discussions. This work was supported by the Hessian Initiative for
  Excellence (LOEWE) through the Helmholtz International Center for FAIR
  (HIC for FAIR), the Bundesministerium für Bildung und Forschung, Germany (BMBF)
  and the Helmholtz-Gemeinschaft through the Research School for Quark-Matter Studies (H-QM).
\end{acknowledgments}
%%%%%%%%%%%%%%%%%%%%%%%%%%%%%%%%%%%%%%%%%%%%%%%%%%%%%%%%%%%%%%
%%%%%%%%%%%%%%%%%%%%%%%%%%%%%%%%%%%%%%%%%%%%%%%%%%%%%%%%%%%%%%

\bibliography{Bibliothek}% Produces the bibliography via BibTeX.

\end{document}